\documentclass[twocolumn, dvipsnames]{aastex63}
\usepackage{amsmath}
\usepackage{longtable}

\received{}
\revised{}
\accepted{}
\shorttitle{The  ALeRCE  Light  Curve  Classifier }
\shortauthors{S\'anchez-S\'aez et al.}
\graphicspath{{./}{figures/}}

\begin{document}

\title{Alert Classification for the ALeRCE Broker System: The Light Curve Classifier}

\correspondingauthor{P. S\'anchez-S\'aez}
\email{pasanchezsaez@gmail.com}

\author[0000-0003-0820-4692]{P. S\'anchez-S\'aez}
\affiliation{Millennium Institute of Astrophysics (MAS), Nuncio Monse{\~{n}}or S{\'{o}}tero Sanz 100, Providencia, Santiago, Chile} 
\affiliation{Instituto de Astrof{\'{\i}}sica, Facultad de F{\'{i}}sica, Pontificia Universidad Cat{\'{o}}lica de Chile, Casilla 306, Santiago 22, Chile}
\affiliation{Faculty of Engineering and Sciences, Universidad Adolfo Iba\~nez, Diagonal Las Torres 2700, Pe\~nalol\'en, Santiago, Chile}

\author[0000-0003-3627-0216]{I. Reyes}
\affiliation{Millennium Institute of Astrophysics (MAS), Nuncio Monse{\~{n}}or S{\'{o}}tero Sanz 100, Providencia, Santiago, Chile} 
\affiliation{Center for Mathematical Modeling, Universidad de Chile, Beauchef 851, North building, 7th floor, Santiago 8320000, Chile}
\affiliation{Department of Electrical Engineering, Universidad de Chile, Av. Tupper 2007, Santiago 8320000, Chile}

\author{C. Valenzuela}
\affiliation{Center for Mathematical Modeling, Universidad de Chile, Beauchef 851, North building, 7th floor, Santiago 8320000, Chile}
\affiliation{Millennium Institute of Astrophysics (MAS), Nuncio Monse{\~{n}}or S{\'{o}}tero Sanz 100, Providencia, Santiago, Chile} 
\affiliation{Data Observatory, Diagonal Las Torres 2640, Pe\~nalol\'en, Santiago, Chile}
\affiliation{Faculty of Engineering and Sciences, Universidad Adolfo Iba\~nez, Diagonal Las Torres 2700, Pe\~nalol\'en, Santiago, Chile}

\author[0000-0003-3459-2270]{F. F\"orster}
\affiliation{Departmento de Matem\'aticas, Facultad de Ciencia, Universidad de Santiago de Chile, Av. Libertador Bernardo O’Higgins 3663. Estaci\'on Central, Santiago, Chile}
\affiliation{Millennium Institute of Astrophysics (MAS), Nuncio Monse{\~{n}}or S{\'{o}}tero Sanz 100, Providencia, Santiago, Chile} 
\affiliation{Departamento de Astronom\'ia, Universidad de Chile, Casilla 36D, Santiago, Chile}

\author[0000-0003-4723-9660]{S. Eyheramendy}
\affiliation{Faculty of Engineering and Sciences, Universidad Adolfo Iba\~nez, Diagonal Las Torres 2700, Pe\~nalol\'en, Santiago, Chile}
\affiliation{Millennium Institute of Astrophysics (MAS), Nuncio Monse{\~{n}}or S{\'{o}}tero Sanz 100, Providencia, Santiago, Chile}

\author[0000-0002-1835-7433]{F. Elorrieta}
\affiliation{Departmento de Matem\'aticas, Facultad de Ciencia, Universidad de Santiago de Chile, Av. Libertador Bernardo O’Higgins 3663. Estaci\'on Central, Santiago, Chile}
\affiliation{Millennium Institute of Astrophysics (MAS), Nuncio Monse{\~{n}}or S{\'{o}}tero Sanz 100, Providencia, Santiago, Chile}

\author[0000-0002-8686-8737]{F. E. Bauer}
\affiliation{Instituto de Astrof{\'{\i}}sica, Facultad de F{\'{i}}sica, Pontificia Universidad Cat{\'{o}}lica de Chile, Casilla 306, Santiago 22, Chile} 
\affiliation{Centro de Astroingenier{\'{\i}}a, Pontificia Universidad Cat{\'{o}}lica de Chile, Av. Vicu\~{n}a Mackenna 4860, 7820436 Macul, Santiago, Chile} 
\affiliation{Millennium Institute of Astrophysics (MAS), Nuncio Monse{\~{n}}or S{\'{o}}tero Sanz 100, Providencia, Santiago, Chile}
\affiliation{Space Science Institute, 4750 Walnut Street, Suite 205, Boulder, Colorado 80301} 

\author[0000-0002-2720-7218]{G. Cabrera-Vives}
\affiliation{Department of Computer Science, University of Concepc\'on, Edmundo Larenas 219, Concepci\'on, Chile}
\affiliation{Millennium Institute of Astrophysics (MAS), Nuncio Monse{\~{n}}or S{\'{o}}tero Sanz 100, Providencia, Santiago, Chile} 

\author[0000-0001-9164-4722]{P. A. Est\'evez}
\affiliation{Department of Electrical Engineering, Universidad de Chile, Av. Tupper 2007, Santiago 8320000, Chile}
\affiliation{Millennium Institute of Astrophysics (MAS), Nuncio Monse{\~{n}}or S{\'{o}}tero Sanz 100, Providencia, Santiago, Chile} 

\author[0000-0001-6003-8877]{M. Catelan}
\affiliation{Instituto de Astrof{\'{\i}}sica, Facultad de F{\'{i}}sica, Pontificia Universidad Cat{\'{o}}lica de Chile, Casilla 306, Santiago 22, Chile} 
\affiliation{Centro de Astroingenier{\'{\i}}a, Pontificia Universidad Cat{\'{o}}lica de Chile, Av. Vicu\~{n}a Mackenna 4860, 7820436 Macul, Santiago, Chile} 
\affiliation{Millennium Institute of Astrophysics (MAS), Nuncio Monse{\~{n}}or S{\'{o}}tero Sanz 100, Providencia, Santiago, Chile}

\author[0000-0003-0006-0188]{G. Pignata}
\affiliation{ Departamento de Ciencias Fis\'icas, Universidad Andres Bello, Avda. Republica 252, Santiago, Chile}
\affiliation{Millennium Institute of Astrophysics (MAS), Nuncio Monse{\~{n}}or S{\'{o}}tero Sanz 100, Providencia, Santiago, Chile} 

\author[0000-0003-3541-1697]{P. Huijse}
\affiliation{Informatics Institute, Universidad Austral de Chile, General Lagos 2086, Valdivia, Chile}
\affiliation{Millennium Institute of Astrophysics (MAS), Nuncio Monse{\~{n}}or S{\'{o}}tero Sanz 100, Providencia, Santiago, Chile}

\author[0000-0001-7208-5101]{D. De Cicco}
\affiliation{Millennium Institute of Astrophysics (MAS), Nuncio Monse{\~{n}}or S{\'{o}}tero Sanz 100, Providencia, Santiago, Chile} 
\affiliation{Instituto de Astrof{\'{\i}}sica, Facultad de F{\'{i}}sica, Pontificia Universidad Cat{\'{o}}lica de Chile, Casilla 306, Santiago 22, Chile}

\author[0000-0001-5675-6323]{P. Ar\'evalo}
\affiliation{Instituto de F\'isica y Astronom\'ia, Facultad de Ciencias, Universidad de Valpara\'iso, Gran Bretana No. 1111, Playa Ancha, Valparaíso, Chile}

\author[0000-0003-4673-8791]{R. Carrasco-Davis}
\affiliation{Department of Electrical Engineering, Universidad de Chile, Av. Tupper 2007, Santiago 8320000, Chile}

\author{J. Abril}
\affiliation{European Southern Observatory (ESO), Alonso de C\'ordova 3107, Vitacura, Santiago, Chile}
\affiliation{Centro de Estudios de F\'isica del Cosmos de Arag\'on (CEFCA) - Unidad Asociada al CSIC, Plaza San Juan, 1, E-44001, Teruel, Spain}

\author[0000-0002-9740-9974]{R. Kurtev}
\affiliation{Instituto de F\'isica y Astronom\'ia, Facultad de Ciencias, Universidad de Valpara\'iso, Gran Bretana No. 1111, Playa Ancha, Valparaíso, Chile} 
\affiliation{Millennium Institute of Astrophysics (MAS), Nuncio Monse{\~{n}}or S{\'{o}}tero Sanz 100, Providencia, Santiago, Chile} 

\author[0000-0002-5936-7718]{J. Borissova}
\affiliation{Instituto de F\'isica y Astronom\'ia, Facultad de Ciencias, Universidad de Valpara\'iso, Gran Bretana No. 1111, Playa Ancha, Valparaíso, Chile} 
\affiliation{Millennium Institute of Astrophysics (MAS), Nuncio Monse{\~{n}}or S{\'{o}}tero Sanz 100, Providencia, Santiago, Chile} 

\author[0000-0002-2045-7134]{J. Arredondo}
\affiliation{Millennium Institute of Astrophysics (MAS), Nuncio Monse{\~{n}}or S{\'{o}}tero Sanz 100, Providencia, Santiago, Chile} 

\author{E. Castillo-Navarrete }
\affiliation{Millennium Institute of Astrophysics (MAS), Nuncio Monse{\~{n}}or S{\'{o}}tero Sanz 100, Providencia, Santiago, Chile} 
\affiliation{Center for Mathematical Modeling, Universidad de Chile, Beauchef 851, North building, 7th floor, Santiago 8320000, Chile}

\author{D. Rodriguez}
\affiliation{Millennium Institute of Astrophysics (MAS), Nuncio Monse{\~{n}}or S{\'{o}}tero Sanz 100, Providencia, Santiago, Chile} 

\author{D. Ruz-Mieres}
\affiliation{Millennium Institute of Astrophysics (MAS), Nuncio Monse{\~{n}}or S{\'{o}}tero Sanz 100, Providencia, Santiago, Chile} 
\affiliation{Center for Mathematical Modeling, Universidad de Chile, Beauchef 851, North building, 7th floor, Santiago 8320000, Chile}

\author[0000-0002-7003-5087]{A. Moya}
\affiliation{Center for Mathematical Modeling, Universidad de Chile, Beauchef 851, North building, 7th floor, Santiago 8320000, Chile}
\affiliation{Millennium Institute of Astrophysics (MAS), Nuncio Monse{\~{n}}or S{\'{o}}tero Sanz 100, Providencia, Santiago, Chile} 

\author{L. Sabatini-Gacit\'ua}
\affiliation{Center for Mathematical Modeling, Universidad de Chile, Beauchef 851, North building, 7th floor, Santiago 8320000, Chile}
\affiliation{Millennium Institute of Astrophysics (MAS), Nuncio Monse{\~{n}}or S{\'{o}}tero Sanz 100, Providencia, Santiago, Chile} 

\author{C. Sep\'ulveda-Cobo}
\affiliation{Center for Mathematical Modeling, Universidad de Chile, Beauchef 851, North building, 7th floor, Santiago 8320000, Chile}
\affiliation{Millennium Institute of Astrophysics (MAS), Nuncio Monse{\~{n}}or S{\'{o}}tero Sanz 100, Providencia, Santiago, Chile} 

\author{E. Camacho-I\~niguez}
\affiliation{Instituto de Astrof{\'{\i}}sica, Facultad de F{\'{i}}sica, Pontificia Universidad Cat{\'{o}}lica de Chile, Casilla 306, Santiago 22, Chile}



\begin{abstract}

We present the first version of the ALeRCE (Automatic Learning for the Rapid Classification of  Events) broker light curve classifier. ALeRCE is currently processing the Zwicky Transient Facility (ZTF) alert stream, in preparation for the Vera C. Rubin Observatory. The ALeRCE light curve classifier uses variability features computed from the ZTF alert stream, and colors obtained from AllWISE and ZTF photometry. We apply a Balanced Random Forest algorithm with a two-level scheme, where the top level classifies each source as periodic, stochastic, or transient, and the bottom level further resolves each of these hierarchical classes, amongst 15 total classes. This classifier corresponds to the first attempt to classify multiple classes of stochastic variables (including core- and host-dominated active galactic nuclei, blazars, young stellar objects, and cataclysmic variables) in addition to different classes of periodic and transient sources, using real data. We created a labeled set using various public catalogs (such as the Catalina Surveys and {\em Gaia} DR2 variable stars catalogs, and the Million Quasars catalog), and we classify all objects with $\geq6$ $g$-band or $\geq6$ $r$-band detections in ZTF (868,371 sources as of 2020/06/09), providing updated classifications for sources with new alerts every day. For the top level we obtain macro-averaged precision and recall scores of 0.96 and 0.99, respectively, and for the bottom level we obtain macro-averaged precision and recall scores of 0.57 and 0.76, respectively. Updated classifications from the light curve classifier can be found at the \href{http://alerce.online}{ALeRCE Explorer website}.

\end{abstract}



\section{Introduction}\label{intro}

Brightness variations of astrophysical objects offer key insights into their physical emission mechanisms and related phenomena. In stars, pulsations, both radial and non-radial, can result from a thermodynamic engine operating in their partial ionization layers, when stars are located inside one of the several so-called instability strips that are found in the Hertzsprung-Russell diagram. Eruptive events can be generated by material being lost from a star, or occasionally accreted onto it, as is typical in protostars and young stellar objects (YSOs). Explosive events can occur when material is accreted onto compact objects, such as white dwarfs in the case of cataclysmic variables (CVs) or neutron stars in the case of X-ray binaries, or star mergers. Brightness changes can also originate from the rotation of stars, caused by surface features such as starspots, and/or by stars' ellipsoidal shapes. Finally, eclipses can occur, depending on the observer's line-of-sight, due to the presence of binary companions, planets, and/or other circumstellar material. These and other classes of stellar variability are reviewed and summarized, for instance, in \citet{CS15}, where extensive additional references can be found. In addition, there are a wide array of transients such as kilonovae \citep{Metzger10}, supernovae (SNe; \citealt{Woosley02}), and tidal disruption events, which are beacons of destructive episodes in the life of a star \citep{Komossa15}. Galaxies, in turn, can also present a wide array of variability phenomena. In those hosting strongly accreting massive black holes (BHs), for instance, variations develop due to the stochastic nature of the accretion disk, corona, and jet emission, potentially related to both the BH properties and the structure of the material in the immediate vicinity (e.g., \citealt{MacLeod10,Caplar17,Sanchez-Saez18}).

To study the variability of individual objects in detail and use this information to probe different physical models, observations over a wide range of timescales are required. Hence, long and intensive campaigns of a large number of targets are crucial. In recent years surveys covering a significant part of the sky, revisiting the same regions on timescales from days to years, and containing a large sample of serendipitous objects, are now becoming available as predecessors of the Vera C. Rubin Observatory Legacy Survey of Space and Time (LSST; \citealt{LSST}). 

Among these is the Zwicky Transient Facility (ZTF; \citealt{Bellm14,Bellm19}), which had first light in 2017 and employs a powerful 47\,deg$^{2}$ field-of-view camera mounted on the Samuel Oschin 48-inch Schmidt telescope. ZTF is designed to image the entire northern sky every three nights and scan the plane of the Milky Way twice each night to a limiting magnitude of 20.5 in $gri$, thus enabling a wide variety of novel multiband time series studies, in preparation for the LSST.

LSST, which aims for first light in 2022, will revolutionize time domain astronomy, enabling for the first time the study of transient and variable objects over long periods of time ($\sim 10$ years) with $\gtrsim$1000 visits, down to very faint magnitudes ($r \sim24.5$ for single images of the entire sky every 3 days, $\sim26.1$ for yearly stacks, and $\sim27.5$ at full depth; $5 \sigma$), over a large sky area ($>$18,000 deg$^2$). 

Given the large number of sources that ZTF and LSST will observe ($\sim 1$--40 billion objects), it is critical to develop reliable and efficient variability-based selection techniques. This new information allows us to see through degeneracies which might exist from color characterization alone. These selection techniques should ideally take advantage of the multiband light curves provided by surveys like LSST and ZTF, and separate different subclasses of variable and transient objects without the need for optical spectra, which are still quite expensive to obtain for such large samples.

This new generation of large etendue survey telescopes has demonstrated a growing need for sophisticated astronomical alert processing systems  (i.e., systems that are able to detect changes in the sky of an astrophysical origin). These systems involve the real-time processing of data for alert generation, real-time annotation and classification of alerts (up to 40 million events per night) and real-time reaction to interesting alerts using available astronomical resources (e.g., via Target Observation Managers, or TOMs). In order to use these resources intelligently and efficiently, the astronomical community has been developing a new generation of alert filtering systems known as ``brokers''. One such community broker is the project ALeRCE (Automatic Learning for the Rapid Classification of Events; \citealt{Forster20}). ALeRCE is an initiative led by an interdisciplinary and inter-institutional group of scientists from several institutions both in Chile and the United States. The main aim of ALeRCE is to facilitate the study of non-moving variable and transient objects. 

ALeRCE is currently processing the ZTF alert stream, providing classifications of different variable and transient objects, in preparation for the LSST era. Two classification models are currently available in the ALeRCE pipeline: a stamp classifier (or early classifier; \citealt{Carrasco-Davis20}), that uses a Convolutional Neural Network on the first detection stamp of a source to classify it among five broad classes, namely variable star, active galactic nuclei, SN, asteroid, or bogus; and a light curve classifier (or late classifier), that uses variability features computed from the light curves to classify each source into finer (currently 15) subclasses among three of the five broad classes. 

In this work we present the first version of the ALeRCE light curve classifier. This classifier uses several novel features (see Section \ref{features}), and  employs machine learning (ML) algorithms that can deal with the high class imbalance present in the data, following a two-level scheme. A key goal of ALeRCE is to provide fast classification of transient and variable objects in a highly scalable framework, and thus we only include in this model features that can be computed quickly, avoiding features that require more than one second to compute, based on the computational infrastructure currently at our disposal\footnote{Using 1 CPU per light curve with r5a.xlarge AWS instances} (for more details see \citealt{Forster20}). The main advantage of this classifier is that it can separate multiple classes of transient and variable objects, using features computed from real data, that would be measured from LSST data. Particularly, the light curve classifier can deal with multiple classes of stochastic variable objects (including core, host, and jet-dominated active galactic nuclei, YSOs, and CVs), which have been normally not included by previous classifiers that use real data and classify periodic and transient objects (e.g., \citealt{Richards09,Kim14,Nun16,Villar19}).

This work attempts to separate an unprecedentedly large number of classes (15) of both transients and variable objects using real data (as opposed to using only simulated data). Previous works using real data have mostly focused on selecting either a variety of variable stars classes (e.g., \citealt{Debosscher09,Richards12,Kim16,Elorrieta16,Rimoldini19,Hosenie19,Zorich20}), different classes of variable objects, including variable stars and active galactic nuclei (e.g., \citealt{Kim14,Nun16}), or different classes of transients \citep{Villar19}.

To the best of our knowledge three previous works have used real data to classify transients and variable objects, albeit considering a lower number of classes: \cite{Martinez-Palomera18} used data from the HiTS survey \citep{Forster16,Forster18} to classify eight transient, active galactic nuclei and variable star classes; \cite{Narayan18} used data from The Optical Gravitational Lensing Experiment (OGLE; \citealt{Udalski92}) and the Open Supernova Catalog (OSC; \citealt{Guillochon17}) to classify seven transient and variable star classes; and \cite{D'Isanto16} used Catalina Real-Time Transient Survey (CRTS; \citealt{Drake09}) data to classify six transient and variable object classes. Other works have tested techniques to classify different classes of variables and transients using synthetic data (e.g., \citealt{Boone19AJ}), or a combination of synthetic and real data (e.g., \citealt{Carrasco-Davis19}). 

In addition, this work is the first attempt to separate three different classes of active galactic nuclei (core-dominated or quasi-stellar objects, hereafter ``QSO''; host-dominated, hereafter ``AGN''; and jet-dominated, hereafter ``Blazar''). Previous works have mostly focused on separating active galactic nuclei from the rest (e.g., \citealt{Butler11,Peters15,PalanqueDelabrouille16,Sanchez-Saez19,DeCicco19}).

The paper is organized as follows. In Section \ref{data_section} we describe the data used for this work, the procedure for the light curve construction, as well as the taxonomy and the labeled set used to train the classifier. In Section \ref{features} we define the set of features used by the light curve classifier. In Section \ref{classifiers} we describe the different ML algorithms tested for the classifier. In Section \ref{results} we compare the performance of the different models, and report the results obtained for the labeled and unlabeled ZTF sets. Finally in Section \ref{discussion} we summarize the paper, provide conclusions, and discuss the challenges found during the development of the classifier and the future work.

\section{Data}\label{data_section}

\subsection{Reference Data}\label{data}

\begin{figure*}[ht!]
\begin{center}

 \includegraphics[width=0.9\textwidth]{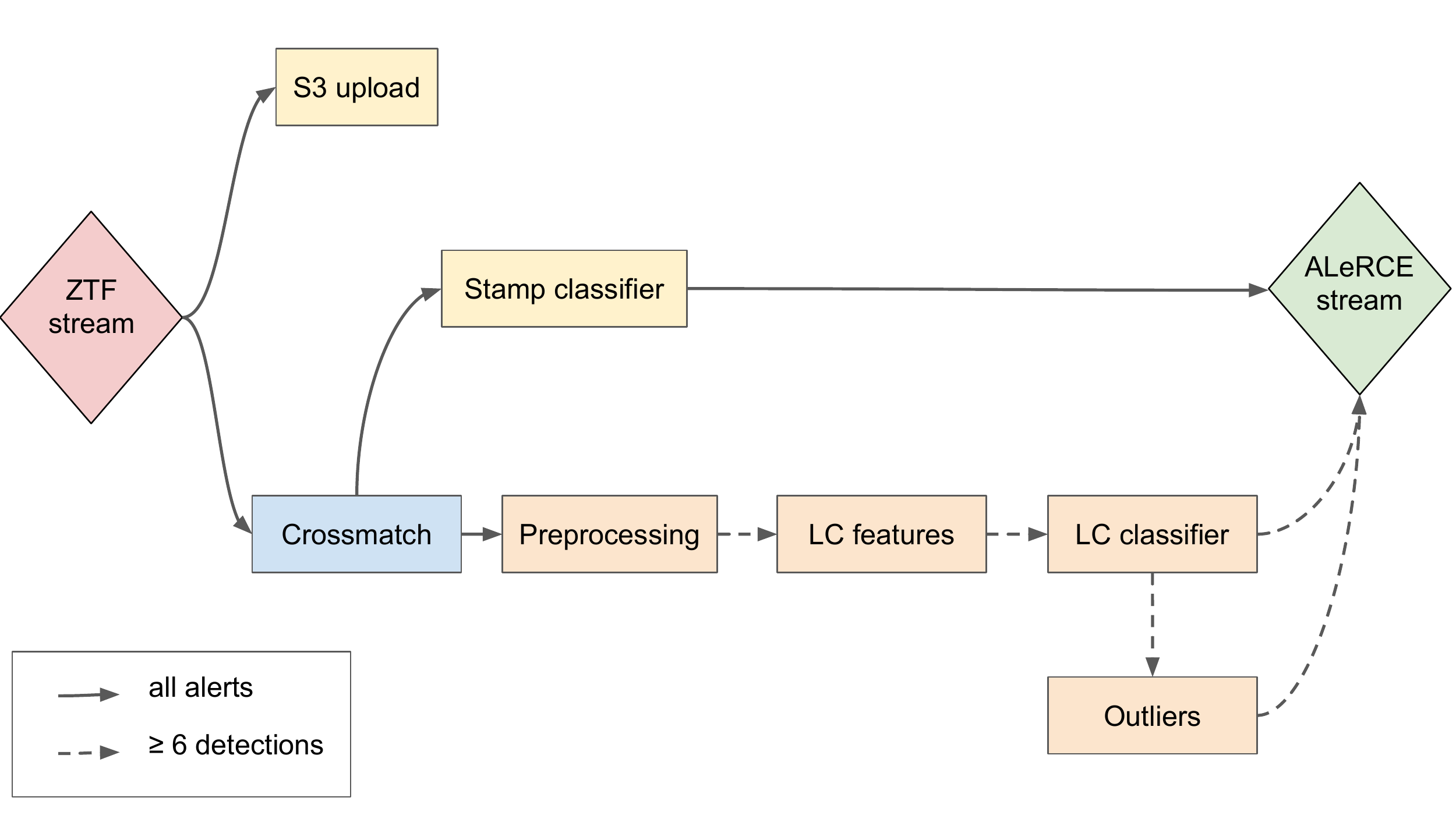}

\caption{A scheme of the ALeRCE pipeline. ZTF alerts are ingested using Kafka and a series of sequential and parallel steps are initiated. Alerts are stored in AWS S3, classified based on its image stamps, crossmatched with other catalogs, and their photometry corrected to take into account difference fluxes. Aggregated light curves are used to compute basic statistics (for internal use) and, if enough data points exist, features are computed, and a light curve and outlier classifiers are applied before sending an output stream. A PostgreSQL database is populated along the way, which can then be queried.   \label{figure:pipeline}}

\end{center}
\end{figure*}

ALeRCE  has been processing the public ZTF alert stream since May 2019, which includes $g$ and $r$ photometry. The ALeRCE pipeline is described in detail by \cite{Forster20}; for clarity, we provide a brief description of the light curve construction process.

The ALeRCE pipeline processes the ZTF Avro alert files.\footnote{For details, see \url{https://zwickytransientfacility.github.io/ztf-avro-alert/}.} These files contain metadata and contextual information for a single event, which are defined as a flux-transient, a reoccurring flux-variable, or a moving object \citep{Masci19}.  To construct light curves, the ALeRCE pipeline uses: the photometry of the difference-image and reference-image (detections);  possible non-detections associated with the target during the previous 30 days of the event (5$\sigma$ magnitude limit in the difference image based on PSF-fit photometry, called \texttt{diffmaglim} by ZTF); the real-bogus quality score reported by ZTF ($rb$, which ranges from 0 to 1, with values closer to 1 implying more reliable detections); and the morphological classification of the closest object obtained from PanSTARRS1 \citep{Tachibana18}. An overview of the pipeline is presented in Figure \ref{figure:pipeline}. In summary, the different stages of the pipeline are: 

\begin{itemize}
\item[1)] Ingestion: the ZTF public stream is ingested using Kafka.
\item[2)] S3 upload: the alert Avro packets are stored in AWS S3 for later access.
\item[3)] Crossmatch: the position of the alert is used to query external catalogs.
\item[4)] Stamp classifier: alerts from new objects are classified using their image cutouts (stamps).
\item[5)] Preprocessing: the photometry associated with a given alert is corrected to take into account the use of difference image fluxes (see details below), and simple statistics associated with the aggregated light curve are computed.
\item[6)] Light curve features: advanced light curve statistics (features) are computed when there are at least six detections in a given band.
\item[7)] Light curve classifier: the light curve classifier described in this work is applied.
\item[8)] Outliers: an outlier detection algorithm is applied. 
\item[9)] ALeRCE stream: the aggregated, annotated and classified light curves are reported in a Kafka stream.

\end{itemize}

 \begin{figure*}[ht]
\begin{center}

 \includegraphics[width=1\textwidth]{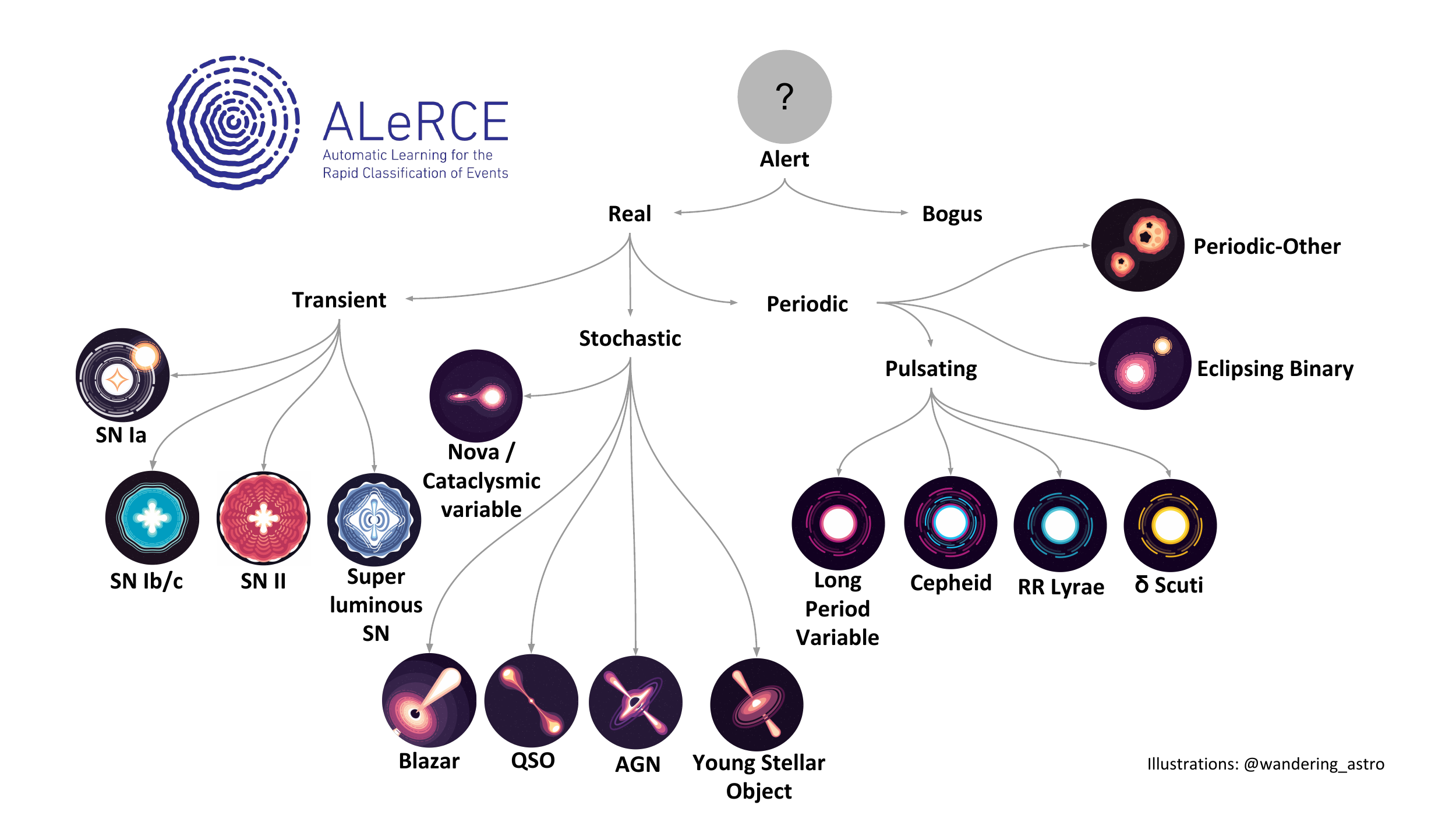}

\caption{Taxonomy tree used in the current version of the ALeRCE light curve classifier. \label{figure:taxonomy}}
\end{center}
\end{figure*}

In step 3) we are experimenting with several catalogs, but for this work we use the AllWISE\footnote{\url{http://wise2.ipac.caltech.edu/docs/release/allwise/}} public Source Catalog \citep{Wright10,Mainzer11}, invoking a match radius of 2 arcseconds, to obtain W1, W2, and W3 photometry (using magnitudes measured with profile-fitting photometry, e.g., \texttt{w1mpro}). 

The preprocessing procedure (step 5) is described in detail in  \cite{Forster20} (see Section A of their appendix). In particular, for the light curve classifier we use the corrected light curves (\texttt{lc\_corr}; $\hat m_{\rm sci}$ in \citealt{Forster20}) for sources whose closest source in the reference image coincides with the location of the alert (in a radius of 1.4 arcseconds). It is important to use the corrected light curves for variable sources, in order to take into account changes in the sign of the difference between the reference and the science images, or possible changes of the reference image. For the rest of the sources, the correction is not possible to perform, and thus  we use the light curves obtained using the difference images (\texttt{lc\_diff}; $m_{\rm diff}$ in \citealt{Forster20}), which correspond in general to transient sources. Note that this criteria does not require prior knowledge about the class of the source. Therefore, in this work we use \texttt{lc\_corr} for sources with available corrected light curves, otherwise we use \texttt{lc\_diff}, except for the Supernova parametric model features and some optical colors, for which we use \texttt{lc\_diff} for all the sources (for mode details see Section \ref{detections} and Appendix \ref{features_appendix}).

\begin{figure*}[htbp]
\begin{center}
\begin{tabular}{ccc}
\includegraphics[scale=0.47]{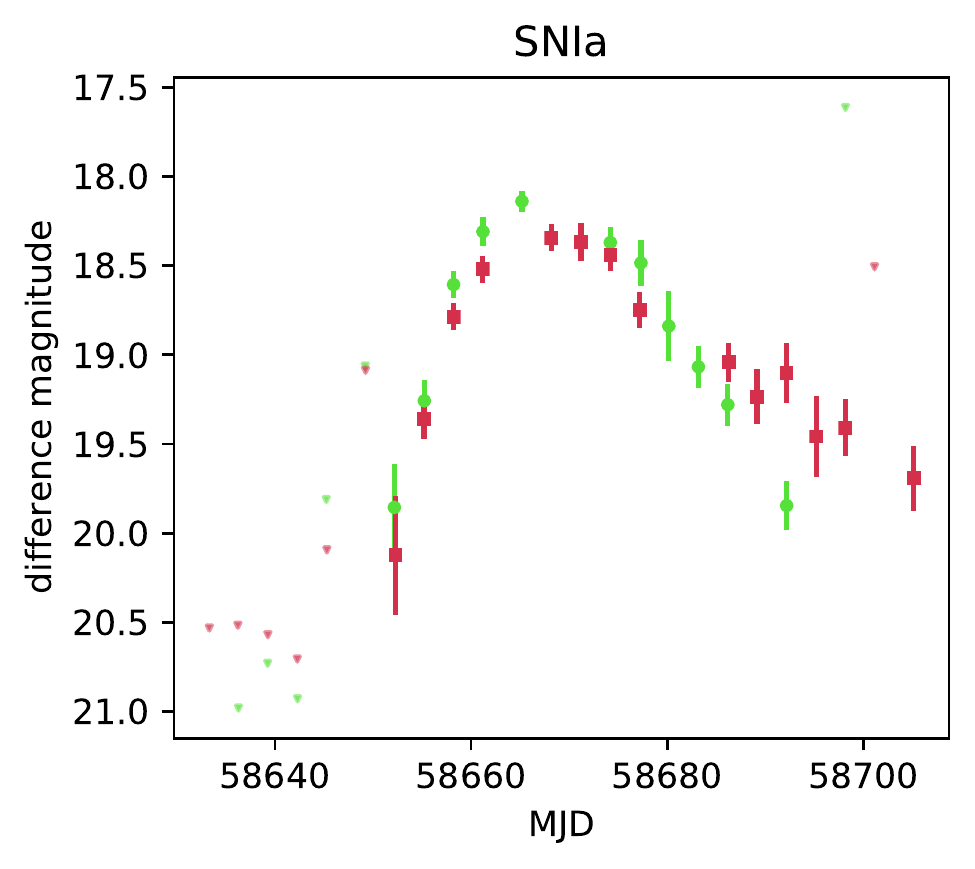} 
& \includegraphics[scale=0.47]{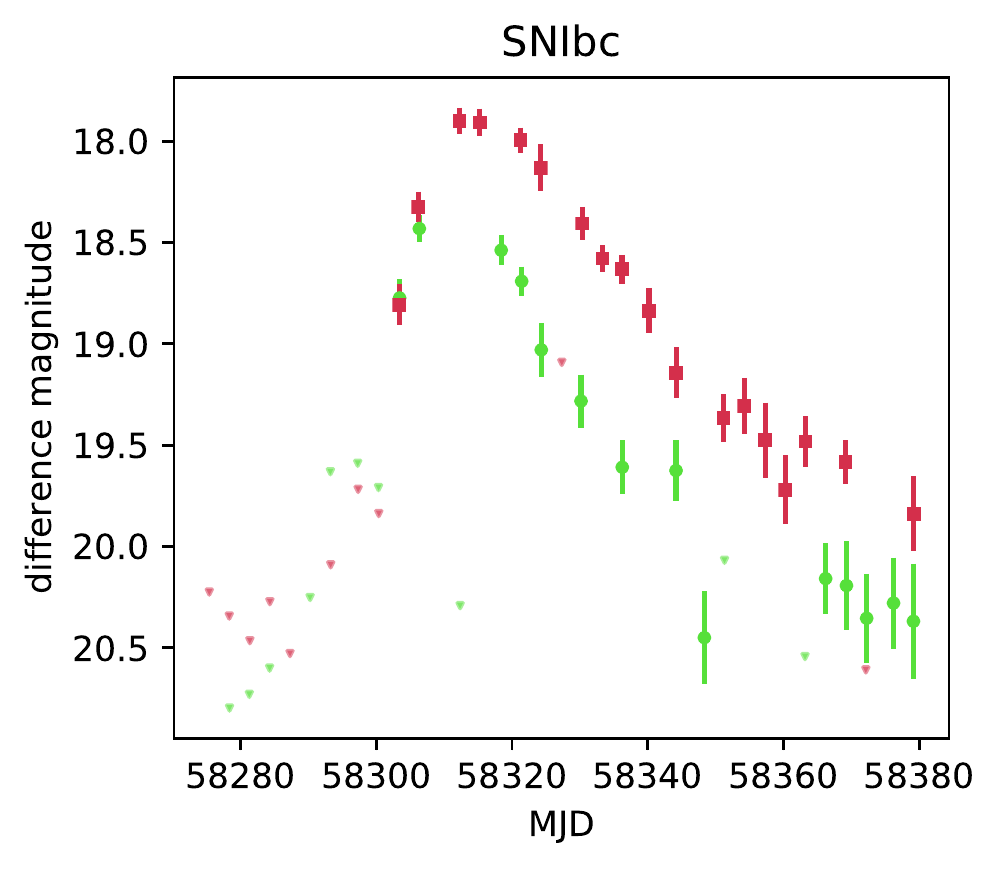} 
& \includegraphics[scale=0.47]{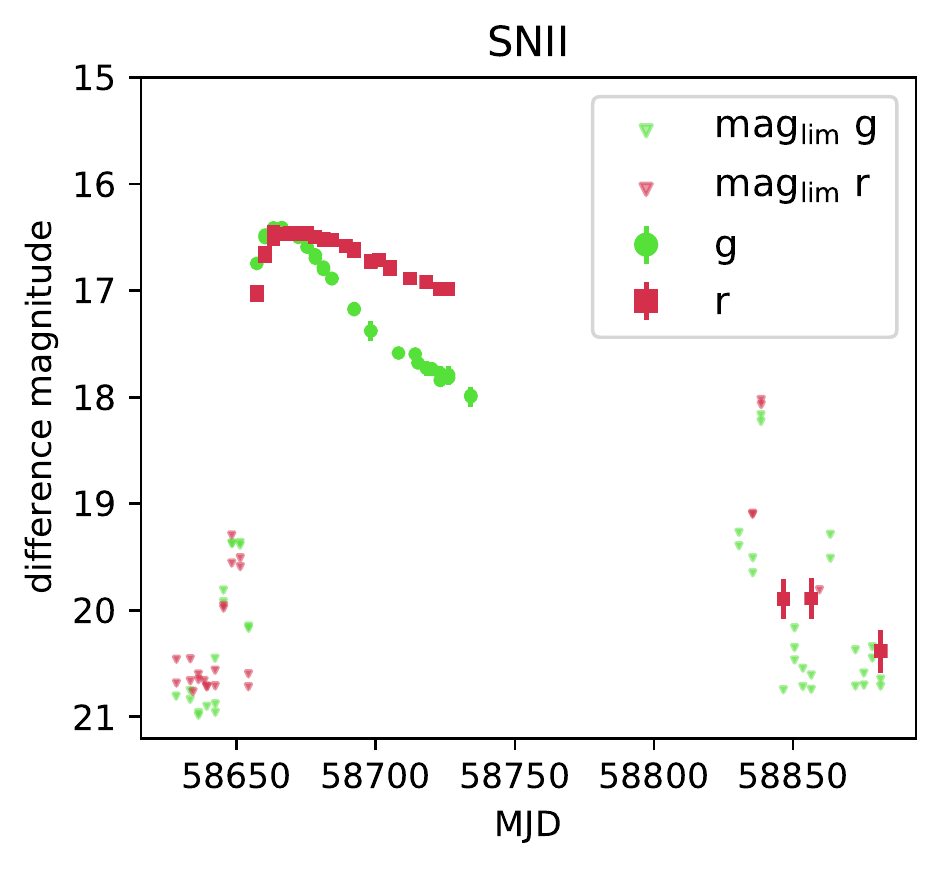} \\
\includegraphics[scale=0.47]{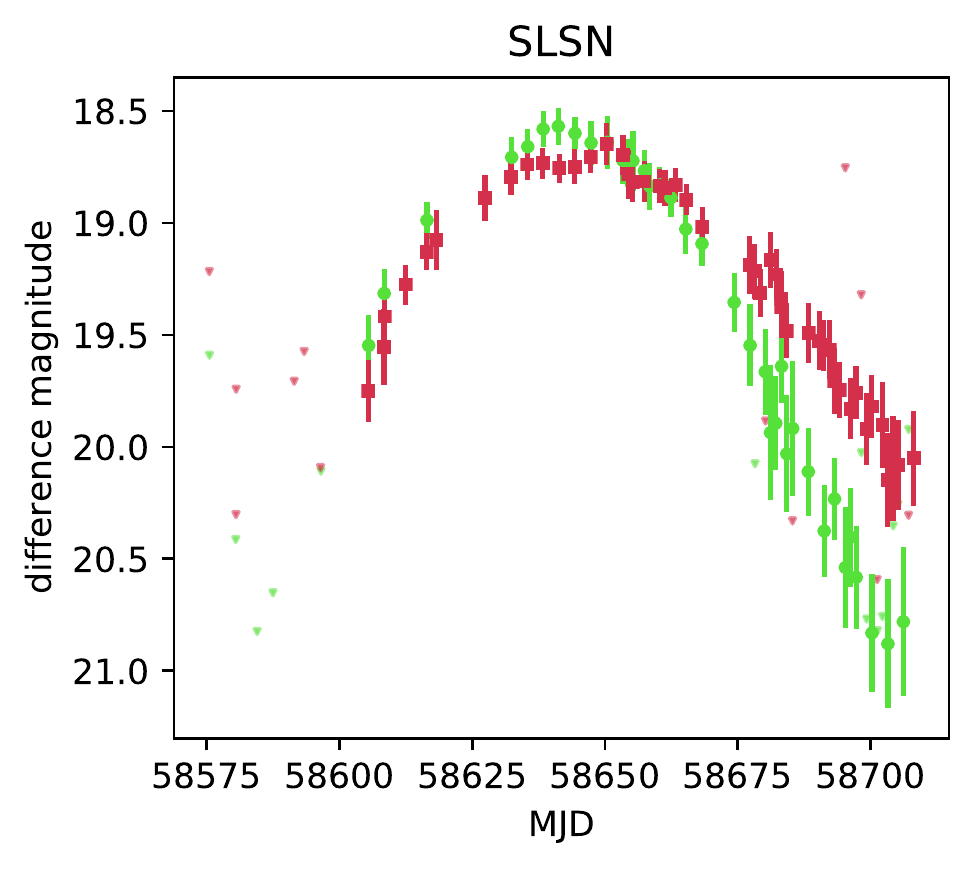} 
& \includegraphics[scale=0.47]{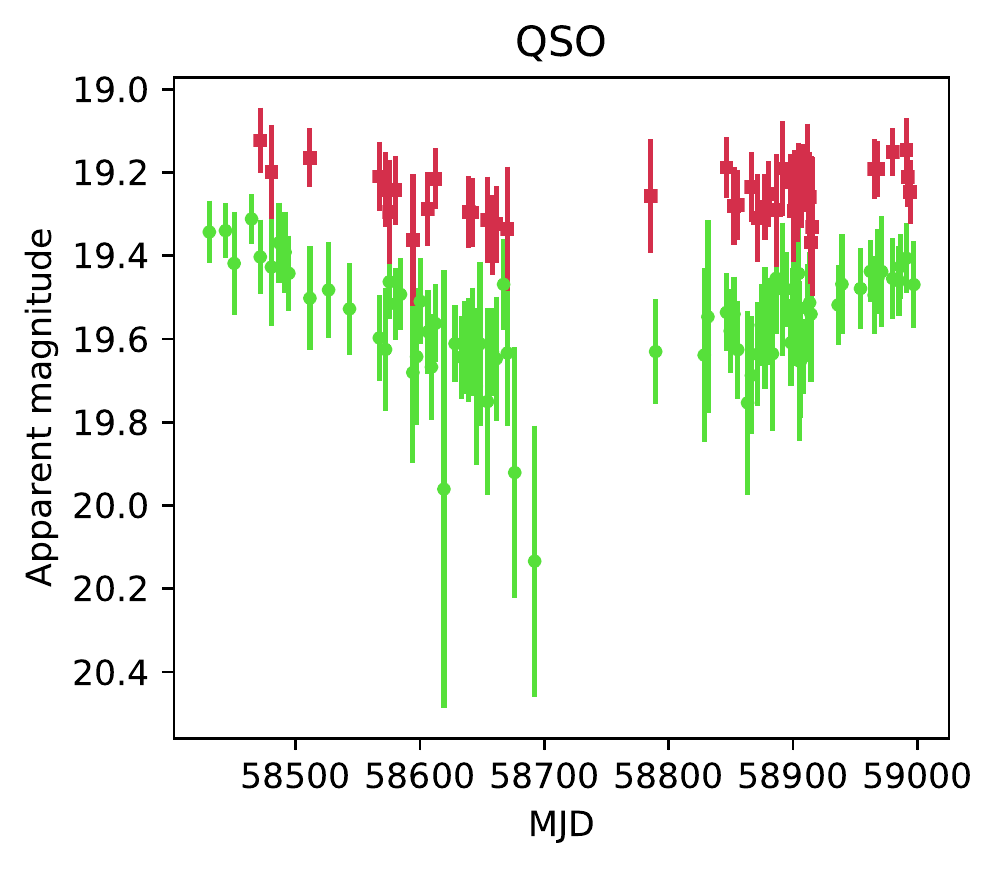} 
&\includegraphics[scale=0.47]{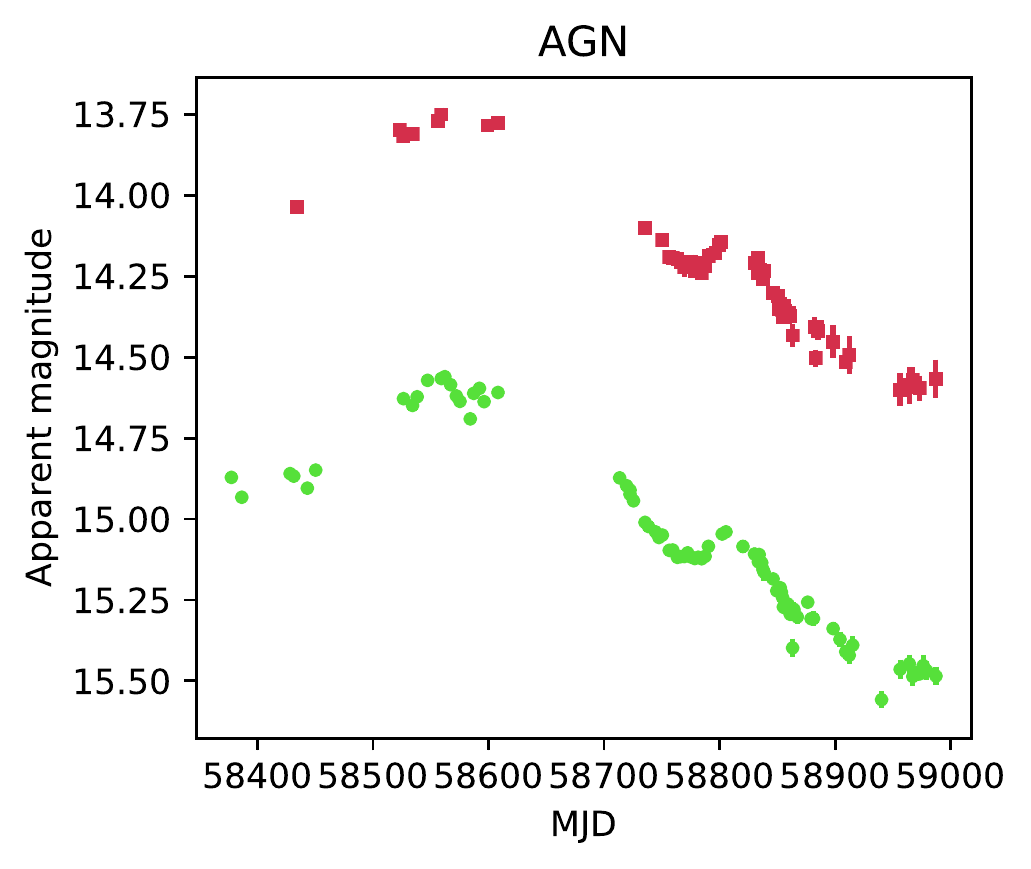} \\
\includegraphics[scale=0.47]{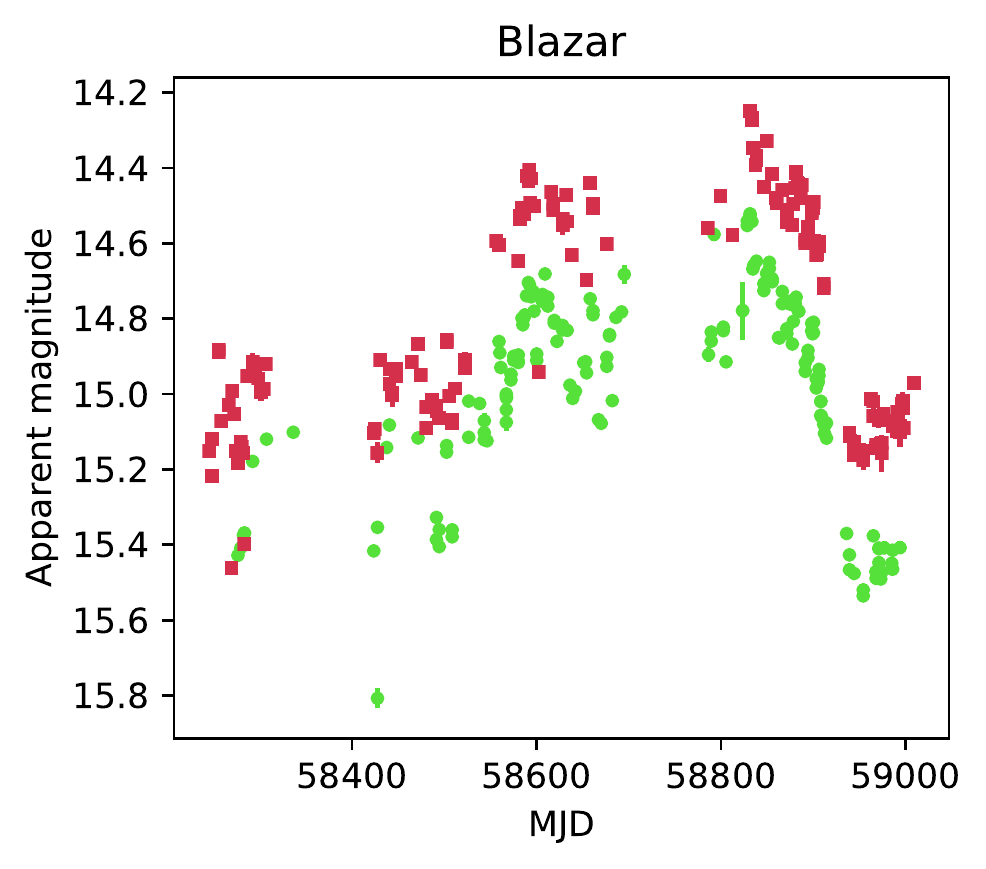} 
& \includegraphics[scale=0.47]{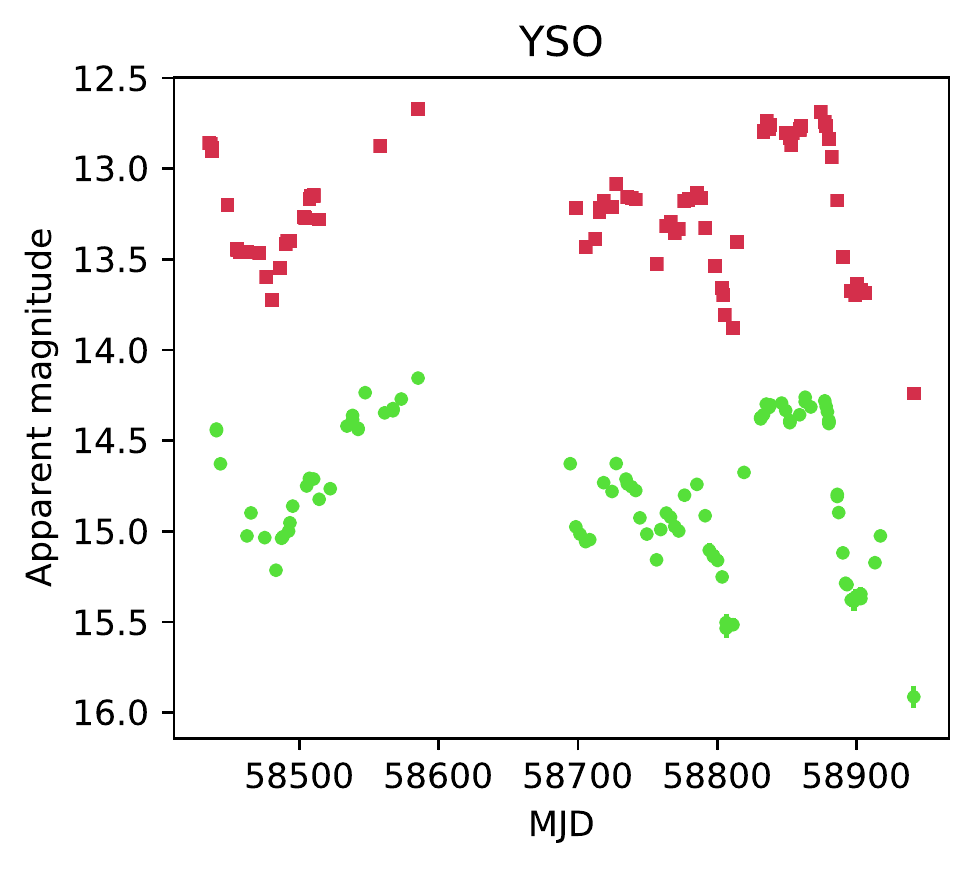}
& \includegraphics[scale=0.47]{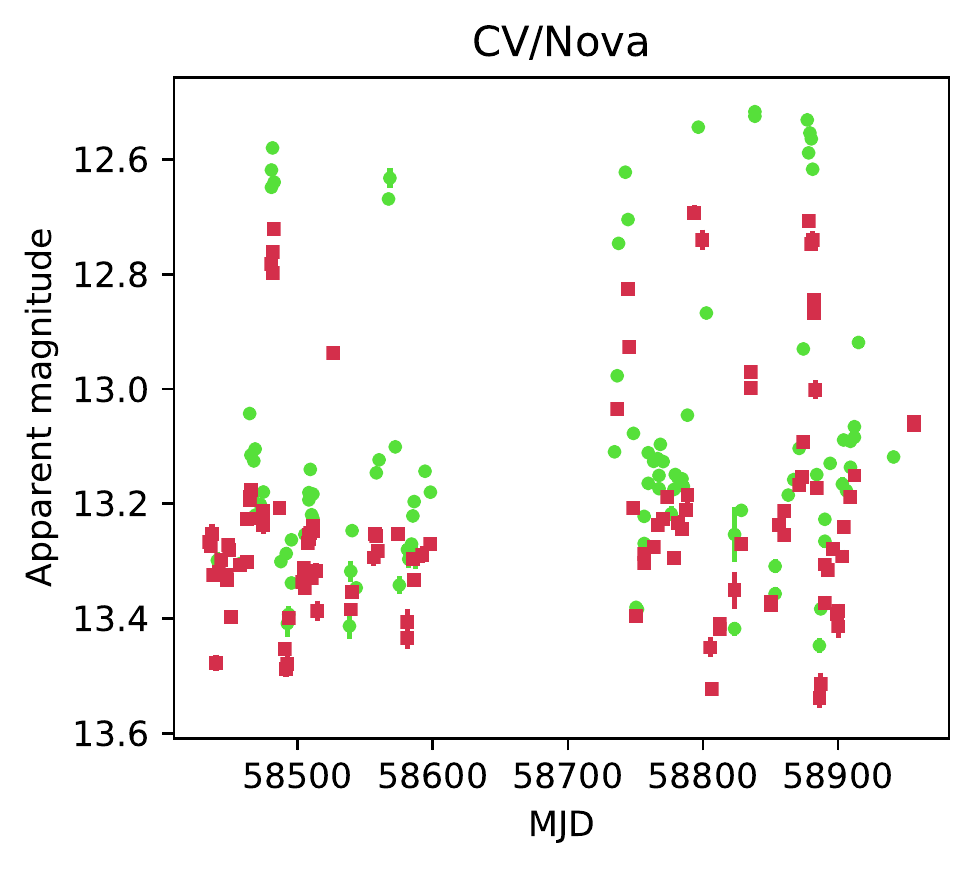} \\
 \includegraphics[scale=0.47]{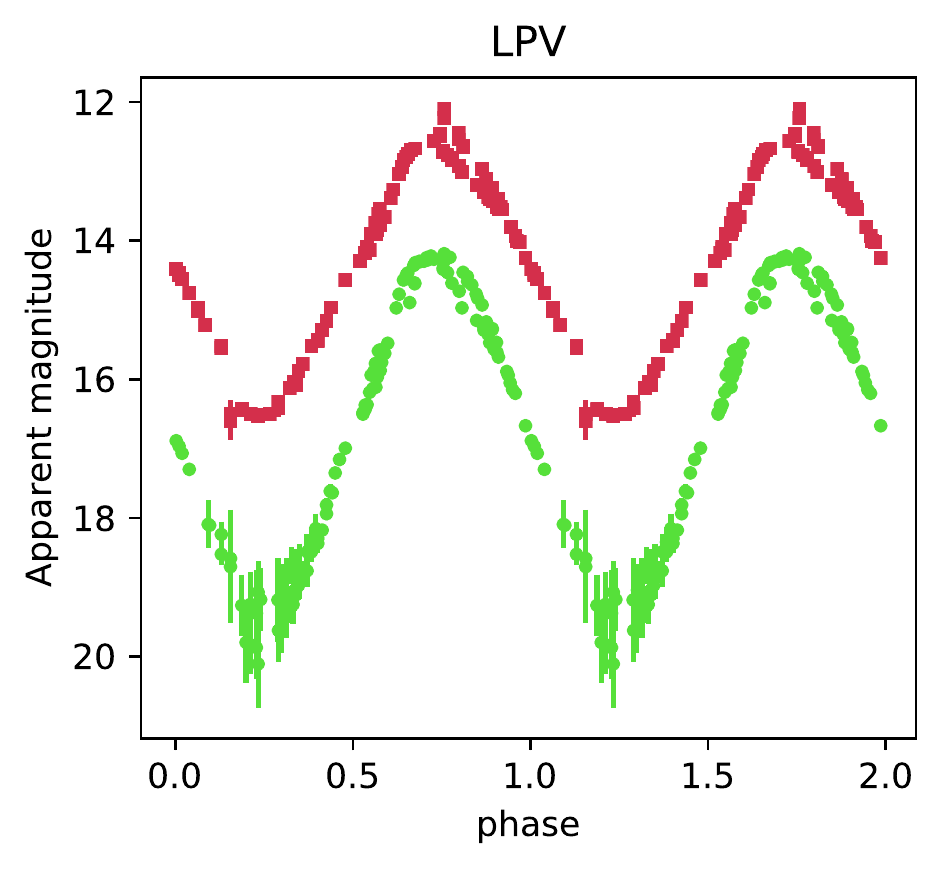} 
&\includegraphics[scale=0.47]{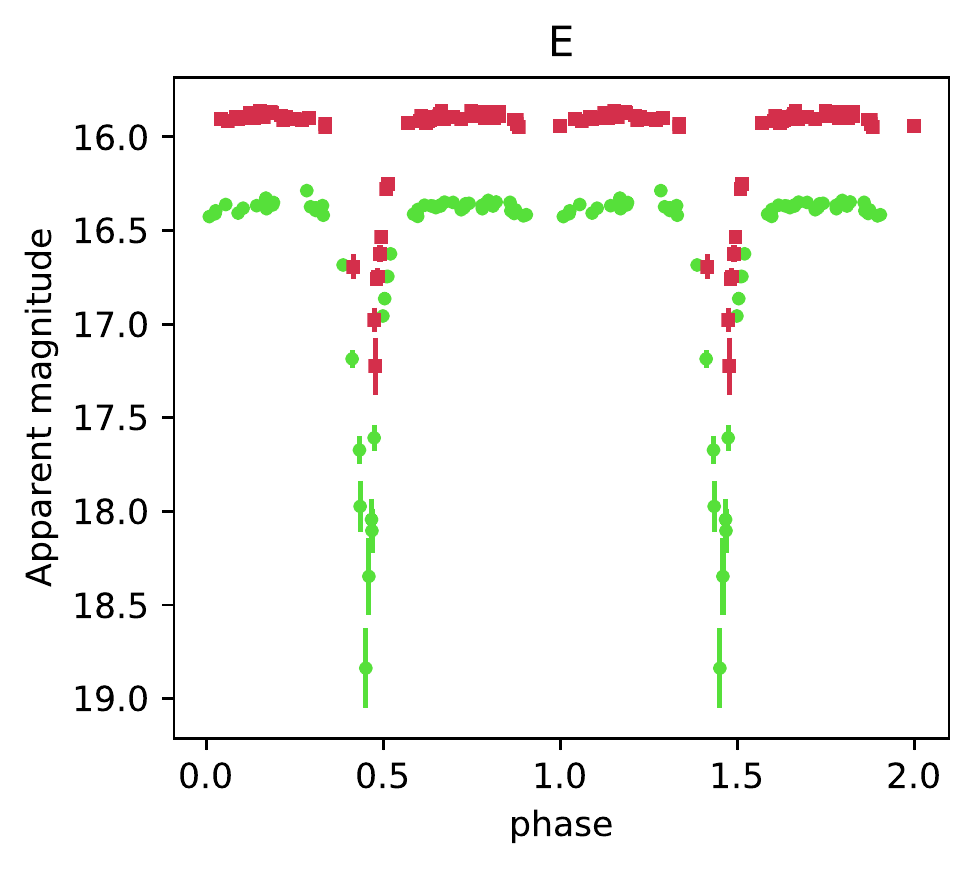} 
& \includegraphics[scale=0.47]{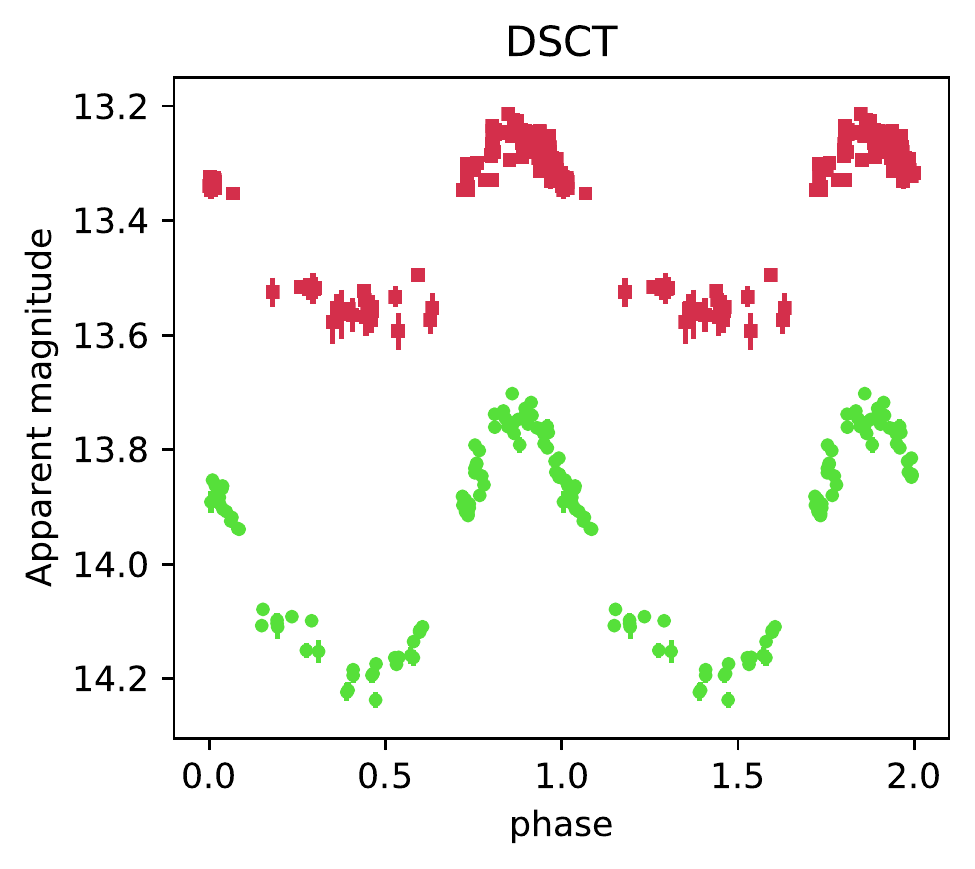} \\
\includegraphics[scale=0.47]{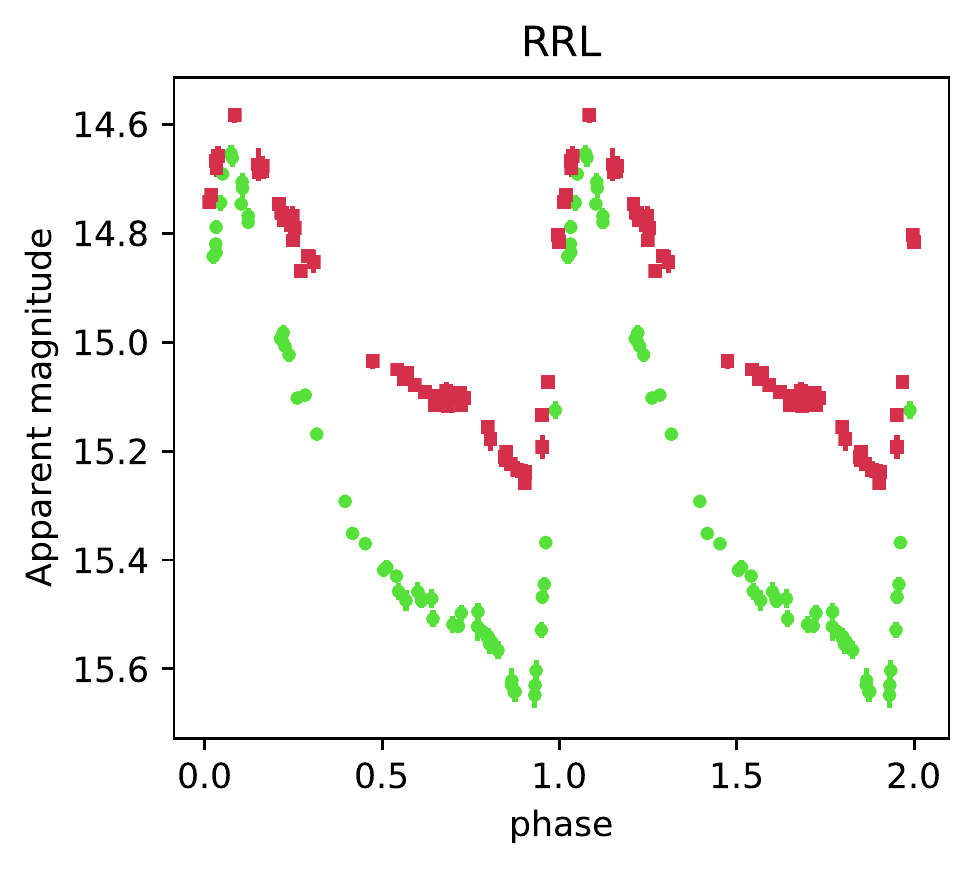} 
& \includegraphics[scale=0.47]{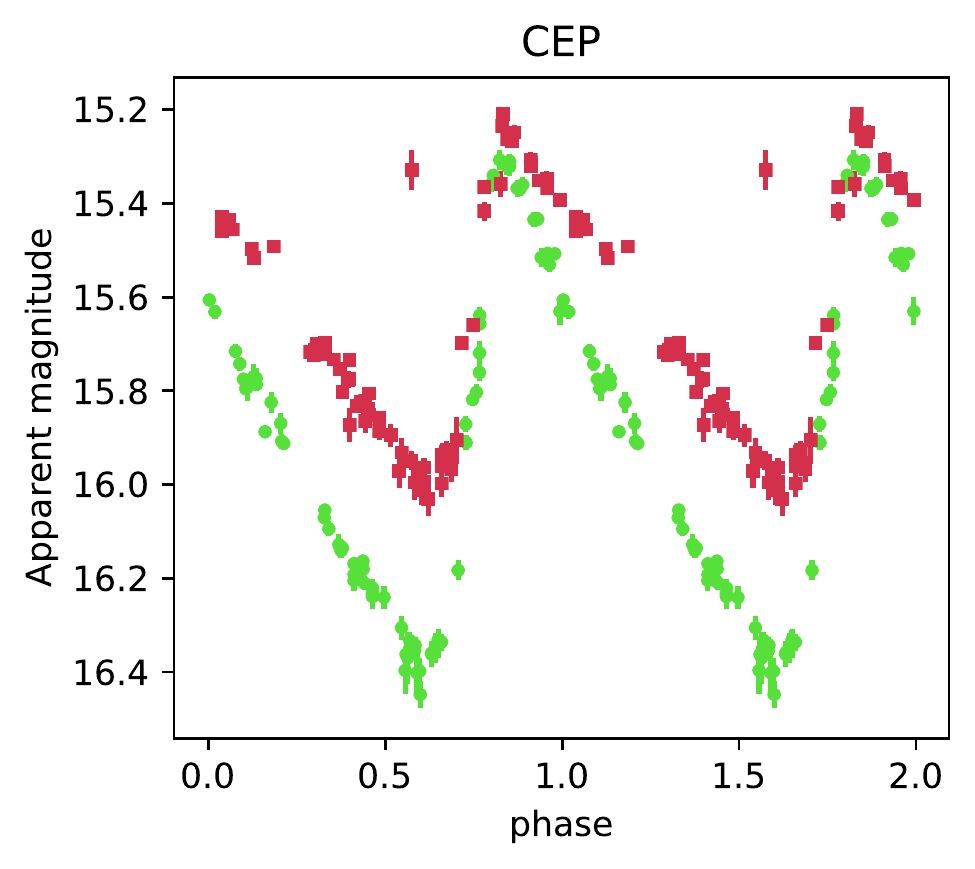} 
& \includegraphics[scale=0.47]{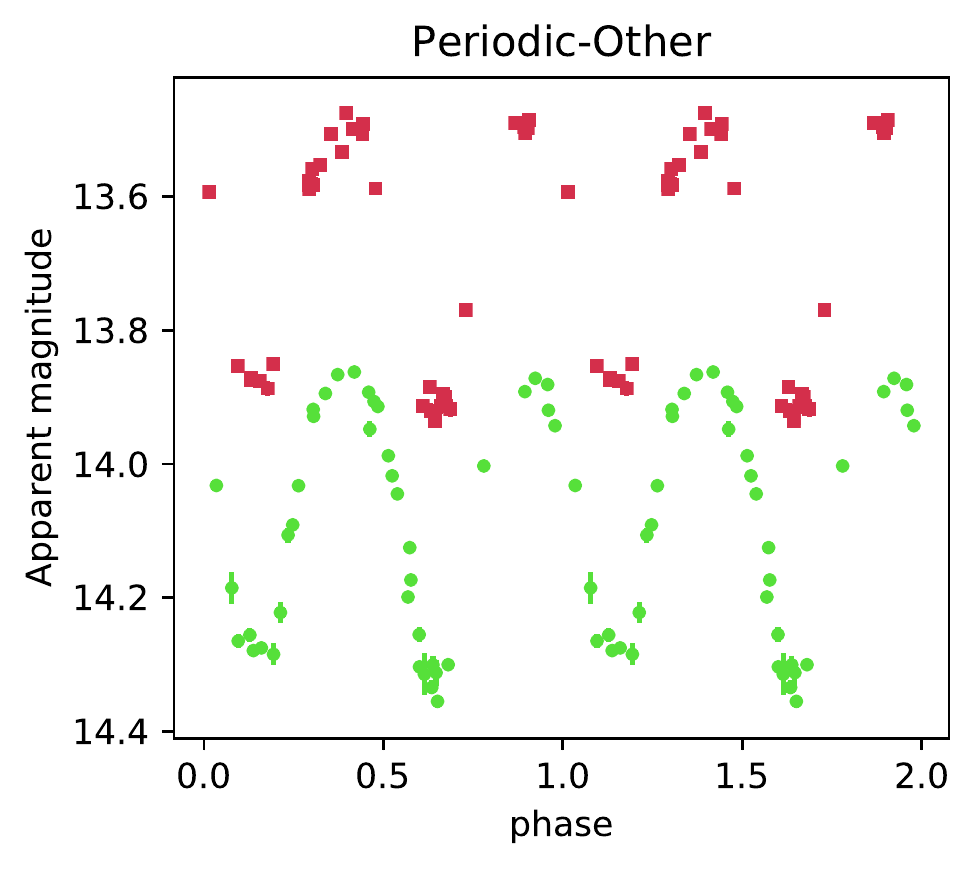} \\
\end{tabular}
\caption{Examples of ZTF light curves of the different classes considered by the light curve classifier. For the transient classes we show the difference magnitude light curves, for the stochastic classes we show the apparent magnitude light curves, and for the periodic classes we show the folded apparent magnitude light curves. Green circles and red squares indicate the $g$ and $r$ bands, respectively. Error bars indicate photometry associated with detections. Triangles denote limiting magnitudes and are shown only for the difference magnitude light curves.    \label{figure:lc_example}}
\end{center}
\end{figure*}

\subsection{Classification Taxonomy}\label{taxonomy}

The first version of the ALeRCE light curve classifier considers 15 subclasses of variable and transient objects, presented as a taxonomy tree defined by the ALeRCE collaboration in Figure \ref{figure:taxonomy}. The taxonomy is subdivided in a hierarchical fashion according to both the physical properties of each class and the empirical variability properties of the light curves, as follows (in parenthesis we indicate the class name used by the classifier):

\begin{itemize}

\item Transient: Type Ia supernova (SNIa), Type Ibc supernova (SNIbc), Type II supernova (SNII), and Super Luminous Supernova (SLSN);
\item Stochastic: Type 1 Seyfert galaxy (AGN; i.e., host-dominated active galactic nuclei), Type 1 Quasar (QSO; i.e., core-dominated active galactic nuclei), blazar (Blazar; i.e, beamed jet-dominated active galactic nuclei), Young Stellar Object (YSO), and Cataclysmic Variable/Nova (CV/Nova);  
\item Periodic: Long-Period Variable (LPV; includes regular, semi-regular, and irregular variable stars), RR Lyrae (RRL), Cepheid (CEP), eclipsing binary (E), $\delta$ Scuti (DSCT), and other periodic variable stars (Periodic-Other; this includes classes of variable stars that are not well represented in the labeled set, e.g., sources classified as miscellaneous, rotational or RS Canum Venaticorum-type systems in CRTS).

\end{itemize}

Figure \ref{figure:lc_example} shows examples of light curves of the different classes considered by the light curve classifier, obtained using ZTF data.

It is important to note that there are a number of less common classes which have not been separated out yet in the ALeRCE taxonomy tree, because the number of cross-matched objects in these classes is too low to train a good classification model (e.g. SNe IIb, TDEs, KNe, among others). There is a catch-all ``Periodic-Other'' class for periodic classes excluded in the taxonomy tree, but not for transient or stochastic classes, and thus, for the moment, these missing classes are being grouped into one or more of the existing ones.

\begin{table*}[htpb]
  \begin{center}
    \caption{Labeled set definition  }
    \label{table:trainingset}
    \begin{tabular}{|c|c|c|c|} 
   
   \hline

Hierarchical Class & Class & \# of sources\tablenotemark{$\dagger$} & Source Catalogs \\

\hline

\hline

&SNIa  &1272 (74.0\%)& TNS\\
 &SNIbc  &94 (5.5\%) & TNS \\
Transient &SNII  &328 (19.1\%) & TNS\\
&SLSN & 24 (1.4\%) & TNS \\

\cline{2-4}

& Total & 1718 & \\

\hline

& QSO &  26168 (75.4\%) & MILLIQUAS (sources with class ``Q'')\\
&AGN&  4667 (13.4\%) & Oh2015, MILLIQUAS (sources with class ``A'')\\
Stochastic &Blazar& 1267 (3.6\%)& ROMABZCAT, MILLIQUAS (sources with class ``B'') \\
&YSO &1740 (5.0\%) & SIMBAD\\
&CV/Nova&  871 (2.5\%) & TNS, ASASSN, JAbril \\

\cline{2-4}

& Total  & 34713 & \\

\hline
&LPV & 14076 (16.2\%) &  CRTS, ASASSN, {\em Gaia}DR2 \\
&E  &37901 (43.5\%)& CRTS, ASASSN, LINEAR \\
 &DSCT & 732 (0.8\%)&  CRTS, ASASSN, LINEAR, {\em Gaia}DR2 \\
Periodic &RRL  &32482 (37.3\%)& CRTS, ASASSN, LINEAR, {\em Gaia}DR2\\
&CEP & 618 (0.7\%)& CRTS, ASASSN \\
&Periodic-Other & 1256 (1.4\%)& CRTS, LINEAR\\

\cline{2-4}

& Total  & 87065 & \\

\hline
  \end{tabular}

\small{
\item $\dagger$ Values in parentheses correspond to the  fraction of sources of a given class (second column) within its corresponding hierarchical class (first column).}
 \end{center}

\end{table*}

\subsubsection{Labeled Set}\label{trainingset}

\begin{figure}[htbp]
\begin{center}

 \includegraphics[scale=0.65]{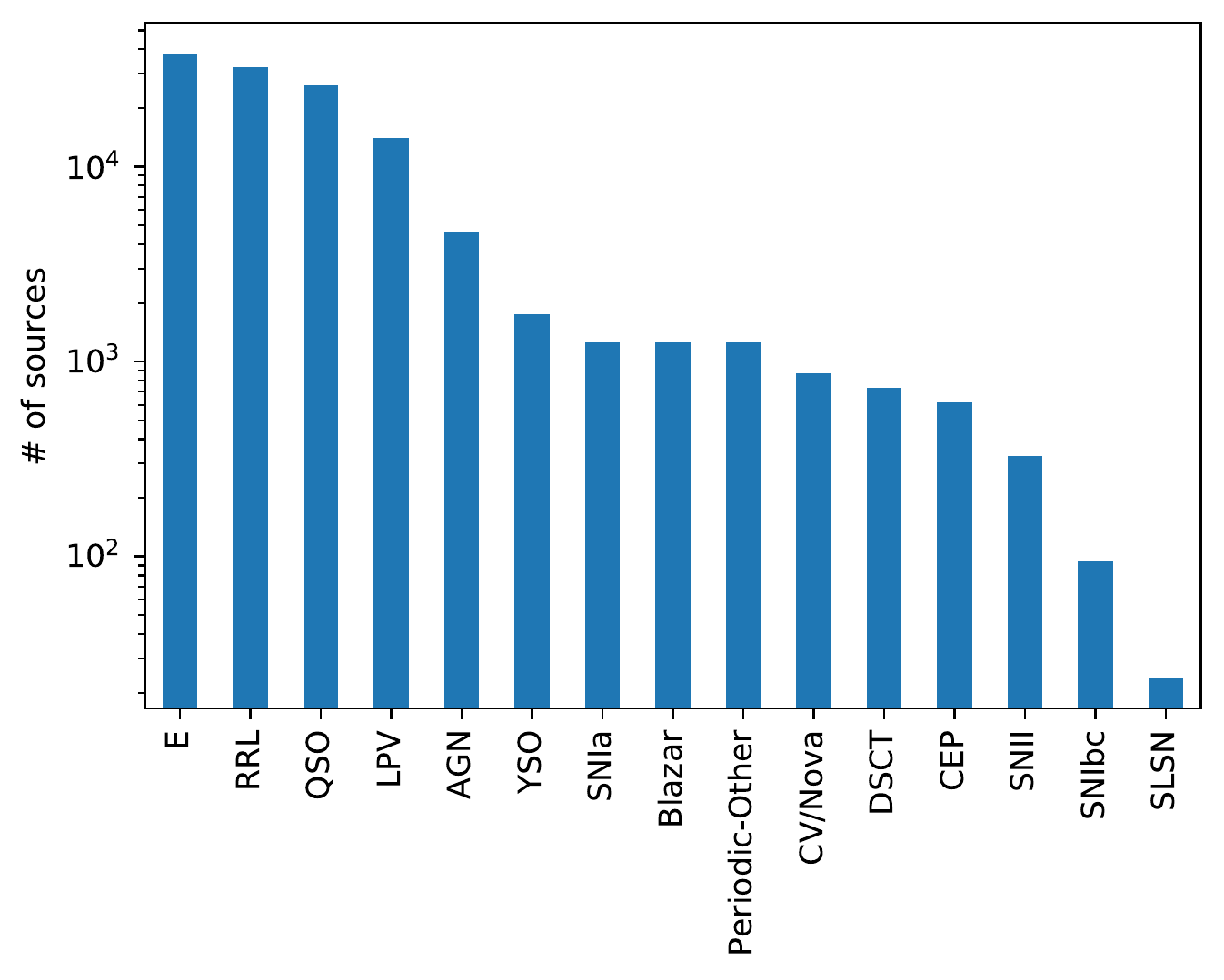}

\caption{Number of sources per class for the labeled set, as reported in Table \ref{table:trainingset}.\label{figure:num_classes_ls}}
\end{center}
\end{figure}

\begin{figure*}[htbp]
\begin{center}
 \includegraphics[scale=0.45]{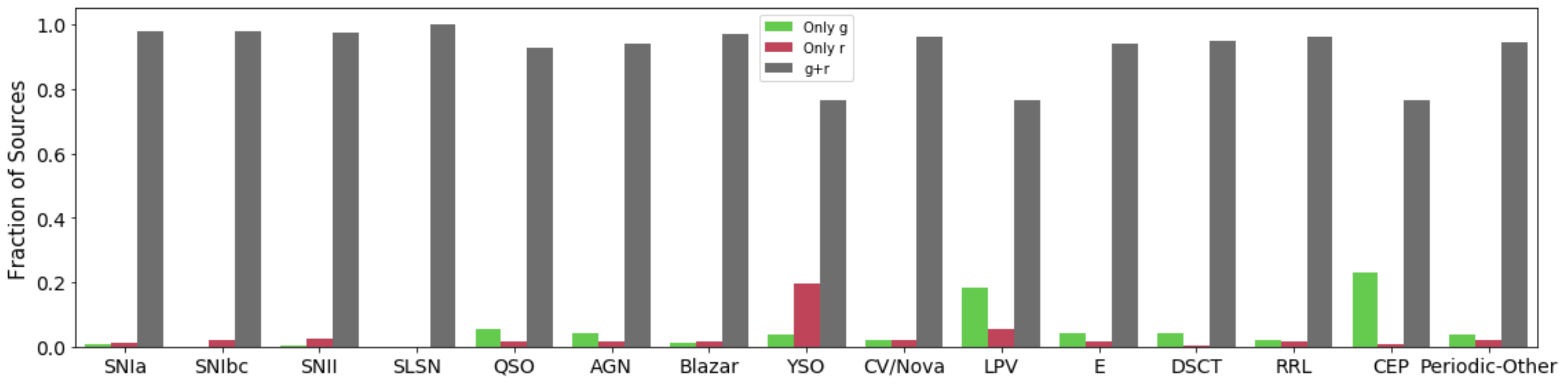}
\caption{For the sources in the labeled set, this figure shows the fraction of sources in each class with photometry: only in the $g$ band (green); only in the $r$ band (red); or in both bands (grey). The reasons for the non-uniformity of coverage may be physical (strongly red or blue source) or organizational (survey focused on one band only). For most classes, the vast majority of the sources ($\gtrsim$92\%) have photometric detections in both $g$ and $r$; the exceptions are the YSO, LPV, and CEP classes, where only 76\% of the sources have photometry in both bands. \label{figure:band_dist}}

\end{center}
\end{figure*}

The labeled set (i.e., the set of sources used to define the training and testing sets) for the light curve classifier was built using sources observed by ZTF (i.e., with ZTF light curves), with known labels obtained via spectroscopic and/or photometric analysis by previous works. Further description of the labeled set construction strategy can be found in \cite{Forster20}. We obtained labels from the following catalogs: the ASAS-SN catalogue of variable stars (ASASSN; \citealt{Jayasinghe18,Jayasinghe19,Jayasinghe19b,Jayasinghe20}), the Catalina Surveys Variable Star Catalogs (CRTS; \citealt{Drake14,Drake17}), LINEAR catalog of periodic light curves (LINEAR; \citealt{Palaversa13}), {\em {\em Gaia}} Data Release 2 ({\em Gaia}DR2; \citealt{Mowlavi18,Rimoldini19}), the Transient Name Server database (TNS)\footnote{\url{https://wis-tns.weizmann.ac.il/}}, the Roma-BZCAT Multi-Frequency Catalog of Blazars (ROMABZCAT; \citealt{Massaro15}), the Million Quasars Catalog (MILLIQUAS, version 6.4c, December 2019; \citealt{Flesch15,Flesch19}), the New Catalog of Type 1 AGNs (Oh2015; \citealt{Oh15}), and the SIMBAD database \citep{Wenger00}. Some additional CV labels were obtained from different catalogs (including \citealt{Ritter03}), compiled by \cite{Abril20} (JAbril). It is worth to mention that we only use the labels provided by these catalogs to build our datasets, and not any other information, such as periods, colors or redshifts. A catalog containing the labeled set can be downloaded at Zenodo: \dataset[10.5281/zenodo.4279623]{https://doi.org/10.5281/zenodo.4279623}.

Table~\ref{table:trainingset} lists the number of sources in the labeled set belonging to each class (with their correspondent percentages according to their hierarchical group), and the catalogs from which the classifications were obtained. Only sources with $\geq6$ detections in $g$ or $\geq6$ detections in $r$ were included (considering data obtained until 2020/06/09). It is clear from the table that there is a high imbalance in the labeled set, with some classes representing less than 5\% of their respective hierarchical group. Figure \ref{figure:num_classes_ls} shows the (ordered) number of sources per class for the labeled set, and Figure \ref{figure:band_dist} shows the fraction of sources in each class with photometry only in the $g$ band, only in the $r$ band, or in both bands.

\section{Features used by the classifier}\label{features}

The light curve classifier uses a total of 152 features. We avoid including features that require a long time to compute, for example features that require the use of  Markov chain Monte Carlo techniques, since one of the goals of the light curve classifier is to provide a fast and highly scalable classification. 142 of these features are computed using solely the public ZTF $g$ and $r$ data. We excluded the mean magnitude as a feature to avoid that any bias in the labeled set magnitude distribution affects the classification of sources that are fainter (or brighter). Features obtained using the ZTF observed magnitudes are called detection features (56 features in the $g$ band, 56 features in the $r$ band, and 12 multi-band features, giving a total of 124 features), and features computed using the ZTF non-detection 5$\sigma$ magnitude limits {\texttt{diffmaglim}'s} are called non-detection features (nine features for each $g$ and $r$ bands, giving a total of 18 features). These features are described in the following sections (\ref{detections} and \ref{nondetections}), as well as in Appendix \ref{features_appendix}. Considering the LSST Data Products Definition Document \citep{LSSTDataProducs}, we expect that all these features would be measured using LSST data.

We also included as features the galactic coordinates of each target (\texttt{gal\_b} and \texttt{gal\_l}), the \texttt{W1-W2} and \texttt{W2-W3} AllWISE colors, and the \texttt{$g-$W2}, \texttt{$g-$W3}, \texttt{$r-$W2}, and \texttt{$r-$W3} colors, where $g$ and $r$ are computed as the mean magnitude of the $g$ band and $r$ band light curves for a given source. In addition, we use information included in the Avro files metadata: the \texttt{sgscore1} parameter, which corresponds to a morphological star/galaxy score of the closest source from PanSTARRS1 \citep{Tachibana18} reported in the ZTF Avro files, with $0 \leq \texttt{sgscore1} \leq 1$, where values closer to 1 imply a higher likelihood of the source being a star; and the median \texttt{rb} (real-bogus) parameter. With these 10 extra features, the total number of features used by the classifier sum to 152.

As we mentioned in Section \ref{data}, in this work we only consider light curves with $\geq6$ epochs in $g$ or $\geq6$ epochs in $r$. If a given source has $\geq6$ epochs just in one band, it is included in the analysis, and the features associated with the missing band are considered as -999 values. This rule applies to all the features used by the classifier; whenever a feature is not available for a given target, we assume a value equal to -999.

\subsection{Detection Features}\label{detections}

Most of the features used by the light curve classifier are computed using the observed magnitudes in the $g$ and $r$ bands (i.e., the detections). There are 56 different features computed for each band, and 12 features computed using a combination of both bands, yielding a total of 124 detection features. The definition of all these features can be found in Table \ref{table:det_features}. We split the table in three blocks. The first block contains new features defined by this work (i.e., novel features). Some of these features are further described in Section \ref{further_feat_descrip}. The second block contains features that correspond to new variants of descriptors included in other works. Some of them are further described in Appendix \ref{features_appendix}. Finally, the third block includes 22 features that come from the Feature Analysis for Time Series (FATS; \citealt{Nun15}) Python package. Hereafter, features ending with ``\_1'' are computed for the $g$ band, and features ending with ``\_2'' are computed for the $r$ band, following the notation used in the ZTF Avro files.

\setlength{\LTcapwidth}{\textwidth}
\begin{longtable*}{p{3.5cm}p{9.5cm}p{4cm}}

\caption{List of detection features used by the light curve classifier. Features marked with $\blacklozenge$ are computed using both $g$ and $r$ bands at the same time. Features marked with * and ** are further described in Section \ref{further_feat_descrip} and Appendix \ref{features_appendix}, respectively.}\label{table:det_features} \\

\hline 

\hline 

Feature & Description & Reference\\

\hline

\texttt{delta\_period} & Absolute value of the difference between the \texttt{Multiband\_period} and the MHAOV period obtained using a single band  & This work \\ 

\texttt{IAR\_phi}* & Level of autocorrelation using a  discrete-time representation of a DRW model & \citet{Eyheramendy18}\\

MHPS parameters* & Obtained from a MHPS analysis (three in total) & \cite{Arevalo12} \\

\texttt{positive\_fraction} & Fraction of detections in the difference-images of a given band which are brighter than the template image & This work \\

\texttt{Power\_rate}* $\blacklozenge$ & Ratio between the power of the multiband periodogram obtained for the best  period candidate ($P$) and $2\times P$, $3\times P$, $4\times P$, $P/2$, $P/3$ or $P/4$ & This work \\

\texttt{PPE}* $\blacklozenge$ & Multiband Periodogram Pseudo Entropy & This work \\

\hline

\texttt{($g$-$r$)\_max} $\blacklozenge$ & $g-r$ color obtained using the brightest \texttt{lc\_diff} magnitude in each band & This work \\

\texttt{($g$-$r$)\_max\_corr} $\blacklozenge$ & $g-r$ color obtained using the brightest \texttt{lc\_corr} magnitude in each band & This work  \\

\texttt{($g$-$r$)\_mean}  $\blacklozenge$& $g-r$ color obtained using the mean \texttt{lc\_diff} magnitude of each band & This work  \\

\texttt{($g$-$r$)\_mean\_corr} $\blacklozenge$ & $g-r$ color obtained using the mean \texttt{lc\_corr} magnitude of each band& This work \\

\texttt{delta\_mag\_fid} & Difference between maximum and minimum observed magnitude in a given band & This work\\

\texttt{ExcessVar}** & Measure of the intrinsic variability amplitude & \cite{Allevato13}\\

\texttt{GP\_DRW\_tau}** & Relaxation time $\tau$ from DRW modeling & \citet{Graham17}\\

\texttt{GP\_DRW\_sigma}** & Amplitude of the variability at short timescales ($t << \tau$), from DRW modeling & \citet{Graham17}\\

Harmonics parameters** & Obtained by fitting a harmonic series up to the seventh harmonic (14 in total) & \citep{Stellingwerf86} \\

\texttt{Multiband\_period}** $\blacklozenge$ & Period obtained using the multiband MHAOV periodogram & \citet{Mondrik15}\\

\texttt{Pvar}** & Probability that the source is intrinsically variable & \cite{McLaughlin96}\\

\texttt{SF\_ML\_amplitude}**& rms magnitude difference of the SF, computed over a 1 yr timescale &  \cite{Schmidt10} \\

\texttt{SF\_ML\_gamma}** & Logarithmic gradient of the mean change in magnitude &  \cite{Schmidt10} \\

SPM features** & Supernova parametric model features (seven in total) & \cite{2019ApJ...884...83V} \\ 

\hline

\texttt{Amplitude} & Half of the difference between the median of the maximum 5\% and of the minimum 5\% magnitudes & \citet{Richards11}\\

\texttt{AndersonDarling} & Test of whether a sample of data comes from a population with a specific distribution & \citet{Nun15}\\

\texttt{Autocor\_length} & Lag value where the auto-correlation function becomes smaller than \texttt{Eta\_e} & \citet{Kim11}\\

\texttt{Beyond1Std} & Percentage of points with photometric mag that lie beyond 1$\sigma$ from the mean & \citet{Richards11}\\

\texttt{Con} & Number of three consecutive data points brighter/fainter than 2$\sigma$ of the light curve & \citet{Kim11}\\

\texttt{Eta\_e} & Ratio of the mean of the squares of successive mag differences to the variance of the light curve& \citet{Kim14}\\

\texttt{Gskew} & Median-based measure of the skew & \citet{Nun15}\\

\texttt{LinearTrend} & Slope of a linear fit to the light curve & \citet{Richards11}\\

\texttt{MaxSlope} & Maximum absolute magnitude slope between two consecutive observations & \citet{Richards11}\\

\texttt{Meanvariance} & Ratio of the standard deviation to the mean magnitude & \citet{Nun15}\\

\texttt{MedianAbsDev} & Median discrepancy of the data from the median data & \citet{Richards11}\\

\texttt{MedianBRP} & Fraction of photometric points within amplitude/10 of the median mag & \citet{Richards11}\\

\texttt{PairSlopeTrend} & Fraction of increasing first differences minus fraction of decreasing first differences over the last 30 time-sorted mag measures & \citet{Richards11}\\

\texttt{PercentAmplitude} & Largest percentage difference between either max or min mag and median mag & \citet{Richards11}\\

\texttt{Psi\_CS} & Range of a cumulative sum applied to the phase-folded light curve & \citet{Kim11}\\

\texttt{Psi\_eta} & \texttt{Eta\_e} index calculated from the folded light curve
 & \citet{Kim14}\\
 
\texttt{Q31} & Difference between the 3\textsuperscript{rd} and the 1\textsuperscript{st} quartile of the light curve & \citet{Kim14}\\ 
 
\texttt{Rcs} & Range of a cumulative sum & \citet{Kim11}\\

\texttt{Skew} & Skewness measure & \citet{Richards11}\\

\texttt{SmallKurtosis} & Small sample kurtosis of the magnitudes & \cite{Richards11}\\

\texttt{Std} & Standard deviation of the light curve & \citet{Nun15}\\

\texttt{StetsonK} & Robust kurtosis measure & \citet{Kim11}\\

 \hline
 \hline

\end{longtable*}

\subsubsection{Description of a relevant set of detection features}\label{further_feat_descrip}

Table \ref{table:det_features} summarizes the definitions of the detection features used by the light curve classifier. Some of these features are worth describing in more detail since they are novel features. All other relevant features are described in Appendix \ref{features_appendix}:

\begin{itemize}

\item Periodogram Pseudo Entropy: To have an estimate of the confidence of the candidate period (obtained with the multiband MHAOV method), we developed a heuristic based on the entropy of the normalized periodogram peaks, which we denote as periodogram pseudo entropy (\texttt{PPE}). This value is computed by recovering the 100 largest values of the periodogram, normalizing them and computing the entropy of that vector. This feature is computed as

\begin{equation}
 PPE =   1 + \frac{1}{\log(100)}\sum_{i=1}^{100} \left (\frac{p_{i}}{Z} \right ) \log\left (\frac{p_{i}}{Z} \right ),
\end{equation}

where $p_i$ is the value of the $i$-th largest peak of the periodogram and $Z = \sum_{i=1}^{100}p_{i}$. This feature takes values between zero (no clear period stands out) and one (periodogram has a single large peak).

\item Power Rate: Ratio between the power of the multiband periodogram obtained for the best period candidate ($P$) versus the power of the multiband periodogram obtained for $n$ times this period $[\texttt{Power\_rate\_n}=Power(P)/Power(n\times P)]$, where $n$ can take values of 2, 3, 4, $1/2$, $1/3$, and $1/4$. We computed these ratios in order to detect cases where we measure an aliased multiple of the period instead of the true period, which is particularly a common issue for some classes of eclipsing binaries (e.g., \citealt{Catelan13,Graham13,McWhirter18,VanderPlas2018}, and references therein).

\item Irregular autoregressive (IAR) model: \cite{Eyheramendy18} introduced this model. It is a discrete-time representation of the continuous autoregressive model of order 1 [CAR(1)], which has desirable statistical properties such as strict stationarity and ergodicity without a distributional assumption. The IAR model is defined by

\begin{equation}  \label{IAR} 
y_{t_j}=\phi^{t_j-t_{j-1}} \, y_{t_{j-1}} + \sigma \, \sqrt{1-\phi^{2(t_j-t_{j-1})}}  \, \varepsilon_{t_j}, 
\end{equation} 

where $\varepsilon_{t_j}$ is a white noise sequence with zero mean and unit variance, $\sigma$ is the standard deviation of $y_{t_j}$, and $\{t_j\}$ are the observational times for $j=1,\ldots,n$. We used a modified version of the IAR model, which considers the estimated variance of the measurement errors $\delta^2_{t_j}$ in the likelihood of the model. Thus, by assuming a Gaussian distribution, the negative log-likelihood of the process is given by

\begin{equation}\label{eq:IARloglik} 
\ell(\phi,\sigma^2)=\frac{n}{2}\log (2\pi)+\frac{1}{2}\sum_{j=1}^n \log \nu_{t_j} + \frac{1}{2}\sum_{j=1}^n \frac{e_{t_j}^2}{\nu_{t_j}},
\end{equation} 

where $e_{t_1}=y_{t_1}$, $\nu_{t_1}=\sigma^2 + \delta_{t_1}^2$,  and $\hat{y}_{t_1}=0$ are the initial values, while $\hat{y}_{t_j} = \phi^{t_j-t_{j-1}}\, y_{t_{j-1}}$,  $e_{t_j}=y_{t_j}-\hat{y}_{t_j}$, and $\nu_{t_j}=\sigma^2 (1-\phi^{2(t_j-t_{j-1})}) + \delta_{t_j}^2$ for $j=2,\ldots,n$. 
Particularly, $\phi$ describes the autocorrelation function of order 1 for a given light curve. We computed the maximum likelihood estimation of the parameter $\phi$ (obtained directly from the light curves), and we used this as a feature for our classifier. We denoted this parameter as \texttt{IAR\_phi}.

\item Mexican Hat Power Spectrum (MHPS): \cite{Arevalo12} proposed a method to compute low-resolution power spectra from data with gaps, where the light curves are convolved with a Mexican hat filter: $F(x)\propto \left[ 1-\frac{x^2}{\sigma^2}\right] e^{-x^2/2\sigma^2}$. Gaps, or generally uneven sampling, are corrected for by convolving a unit-valued mask with the same sampling as the light curve and dividing the convolved light curve by it. This method can be used to isolate structures  with a characteristic timescale ($t\sim\sigma/\sqrt{2\pi^2}$) in a given light curve, in order to estimate the light curve variance associated with that timescale. We compute the light curve variance at two different timescales of 10 and 100 days. The variance associated with the 10 day timescale (``high'' frequency) is denoted \texttt{MHPS\_high}, while the variance associated with the 100 day timescale (``low'' frequency) is denoted \texttt{MHPS\_low}. We also compute the ratio between the low and high frequency variances for a given band, denoted as \texttt{MHPS\_ratio}. The logarithm of \texttt{MHPS\_ratio} is therefore an estimate of the power law slope of the power spectrum of the light curve.

\end{itemize}

\subsection{Non-detection Features}\label{nondetections}

For each detection, the ZTF alert stream includes 5-$\sigma$ magnitude limits (\texttt{diffmaglim}), which are computed from the $g$ and $r$ difference images of the same area of the sky obtained in the previous 30 days, where the target associated with the alert was not detected (non-detections). These non-detections are very informative, since they can, for instance, inform us whether a transient has not been detected before; whether a non-variable source has begun to exhibit a variable behavior; or which range of observed magnitudes we should expect to measure when there are not significant differences between the science and template images, and an alert is not generated. The light curve classifier uses nine different features defined using all the non-detections associated with a given source, computed for both $g$ and $r$ bands, yielding a total of 18 non-detection features. Note that all these features are new, and have not been used before for classification. Table \ref{table:nondet_features} lists the non-detection features used by the light curve classifier. As before, non-detection features ending with ``\_1'' are computed for the $g$ band, and features ending with ``\_2'' are computed for the $r$ band.

\begin{table*}[tb]
\caption{List of non-detection features used by the light curve classifier. Note that ``x'' stands for either $g$ or $r$ bands.} 
\label{table:nondet_features}      
 \renewcommand\arraystretch{1.4}
 \footnotesize
 \resizebox{\textwidth}{!}{
 \begin{tabular}{l l}
\hline

\hline Feature & Description \\
\hline

\texttt{dmag\_first\_det\_fid} &  Difference between the last non-detection \texttt{diffmaglim} in band ``x'' before the first detection \\ &  in any band and the first detected magnitude in band ``x'' \\ 

\texttt{dmag\_non\_det\_fid} &  Difference between the median non-detection \texttt{diffmaglim} in the ``x'' band before the first    \\  & detection and in any band the minimum detected magnitude (peak) in the ``x'' band\\

\texttt{last\_diffmaglim\_before\_fid} & Last non-detection \texttt{diffmaglim} in the ``x'' band before the first detection in any band\\

\texttt{max\_diffmaglim\_before\_fid} &  Maximum non-detection \texttt{diffmaglim} in the ``x'' band before the first detection in any band\\

\texttt{max\_diffmaglim\_after\_fid} & Maximum non-detection \texttt{diffmaglim} in the ``x'' band after the first detection in any band\\

\texttt{median\_diffmaglim\_before\_fid} &  Median non-detection \texttt{diffmaglim} in the ``x'' band before the first detection in any band\\

\texttt{median\_diffmaglim\_after\_fid} & Median non-detection \texttt{diffmaglim} in the ``x'' band after the first detection in any band\\

\texttt{n\_non\_det\_before\_fid} & Number of non-detections in the ``x'' band before the first detection in any band\\

\texttt{n\_non\_det\_after\_fid} &  Number of non-detections in the ``x'' band after the first detection in any band\\

\hline

\end{tabular}

}
\end{table*}

\section{Classification Algorithms}\label{classifiers}

The labeled set used in this work presents a very high imbalance (see Table \ref{table:trainingset}). For instance, QSOs represent 75.4\% of the stochastic sources, while CV/Novae represent just 2.5\%. To deal with this issue, we looked for ML algorithms available in the literature that are designed to mitigate the imbalance problem. In particular, we worked with the \texttt{imbalanced-learn} Python package \citep{imblearn}. \texttt{Imbalanced-learn} includes implementations of several re-sampling algorithms that are commonly used to handle data sets with strong between-class imbalance. The algorithms available in this package are fully compatible with \texttt{scikit-learn} methods.

In the following sections we describe the Random Forest (RF) algorithm used by the light curve classifier, as well as other tested ML algorithms. To train each classifier we randomly split the labeled set into a training set (80\%) and testing set (20\%) in a stratified fashion, preserving the percentage of samples for each class.

\subsection{Balanced Random Forest}\label{RF}

A Decision Tree \citep{rokach08} is a predictive algorithm that uses a tree structure to perform successive partitions on the data according to a certain criterion (e.g., a cut-off value in one of the descriptors or features) and produces possible decision paths, providing a final outcome for each path (the leaves of the tree). Decision Trees are commonly used for classification, where each final leaf is associated with a given class. RFs \citep{randomforest} are algorithms that build multiple Decision Trees, where each tree is trained using a random sub-sample of elements from a given training set, selected allowing repetition (bootstrap sample of the training set), and using a random selection of features. The final classification is obtained by averaging the classifications provided by each single tree. This average score can be interpreted as the probability ($P_{\text{RF}}$) that the input element belongs to a given class. One of the main advantages of RF is that it naturally provides a ranking of features for the classification, by counting the number of times each feature is selected to split the data.

\cite{Chen04} proposed a modified RF that can deal with the imbalanced data classification. In their model each individual tree is trained using a sub-sample of the training set that is defined by generating a bootstrap sample from the minority class, and then randomly selecting the same number of cases, with replacement, from the majority classes. \texttt{Imbalanced-learn} implements the balanced RF classifier proposed by \cite{Chen04}. For the ALeRCE light curve classifier we use their \texttt{BalancedRandomForestClassifier} method, selecting the hyper-parameters (number of trees, maximum number of features per tree, and maximum depth of each tree) with a K-Fold Cross-Validation procedure available in \texttt{scikit-learn}, with $k=5$ folds and using the ``macro-recall'' as target score (see its definition in Section \ref{results_RF}). 

\subsubsection{The two-level classifier approach}\label{hier_classifier}

Considering the hierarchical structure of the taxonomy (see Section \ref{taxonomy}), we decided to construct a balanced RF (BRF) classifier with two-level scheme. The first level consists of a single classifier that separates the sources into three broad classes. The second level consists of three distinct classifiers, which further resolve each class in the first level into subclasses. We then use the probabilities obtained for each independent classifier to obtain the final classification.

In more detail, the first level (top level hereafter) consists of a single classifier which classifies every source as periodic, stochastic, or transient. The second level (bottom level hereafter) consists of three distinct classifiers: Transient, Stochastic, and Periodic. The classes considered by each of these three classifiers are the ones shown in Table \ref{table:trainingset} and Figure \ref{figure:taxonomy}. Each classifier in the bottom level is trained using a training subset having only those classes included in the primary top class (for instance, the Transient classifier only includes sources classified as SNIa, SNIbc, SNII, and SLSN). It is important to note that these four classifiers are independent and process the same input features set described in Section \ref{features}. The final classification is constructed by multiplying the probabilities obtained for each class of the top level [$P_{top}(transient)$, $P_{top}(stochastic)$, and $P_{top}(periodic)$] with the individual probabilities obtained by their correspondent classifier in the bottom level. Namely, the probabilities of the Transient classifier ($P_{T}$) are multiplied by $P_{top}(transient)$, the probabilities of the Stochastic classifier ($P_{S}$) are multiplied by $P_{top}(stochastic)$, and the probabilities of the Periodic classifier ($P_{S}$) are multiplied by $P_{top}(periodic)$. We denote the product of these probabilities as $P$. For instance, the probability of a given source being an RRL corresponds to the product of its probability of being periodic (according to the top level) and its probability of being an RRL (according to the Periodic classifier): 
\begin{equation}
P(RRL) = P_{top}(periodic) \times P_P (RRL),
\end{equation}
while the probability of being a Blazar is computed as: 
\begin{equation}
P(Blazar) = P_{top}(stochastic) \times P_S(Blazar).
\end{equation}
Following this, the sum of the probabilities of the 15 classes for a given source adds up to one. Finally, the class of a given object is determined by selecting the class with the maximum $P$. Hereafter, we refer to the results presented for the bottom level of the classifier as the final predictions.

The best cross-validation performance was obtained with the following hyper-parameter setting: 500 trees in each classifier, maximum depth trees (the nodes are expanded until all leaves are pure), and a maximum number of features equal to the square root of the total number of features, except for the Stochastic classifier, where we used 20\% of the features. In Section \ref{results} we present the results obtained when applying the BRF classifier to the ZTF data.

\subsection{Additional ML algorithms tested}

In addition to RF, we also tested two other supervised classification algorithms: Gradient Boosting and Multilayer Perceptron. These tests were done as a complementary analysis, with the purpose of guiding future efforts in improving the light curve classifier.  

None of these methods has a Python implementation particularly designed to handle imbalanced data sets; however, using \texttt{imbalanced-learn} we can generate balanced training sets. We present the results obtained using both classifiers in Section \ref{results}.

\subsubsection{Gradient Boosting}\label{GB}

Gradient Boosting (GBoost; \citealt{Friedman01}) is an ML algorithm that uses an ensemble of weak prediction models (e.g., Decision Trees) to produce a more robust classification. The method implements a boosting algorithm (using a Gradient Descent algorithm) that trains a sequence of weak models, each compensating the weaknesses of its predecessors. \texttt{eXtreme Gradient Boosting} (\texttt{XGBoost}; \citealt{Chen16}) is a package available in several computing languages (including Python) that implements GBoost algorithms for classification and regression in an efficient and scalable way. It has become one of the most used packages for regression and classification in recent years. 

For the case of GBoost we followed the same two-level strategy described in Section \ref{hier_classifier}. However, since the current version of the \texttt{XGBoost} multi-class classifier was not designed to deal with highly imbalanced data sets (e.g., \citealt{Wang19}), we tested a model that uses \texttt{XGBoost} and is trained with a balanced training set. We constructed this balanced training set using the \texttt{RandomUnderSampler} and \texttt{RandomOverSampler} methods available in \texttt{imbalanced-learn}. For the case of the top level, Periodic, and Stochastic classifiers, we constructed a balanced training set by generating 10 random samples using the \texttt{RandomUnderSampler} method, resampling all classes, and concatenating them, in order to obtain a training set with more than 10,000 objects in total for each classifier. For the case of the Transient classifier we used the \texttt{RandomOverSampler} method, resampling all classes, to generate one random sample with $\sim600$ objects. Each classifier uses the default hyper-parameters defined by the \texttt{XGBoost} Python package, with the exception of the boosting rounds, where we used 500, and the objective function, which was set up to do multi-class classification, using the softmax function for multi-class predictions. As in the case of the BRF model, the class of a given object is determined by selecting the class with the maximum probability (obtained by multiplying the probabilities of the top and bottom levels).  

\subsubsection{Multilayer Perceptron}\label{MLP}

Artificial neural networks (ANNs) are mathematical models inspired by the human brain. ANNs are composed of elemental computational units called neurons \citep{Haykin94}. ANNs can be used to perform complex tasks such as classification or regression. A Multilayer Perceptron (MLP) corresponds to an ANN whose neurons are ordered by layers, where all neurons belonging to a given layer receive the same input vector and each unit processes this vector independently according to its own parameters. The outputs of all neurons in a layer are grouped and form the input vector for the next layer. For the case of classification, when using the softmax activation function, the final layer provides the probabilities that a given element belongs to a given class. One way of obtaining the final class is to assign the label with the maximum probability in the output layer.

For this model, we also followed the two-level strategy described in Section \ref{hier_classifier}. We tested different MLP architectures, changing the number of layers and the number of neurons per layer. We used the \texttt{Keras API} provided by the Python version of \texttt{TensorFlow 2.0} \citep{Tensorflow}. We split the original training set defined above into a new training set (80\%) and a validation set (20\%). In order to deal with the high imbalance of the training set we used the balanced mini-batches generator for Keras provided by the \texttt{imbalanced-learn} Python package. The best performance (considering the categorical cross-entropy loss and accuracy curves for the training and validation sets) was obtained using MLPs with two hidden layers with 256 and 128 neurons for all the classifiers (top level, Transient, Stochastic, and Periodic). Regularization via the dropout method \citep{Srivastava14} is used to prevent overfitting. The dropout fraction is set at 0.50.

\section{Results}\label{results}

\subsection{Results for the BRF classifier}\label{results_RF}

In order to test the performance of our BRF classifier we generated 20 different training and testing sets using the \texttt{ShuffleSplit} iterator provided by \texttt{scikit-learn}, which uses random permutations to split the data, using each time 80\% of the labeled set as training set and 20\% as testing set, preserving the percentage of samples for each class in the original labeled set. Then, we trained 20 different BRF models using each training set, and tested their performance using the corresponding testing sets. We emphasize that the testing sets are never used in training their respective models.

Table \ref{table:scores} lists three different scores: precision, recall, and F1-score. For an individual class these scores are defined as:

\begin{equation}\label{eq:sing_precision} 
\text{Precision}_{i} = \frac{TP_i}{TP_i+FP_i},
\end{equation} 

\begin{equation}\label{eq:sing_recall}
\text{Recall}_i =  \frac{TP_i}{TP_i+FN_i},
\end{equation} 

\begin{equation}\label{eq:sing_f1}
\text{F1-score}_i = 2 \times \frac{\text{Precision}_i \times \text{Recall}_i}{\text{Precision}_i + \text{Recall}_i},
\end{equation}
where $n_{cl}$ is the total number of classes, $TP_i$ is the number of true positives, $FP_i$ is the number of false positives, and $FN_i$ is the number of false negatives, for a given class $i$. Despite the high imbalance present in the labeled set, all classes are equally important, and thus we compute macro-averaged scores:

\begin{equation}\label{eq:precision} 
\text{Precision}_{\text{macro}} = \frac{1}{n_{cl}} \sum_{i=1}^{n_{cl}} \text{Precision}_{i},
\end{equation} 

\begin{equation}\label{eq:recall}
\text{Recall}_{\text{macro}} =  \frac{1}{n_{cl}} \sum_{i=1}^{n_{cl}} \text{Recall}_{i},
\end{equation} 

\begin{equation}\label{eq:f1}
\text{F1-score}_{\text{macro}} = \frac{1}{n_{cl}} \sum_{i=1}^{n_{cl}} \text{F1-score}_{i}.
\end{equation} 

\noindent
For the particular case of the BRF classifier, Table \ref{table:scores} reports the mean and the standard deviation of the macro-averaged scores obtained by the 20 models when applying them to their  respective testing sets.

\begin{table*}[htpb]
  \begin{center}
    \caption{Macro-averaged scores obtained for the BRF classifier in the testing set. The reported values correspond to the mean and standard deviation obtained from the trained 20 models.}
    \label{table:scores}
    \begin{tabular}{|c|c|c|c|} 
   
   \hline
 \hline
Classifier &  Precision & Recall & F1-score \\
 \hline
BRF - top  &  0.96 $\pm$ 0.01 & 0.99 $\pm$ 0.01  & 0.97 $\pm$ 0.01 \\

BRF - bottom  & 0.57 $\pm$ 0.01 & 0.76 $\pm$ 0.02 & 0.59 $\pm$ 0.01 \\

\hline
\hline

 \end{tabular}
  \end{center}
\end{table*}

\begin{figure}[h!]
\begin{center}

 \includegraphics[scale=0.36]{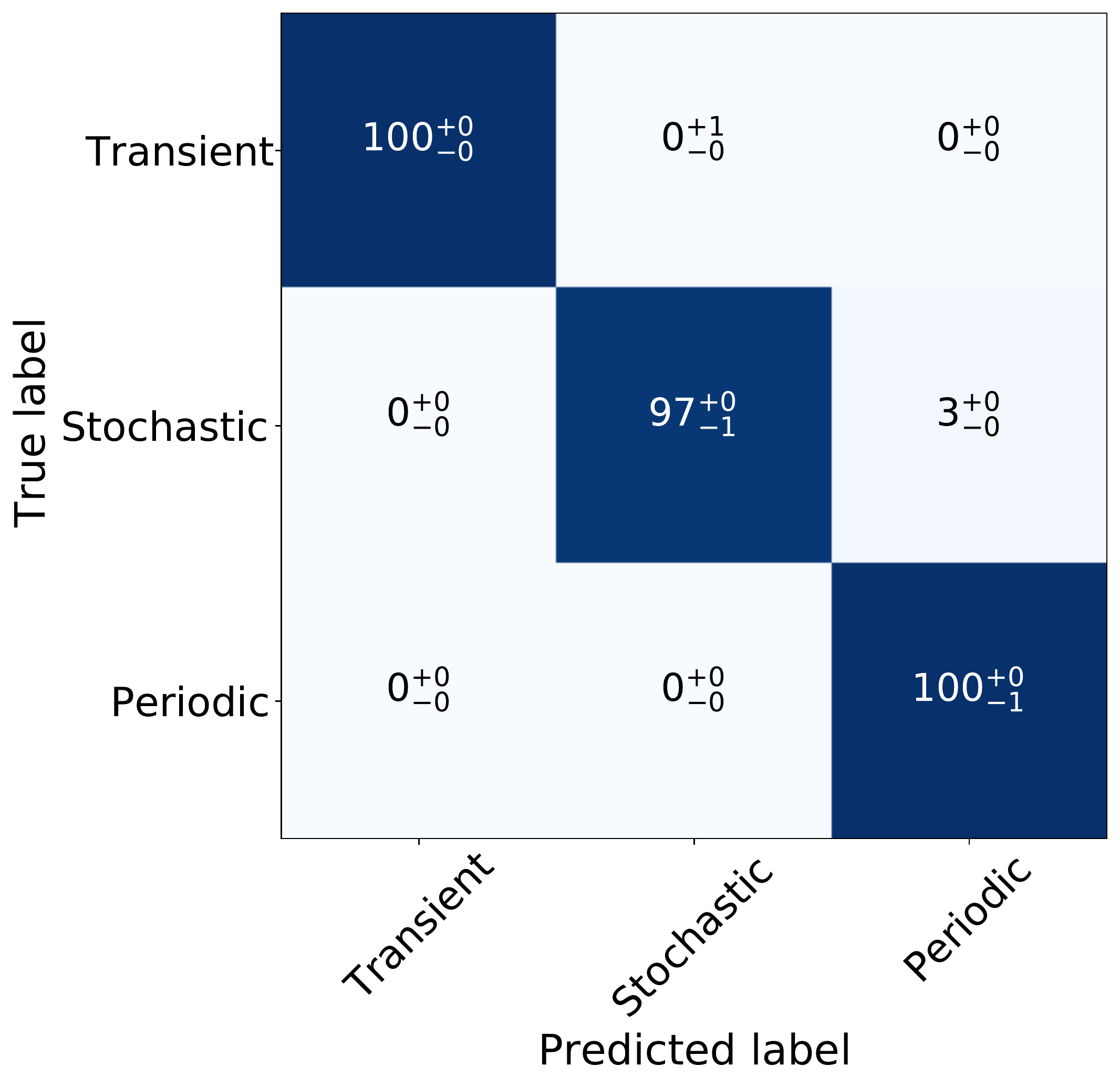}

\caption{Confusion Matrix for the top level BRF classifier. The confusion matrix was obtained by generating 20 different training and testing sets, and by training 20 independent models using each training set separately. After the training, each model is applied to their respective testing set. We provide the median and 5 and 95 percentiles of the confusion matrices obtained for the 20 testing sets. To normalize the confusion matrix results as percentages, we divide each row by the total number of objects per class with known labels. We round this percentages to integer values. This level shows a high degree of accuracy with a low percentage of misclassifications. \label{figure:conf_mat_hier_BRF}}
\end{center}
\end{figure}

\begin{figure*}[htbp]
\begin{center}

 \includegraphics[scale=0.55]{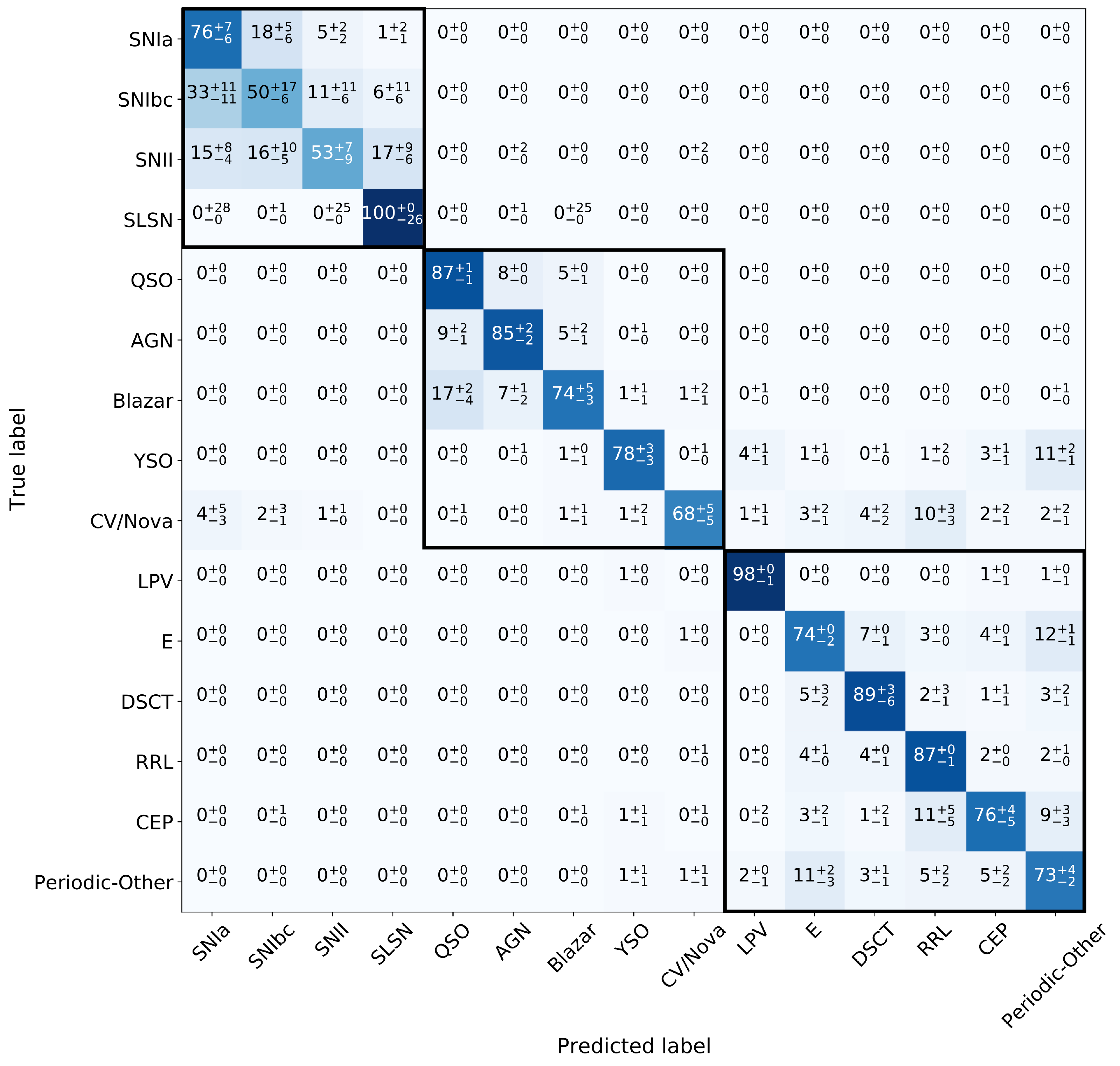}

\caption{As in Figure~\ref{figure:conf_mat_hier_BRF}, but for the bottom level of the BRF classifier. We provide the median and 5 and 95 percentiles of the confusion matrices obtained for the 20 testing sets. The black squares highlight the three classes of the top level (from top to bottom, transient, stochastic, and periodic, respectively). This matrix is quite diagonal, but shows more misclassification among related subtypes compared to the matrix obtained for the top level. }  \label{figure:conf_mat_BRF}
\end{center}
\end{figure*}

In addition, Figures \ref{figure:conf_mat_hier_BRF} and \ref{figure:conf_mat_BRF} show the confusion matrices obtained for the top and bottom levels, respectively. To generate these confusion matrices we used the results obtained when applying each of the 20 BRF models to their corresponding testing sets, providing for each level the median and 5 and 95 percentiles of the confusion matrices obtained by each of the 20 models.

The confusion matrix of the top level shows that the classifier can recover more than 97\% of the true labels, and that the contamination between classes is below 3\%. The scores obtained reflect the good performance of the top level classifier. 

For the case of the bottom level we obtained an F1-score of 0.59, implying significant confusion between classes. From Figure \ref{figure:conf_mat_BRF} we can see that the fraction of true positives in the confusion matrix of the bottom level has values between 50\% and 100\%, with
mean, median and standard deviation of 78\%, 76\%, and 14\%, respectively. In addition, from the figure it can be observed that the confusion is most often observed among classes with similar characteristics, like among the SN classes (particularly among SNIa versus SNIbc and SNII versus SLSN); among Blazar, AGN, and QSO classes; and among various periodic classes. The highest standard deviation of the predictions is observed for the case of SLSN. This is a result of the low number of SLSN in the labeled set.

To complement our analysis, in Appendix \ref{on-level-RF} we provide the results obtained by a one-level multi-class RF model. From its results we can conclude that the light curve classifier improves considerably when a two-level strategy is followed.

\subsubsection{Comparison with the GBoost and MLP classifiers}\label{results_other_class}

In this work we tested two other classifiers: GBoost and MLP. For these models we present the results obtained by using 80\% of the labeled set as a training set, and the remaining 20\% as a testing set, preserving the percentage of samples for each class in the original labeled set.

\begin{figure*}[htbp]
\begin{center}
\begin{tabular}{cc}
\includegraphics[scale=0.33]{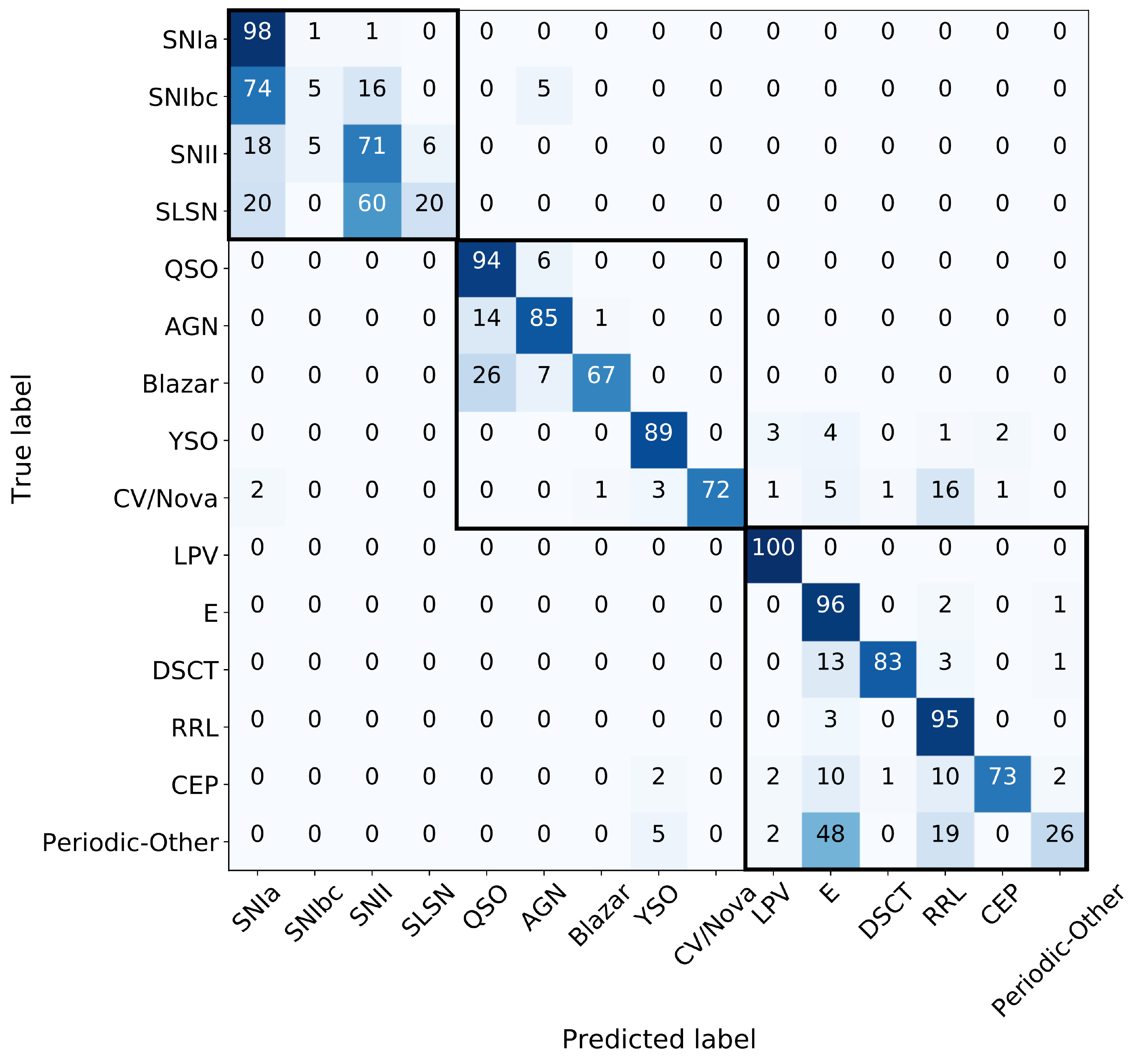} &
  \includegraphics[scale=0.33]{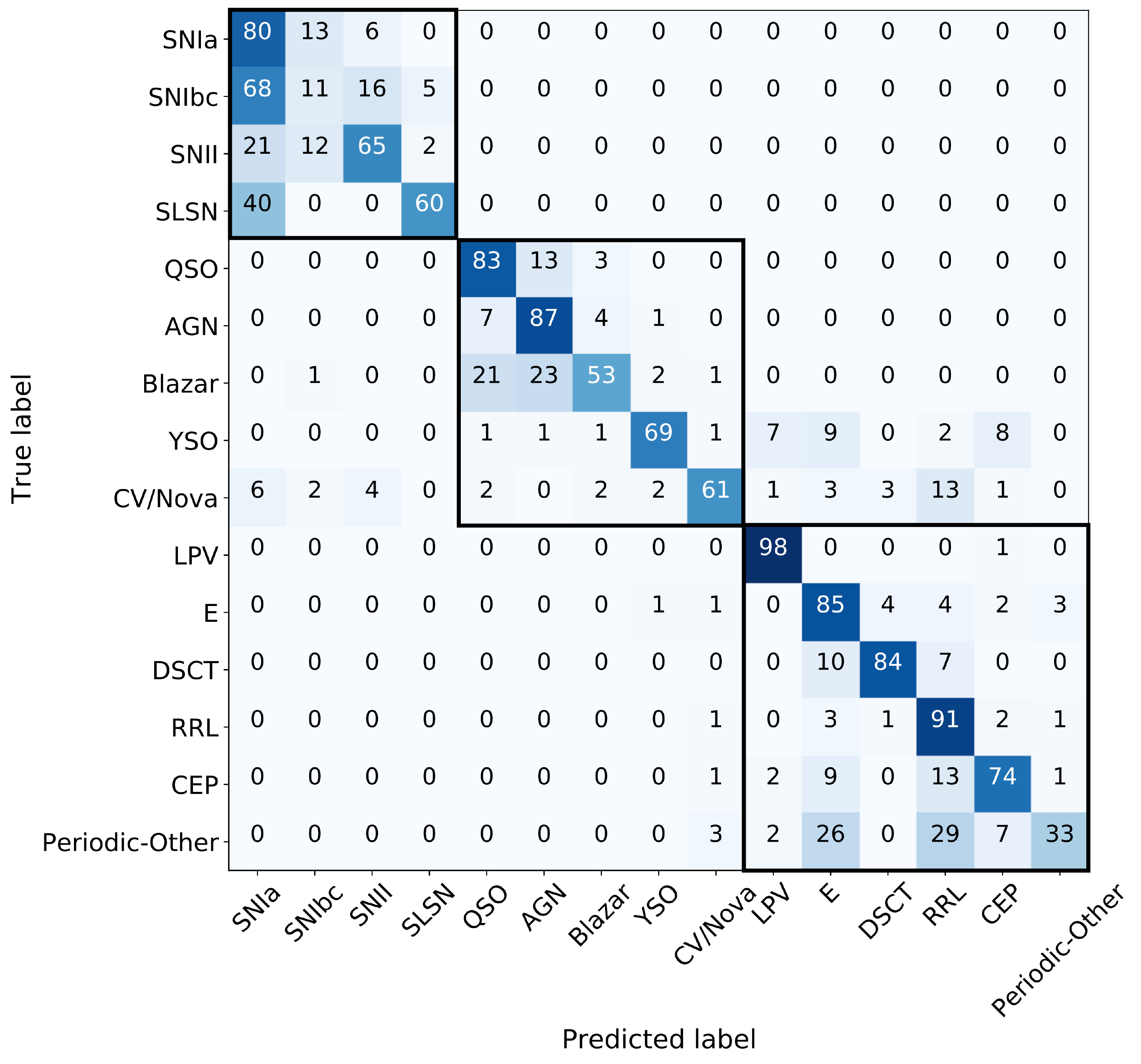} \\
\end{tabular}
\caption{Confusion matrices of the bottom level obtained when applying the GBoost (left) and the MLP (right) classifiers to the testing set. The black squares highlight the three hierarchical classes (from top to bottom, transient, stochastic, and periodic, respectively). To normalize the confusion matrix results as percentages, we divide each row by the total number of objects per class with known labels. We round this percentages to integer values. These matrices present a high percentage of misclassifications compared to the bottom level of the BRF model (Figure \ref{figure:conf_mat_BRF}).  \label{figure:conf_max_others}}
\end{center}
\end{figure*}

For the case of the GBoost classifier, the macro-averaged precision, recall, and F1-score of the top level have a value of 0.99. For the bottom level the macro-averaged precision, recall, and F1-score are, respectively, 0.72, 0.72, and 0.71. On the other hand, for the MLP classifier  the macro-averaged precision, recall, and F1-score of the top level are 0.94, 0.99, and 0.96, respectively. For the bottom level the macro-averaged precision, recall, and F1-score are, respectively, 0.54, 0.69, and 0.58. The confusion matrices obtained for the bottom level of the GBoost and MLP classifiers in the testing set are presented on the left and right sides of Figure \ref{figure:conf_max_others}, respectively. 

The precision and F1-score obtained by GBoost are in general better than the ones obtained by the BRF classifier, with the exception of the recall score of the bottom level. However, the fraction of true positives in the confusion matrix of the bottom level range between 5\% and 100\%, with a mean, median, and standard deviation of 72\%, 83\%, and 29\%, respectively, which explain the lower recall obtained by GBoost, compared to the BRF classifier. In addition, the classes with the largest fraction of true positives in the confusion matrix of GBoost correspond to the most populated classes in the labeled set, like QSOs, which represent 75.4\% of the stochastic sources; SNIa, which represent 74.0\% of the transients; or LPV, E, and RRL, which represent 16.2\%, 43.5\%, and 37.3\% of the periodic sources, respectively. This is not observed in the results obtained by BRF, where there is no evidence of correlation between the representativity of a given class and its fraction of true positives in the confusion matrix shown in Figure \ref{figure:conf_mat_BRF}. 

The results obtained using GBoost are promising. We obtained good scores although the current versions of GBoost available in the literature have not been designed to deal with high imbalance for the case of multi-class classification. Therefore, further efforts should be done to implement GBoost in future versions of the light curve classifier. In particular, new implementations of GBoost for multi-class classification that follow similar approaches to the ones proposed by \cite{Chen04} or \cite{Wang19} should be tested, as should combinations of GBoost with data augmentation techniques (i.e., generating synthetic light curves of less populated classes using physical and/or statistical models).

For the case of the MLP classifier the scores obtained are in general lower compared to BRF and GBoost. Its confusion matrix for the bottom level (see Figure \ref{figure:conf_max_others}) presents the same issues already discussed for the case of GBoost. The fraction of true positives in the confusion matrix ranges between 11\% and 98\%, with a mean, median, and standard deviation of 69\%, 74\%, and 22\%, respectively. Therefore, we conclude that more work should be done in order to obtain better results with MLP. 

From these tests we can conclude that the BRF is the model that currently achieves results that are less biased towards the most populated classes in the labeled set, i.e., it is able to predict all sub-classes, and the fraction of true positives does not correlate with how representative a given class in the labeled set is, compared to GBoost and MLP. Thus, we decided to use BRF as the final model for the first version of the ALeRCE light curve classifier. Future work will further exploit the potential of the GBoost and MLP classifiers. For the rest of this paper, presented results correspond to the BRF classifier.

\subsubsection{Results for the BRF classifier excluding AllWISE data}\label{no_wise}

\begin{figure}[htbp]
\begin{center}

\includegraphics[scale=0.33]{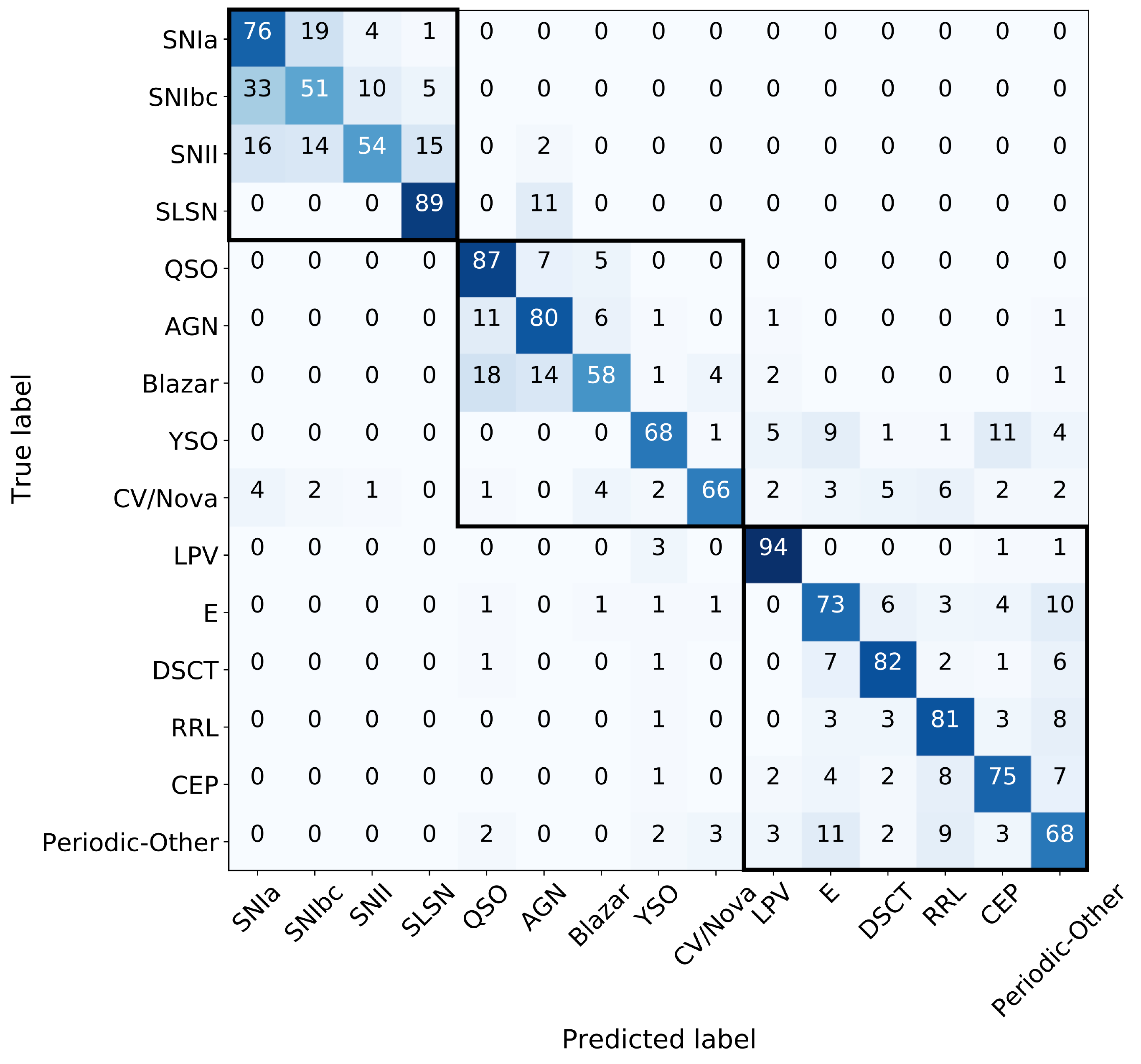}

\caption{As in Figure \ref{figure:conf_mat_BRF}, but for a model that excludes AllWISE data. We drop the errors, which are comparable to those listed in Figure \ref{figure:conf_mat_BRF}, for simplicity. The results obtained for this model are reasonable, although the classification of some stochastic classes are affected by the lack of AllWISE data.  \label{figure:conf_max_noWISE}}
\end{center}
\end{figure}

We tested a version of the BRF classifier that excludes features computed using AllWISE data. The macro-averaged precision, recall, and F1-score of the top level are 0.93, 0.97, and 0.95, respectively. For the bottom level the macro-averaged precision, recall, and F1-score are, respectively, 0.53, 0.72, and 0.55. These scores are slightly smaller than the ones obtained using the original version of the BRF classifier. As can be observed in the confusion matrix shown in Figure \ref{figure:conf_max_noWISE}, the stochastic classes are the most affected by the lack of AllWISE data, particularly YSOs and Blazars, whose fraction of true positives decreased 10\% and 16\%, respectively. This happens because of the similarities observed between the light curves of these and other classes, which are not easily separated using variability features alone. However, the results obtained by this version of the classifier are still good enough to be used in the case that AllWISE data are not available, as occurs, for instance, for faint objects. 

\subsection{Performance of the BRF classifier as a function of magnitude and number of detections}\label{performance}

As can be seen in Figure \ref{figure:band_dist} for some classes, like YSOs, Cepheids, and LPVs, a non-negligible fraction of sources in the labeled set have photometry available only in one band. It is therefore important to know how well the classifier behaves when a single band is available for a given source. 

\begin{figure*}[htbp]
\begin{center}

 \includegraphics[scale=0.5]{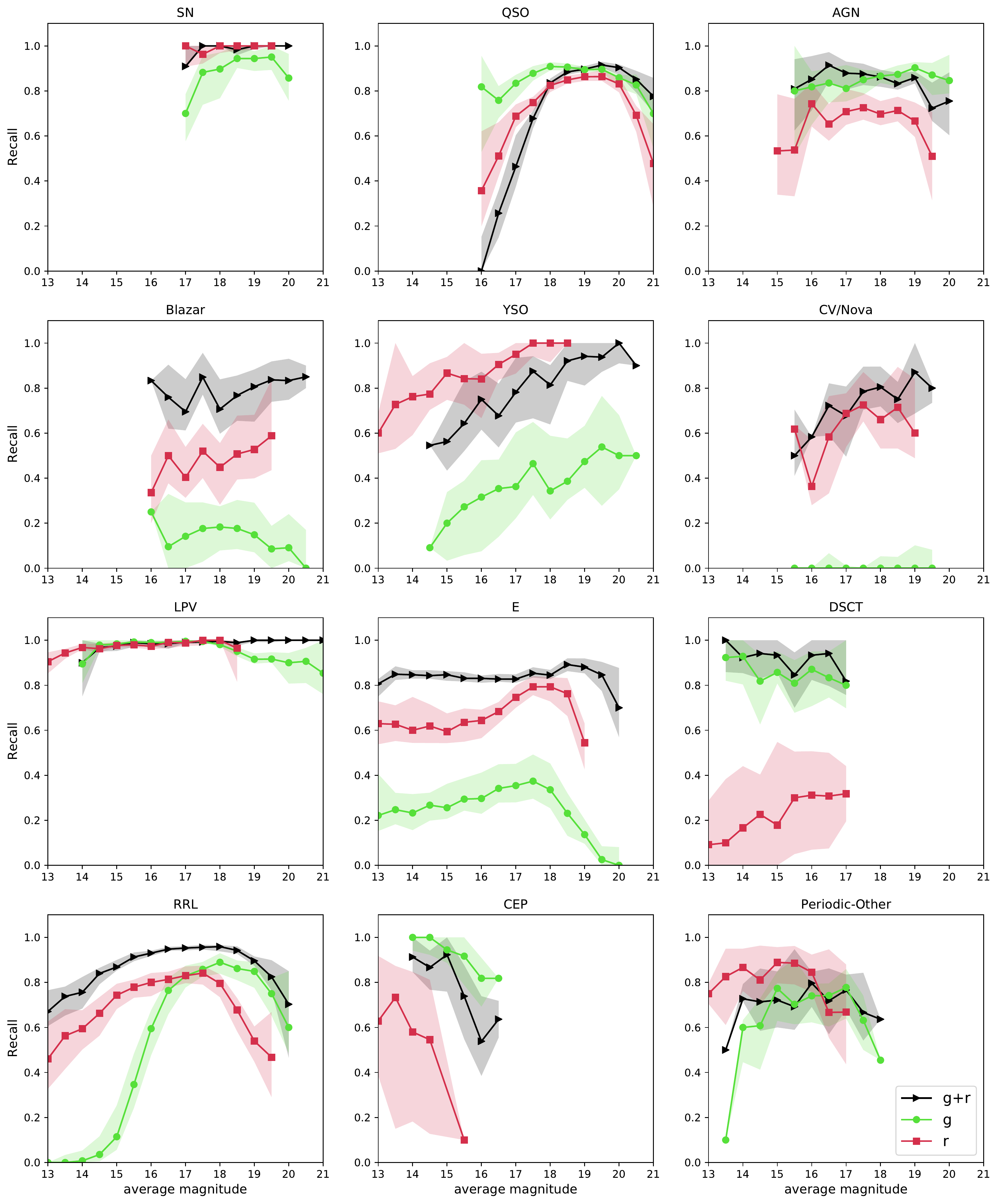}

\caption{Recall for each stochastic and periodic subclass, as well as all transients (grouped as SN), as a function of the average magnitude. The x-axis ranges from 13 to 21 magnitudes, this range includes $\sim$90\% of the sources. In black triangles we show the recall curves obtained when $g$ and $r$ photometries are available (considering the average magnitude in the $g$ band), in green circles when only the $g$ band is available, and in red squares when only the $r$ band is available.  The shaded regions were obtained by generating 20 different training and testing sets, and training 20 independent models using each of these sets. We report the median and 5 and 95 percentile values obtained from the 20 models. There is a truly wide variety of behaviors (see discussion in the text). \label{figure:recall_mag}}
\end{center}
\end{figure*}

\begin{figure*}[htbp]
\begin{center}

 \includegraphics[scale=0.5]{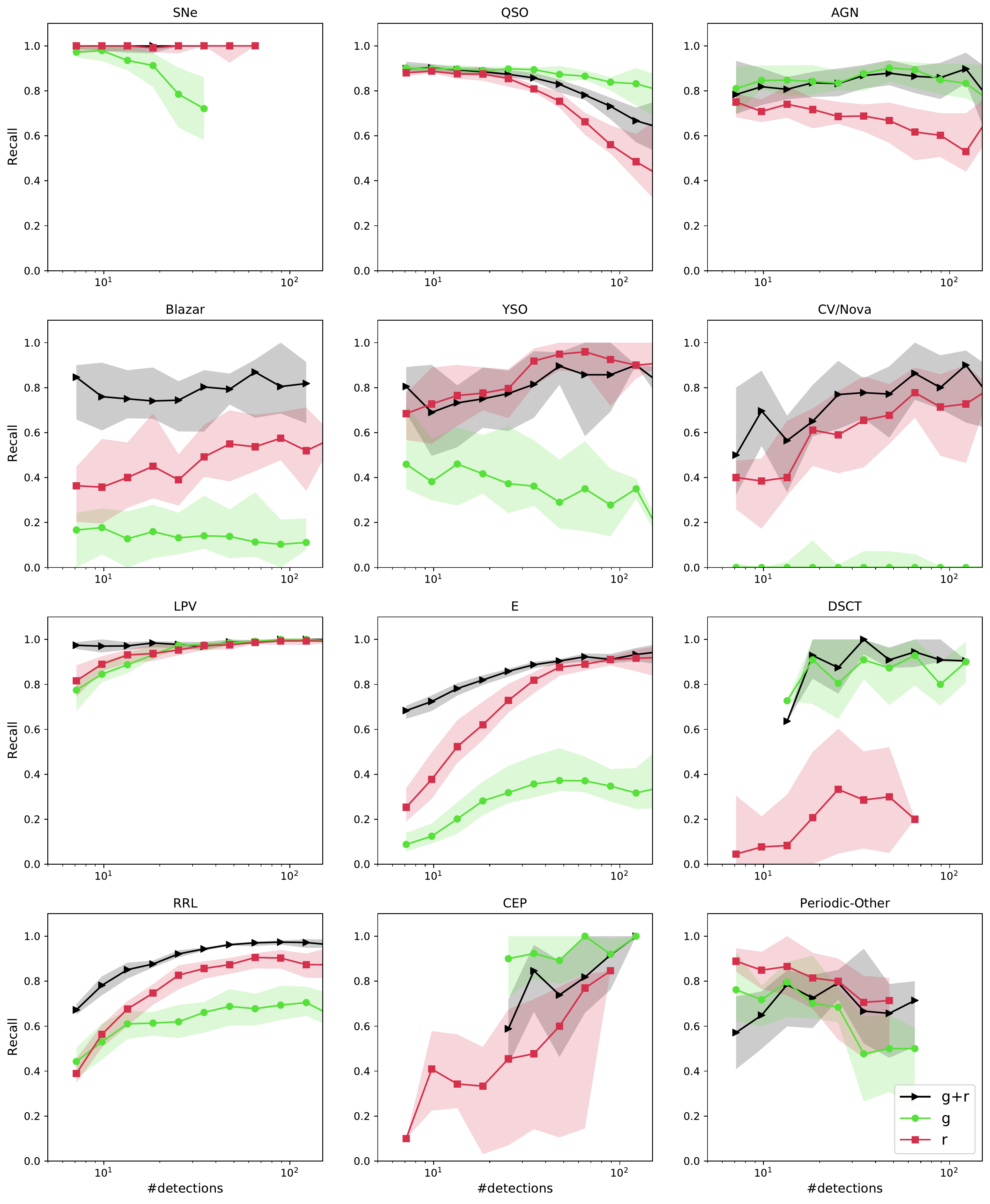}

\caption{ As in Figure~\ref{figure:recall_mag}, but plotting the Recall as a function of the number of detections in the light curve (in logarithmic scale).
The x-axis ranges from 6 to 150 detections, this range includes $\sim$90\% of the sources. Again, there is a wide variety of behaviors (see discussion in the text). \label{figure:recall_ndet}}
\end{center}
\end{figure*}

To evaluate this, we created 20 new testing sets defined considering only those sources with $\geq6$ detections in both $g$ and $r$ bands, from the 20 testing sets previously generated using the \texttt{ShuffleSplit} iterator (see Section \ref{results_RF}). We then classified each new testing set with its respective model, considering: a) the features available for the $g$ and $r$ bands, b) the features available only for the $g$ band (i.e., we hide the features measured using the $r$ band), and c) the features available only for the $r$ band (i.e., we hide the features measured using the $g$ band). Figures \ref{figure:recall_mag} and \ref{figure:recall_ndet} illustrate the results of this analysis. Figure \ref{figure:recall_mag} reports the recall values as a function of the average magnitudes per class (i.e., the recall values are computed by comparing the true and predicted labels for the objects within each magnitude bin), with the SN subtypes grouped in the unique class SN. Figure \ref{figure:recall_ndet} reports the recall values as a function of the number of detections (i.e., the recall values are computed for each bin of the ZTF number of detections). From both figures we can infer that in general the best results are obtained when photometry from both $g$ and $r$ are available, with the exceptions of QSO, CEP and Periodic-Other classes.

From Figure \ref{figure:recall_mag} we can also conclude that the reliability of the classification versus the average magnitude is different for each class. These distributions in general follow the magnitude distribution in the labeled set of each class considered in this model (with the exception of RRL). For instance, for the labeled set, the CEP class corresponds to one of the brightest classes, having in general $r<16$, while the LPV class covers a broader range of magnitudes. On the other hand, from Figure \ref{figure:recall_ndet}, we can infer that in general the classification improves when more detections are available in both bands, with the exceptions of the QSO and Periodic-Other classes.

The results obtained for Periodic-Other are not surprising since this class includes all the periodic classes not considered in the classifier (including several different types of pulsating stars, as well as the rotational variables). The results observed for the CEP class are probably due to the large fraction of Cepheids in the labeled set with photometry only in the $g$ band, which is produced by the saturation limit of the ZTF survey (12.5 to 13.2 magnitudes), and the fact that Cepheids tend to be very bright particularly in the $r$ band, and thus for bright Cepheids the $r$ band light curve is not available. The results obtained for AGNs and QSOs are likely related with incorrect labels, which we discuss further in Appendix \ref{qso_agn}.

Peculiar results of the recall values of some classes are reported in Figures \ref{figure:recall_mag} and \ref{figure:recall_ndet} when only one band is available. For SNe, there is a decrease in the recall curve in the $g$ band, presumably due to the fact that in general the $g$-band light curves of SNe tend to decay faster than the $r$-band light curves, producing shorter (and thus fewer detections) $g$-band light curves. This trend can be seen in the SN shown in Figure \ref{figure:lc_example}, as well as more generally in the light curve statistics for SNe. The average number of detections of SNe light curves is 12 and 16 in the $g$ and $r$ bands, respectively, and the total time length of SNe light curves corresponds to 53 and 64 days in the $g$ and $r$ bands, respectively. The low recall obtained for bright RRL when only the $g$ band is available may be  produced by differences in the variability features measured for different RRL sub-types. This issue is further discussed in Appendix \ref{RRL_case}. The zero value recall curves obtained for the CV/Nova class when only the $g$ band is available is produced by the similarities in the AllWISE+ZTF colors of CV/Novae and some periodic classes. This is discussed in Appendix \ref{CV_case}. Despite this, most of the sources in the labeled set have photometry in both $g$ and $r$ bands (see Figure \ref{figure:band_dist}), and thus this low performance obtained for some classes when only one band is available does not highly affect the presented results.

\subsection{The deployed BRF classifier}\label{final_RF_class}

\begin{figure}[htbp]
\begin{center}

\includegraphics[scale=0.33]{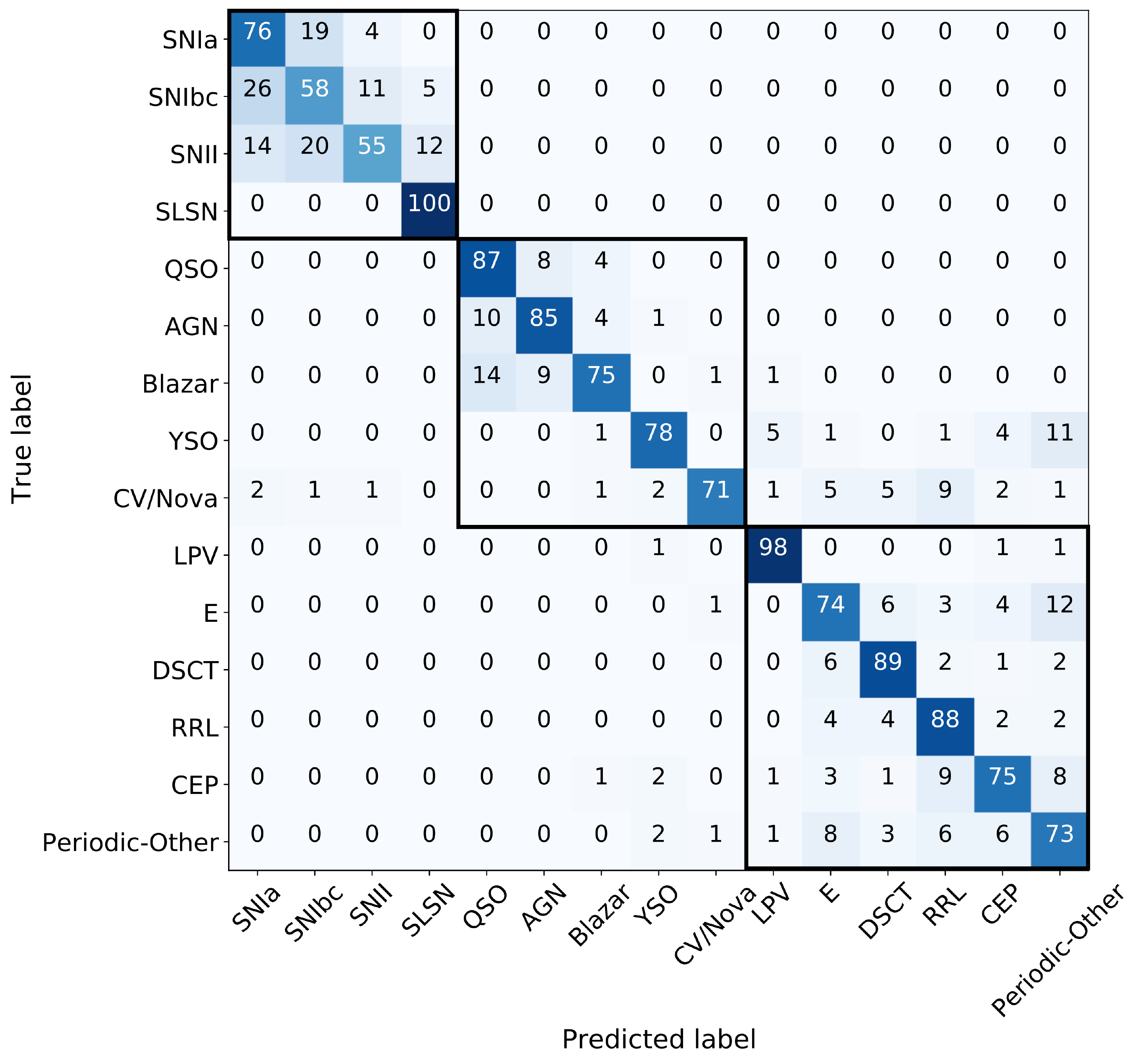}

\caption{As in Figure \ref{figure:conf_max_others}, but for the bottom level of the deployed BRF classifier.  \label{figure:conf_max_deployed}}
\end{center}
\end{figure}

In order to use the BRF classifier to classify the ZTF alert stream we need to train a single BRF model. We call this model ``the deployed BRF classifier''. This  ``deployed'' model corresponds to a totally-independent classifier (i.e., it is different from the previously trained 20 models), and uses the same 152 features presented in Section \ref{features}. As in the previous sections, we trained the deployed BRF model using 80\% of the labeled set as the training set and the remaining 20\% as the testing set. The macro-averaged precision, recall, and F1-score obtained for the top level are 0.96, 0.98, and 0.97, respectively, while for the bottom level they are 0.58, 0.79, and 0.61. The reason for these large differences in the macro-averaged metrics between the top and bottom levels can be understood from Figure \ref{figure:conf_max_deployed}, which shows the confusion matrix of the bottom level of the deployed BRF model. This confusion matrix is in agreement with the results presented in Figure \ref{figure:conf_mat_BRF}, and shows that some classes, such as the the SNIbc, CV/Nova, YSO,  CEP, and Periodic-Other classes, require further work in order to improve the results obtained by the classifier; this may involve data augmentation procedures, and better period estimations, as discussed in the following sections.

\subsubsection{Feature ranking of the deployed BRF classifier}\label{feat_relevance}

In Table \ref{table:features_ranking} we list the feature ranking (top 30) for each classifier within the two-level BRF classifier (top level, Transient, Stochastic, and Periodic). The feature ranking is computed considering which features separate better the subclasses within each classifier, with more informative features having higher ranks (for more details see \citealt{hastie2009elements}). From the table we can see that for all the classifiers, a considerable fraction of the top 30 features correspond to colors computed using the AllWISE and ZTF photometry, as well as new detection features (i.e., features not included in the FATS package) and non-detection features. Moreover, it can be observed that the ranking of features changes for each classifier.

The top level classifier is dominated by different types of features: ZTF and ALLWISE colors, morphological properties of the images (\texttt{sgscore1}), 
variability features related with the amplitude of the variability at short and long timescales (\texttt{MHPS\_low}, \texttt{GP\_DRW\_sigma}, \texttt{Meanvariance}, \texttt{ExcessVar}, and \texttt{SPM\_A}),
variability features that detect smooth decrease or increase of the luminosity (\texttt{LinearTrend}, \texttt{SPM\_tau\_rise},\texttt{SPM\_tau\_fall}), features related with the quality of a supernova parametric model fitting (\texttt{SPM\_chi}), and features related with transient appearance or disappearance (\texttt{positive\_fraction}, and \texttt{n\_non\_det\_after\_fid}). 

On the other hand, the Transient classifier is dominated by the SPM features (e.g., \texttt{SPM\_gamma}, \texttt{SPM\_beta}, \texttt{SPM\_t0}, \texttt{SPM\_tau\_rise}, and \texttt{SPM\_tau\_fall}). Other relevant features are the optical colors in the peak and the mean of the light curve, measured from the difference image light curves, features that detect smooth increase or decrease of the observed flux (\texttt{LinearTrend}), features that measure the level of correlation in the light curve (\texttt{IAR\_phi}), features related with the amplitude of the variability (\texttt{MHPS\_low}), and features related with the appearance of a transient source (\texttt{dmag\_first\_det\_fid}, \texttt{last\_diffmaglim\_before\_fid\_1}). Note that SN rise related features, such as \texttt{SPM\_t0} and \texttt{SPM\_rise}, are some of the most relevant features for the classification of transients, and are crucial for the early classification of SNe. Also, note that \texttt{SPM\_t0} is not the explosion time, but some characteristic time where the SN has risen significantly.

For the Stochastic classifier, 12 of the top 30 features are related with color, morphology and distance to the Galactic plane, and the rest correspond to features related with the amplitude of the variability observed at different time scales (e.g., \texttt{ExcessVar}, \texttt{SPM\_A},\texttt{Meanvariance}, \texttt{GP\_DRW\_sigma}, and \texttt{Amplitude}), and features related with the time scale of the variability (\texttt{IAR\_phi}, \texttt{GP\_DRW\_tau}).

Finally, the Periodic classifier is clearly dominated by the \texttt{Multiband\_period} feature, but also by different colors, by features related with the amplitude of the variability (e.g., \texttt{delta\_mag\_fid}, \texttt{Amplitude}, \texttt{ExcessVar}, \texttt{Meanvariance}, and \texttt{GP\_DRW\_sigma}), and features related with the timescale of the variations (e.g., \texttt{GP\_DRW\_tau}, and \texttt{IAR\_phi}).

\begin{table*}
\footnotesize
\begin{center}
\caption{Feature ranking (top 30) for each layer of the deployed BRF classifier. Features marked with $\dagger$ correspond to non-detection features, features marked with $\star$ correspond to features computed using AllWISE data, and features marked with $\ddagger$ correspond to metadata features. Subscripts ``\_1'' and ``\_2'' refers respectively to $g$ and $r$ bands.
} \label{table:features_ranking} 
\begin{tabular}{lll|llll|llll|lll} \hline
\hline

\multicolumn{2}{c}{Top level} & & & \multicolumn{2}{c}{Transient}  & & & \multicolumn{2}{c}{Stochastic} & & & \multicolumn{2}{c}{Periodic}  \\

Feature &  Rank & & & Feature &  Rank  & & & Feature &  Rank & & & Feature &  Rank    \\
\hline

\texttt{W1-W2}$\star$ & 0.094 & & &  \texttt{SPM\_t0\_1} & 0.033 & & & \texttt{W1-W2}$\star$ & 0.109 & & & \texttt{Multiband\_period} & 0.089\\

\texttt{sgscore1}$\ddagger$ & 0.053 & & & \texttt{SPM\_gamma\_2} & 0.029 & & & \texttt{sgscore1}$\ddagger$ & 0.058 & & & \texttt{$g$-W2}$\star$ & 0.062\\

\texttt{positive\_fraction\_2} & 0.050 & & & \texttt{SPM\_tau\_rise\_2} & 0.028 & & & \texttt{$r$-W2}$\star$ & 0.049 & & & \texttt{$r$-W2}$\star$ & 0.034\\

\texttt{positive\_fraction\_1} & 0.048 & & & \texttt{SPM\_tau\_rise\_1} & 0.025 & & & \texttt{($g$-$r$)\_mean\_corr} & 0.048 & & & \texttt{($g$-$r$)\_max\_corr} & 0.030\\

\texttt{SPM\_tau\_rise\_1} & 0.035 & & & \texttt{($g$-$r$)\_max} & 0.023 & & & \texttt{$g$-W2}$\star$ & 0.046 & & & \texttt{$g$-W3}$\star$ & 0.028\\

\texttt{LinearTrend\_2} & 0.032 & & & \texttt{SPM\_t0\_2} & 0.022 & & & \texttt{gal\_b}$\ddagger$ & 0.045 & & & \texttt{($g$-$r$)\_mean} & 0.027\\

\texttt{SPM\_chi\_1} & 0.031 & & & \texttt{LinearTrend\_2} & 0.019 & & & \texttt{$g$-W3}$\star$ & 0.037 & & & \texttt{($g$-$r$)\_max} & 0.025\\

\texttt{$g$-W2}$\star$ & 0.031 & & & \texttt{AndersonDarling\_2} & 0.018 & & & \texttt{($g$-$r$)\_max\_corr} & 0.035 & & & \texttt{GP\_DRW\_tau\_1} & 0.023\\

\texttt{$g$-W3}$\star$ & 0.031 & & & \texttt{SPM\_gamma\_1} & 0.017 & & & \texttt{ExcessVar\_2} & 0.033 & & & \texttt{($g$-$r$)\_mean\_corr} & 0.022\\

\texttt{n\_non\_det\_after\_fid\_2}$\dagger$ & 0.026 & & & \texttt{SPM\_tau\_fall\_2} & 0.017 & & & \texttt{Meanvariance\_2} & 0.026 & & & \texttt{IAR\_phi\_1} & 0.022\\

\texttt{W2-W3}$\star$ & 0.025 & & & \texttt{dmag\_first\_det\_fid\_1}$\dagger$ & 0.015 & & & \texttt{($g$-$r$)\_mean} & 0.025 & & & \texttt{Amplitude\_1} & 0.017\\

\texttt{SPM\_gamma\_1} & 0.024 & & & \texttt{MHPS\_low\_1} & 0.013 & & & \texttt{delta\_mag\_fid\_2} & 0.024 & & & \texttt{ExcessVar\_1} & 0.017\\

\texttt{SPM\_tau\_rise\_2} & 0.023 & & & \texttt{LinearTrend\_1} & 0.013 & & & \texttt{$r$-W3}$\star$ & 0.023 & & & \texttt{delta\_mag\_fid\_1} & 0.016\\

\texttt{SPM\_A\_2} & 0.023 & & &\texttt{($g$-$r$)\_mean} & 0.012 & & & \texttt{W2-W3}$\star$ & 0.022 & & & \texttt{Meanvariance\_1} & 0.016\\

\texttt{SPM\_chi\_2} & 0.020 & & & \texttt{MHPS\_ratio\_1} & 0.011 & & & \texttt{Amplitude\_2} & 0.022 & & & \texttt{$r$-W3}$\star$ & 0.016\\

\texttt{SPM\_A\_1} & 0.019 & & & \texttt{SPM\_tau\_fall\_1} & 0.011 & & & \texttt{Std\_2} & 0.015 & & & \texttt{Std\_1} & 0.015\\

\texttt{$r$-W2}$\star$ & 0.018 & & & \texttt{SPM\_beta\_2} & 0.011 & & & \texttt{($g$-$r$)\_max} & 0.015 & & & \texttt{GP\_DRW\_sigma\_1} & 0.015\\

\texttt{ExcessVar\_1} & 0.017 & & & \texttt{MHPS\_ratio\_2} & 0.010 & & & \texttt{SPM\_A\_2} & 0.014 & & & \texttt{GP\_DRW\_tau\_2} & 0.012\\

\texttt{ExcessVar\_2} & 0.016 & & & \texttt{Skew\_2} & 0.009 & & & \texttt{PercentAmplitude\_2} & 0.013 & & & \texttt{PercentAmplitude\_1} & 0.012\\

\texttt{$r$-W3}$\star$ & 0.014 & & & \texttt{sgscore1}$\ddagger$ & 0.009 & & & \texttt{SPM\_A\_1} & 0.012 & & & \texttt{W1-W2}$\star$ & 0.009\\

\texttt{Rcs\_2} & 0.013  & & &  \texttt{SPM\_beta\_1} & 0.009  & & & \texttt{ExcessVar\_1} & 0.011  & & & \texttt{W2-W3}$\star$ & 0.009\\

\texttt{GP\_DRW\_sigma\_2} & 0.013  & & & \texttt{Power\_rate\_2} & 0.009  & & & \texttt{IAR\_phi\_1} & 0.010 & & & \texttt{SF\_ML\_amplitude\_1} & 0.009\\

\texttt{SPM\_tau\_fall\_1} & 0.013  & & & \texttt{IAR\_phi\_2} & 0.009  & & & \texttt{GP\_DRW\_sigma\_2} &0.009 & & & \texttt{SPM\_A\_1} & 0.009\\

\texttt{Meanvariance\_2} & 0.012  & & & \texttt{dmag\_first\_det\_fid\_2}$\dagger$ & 0.009  & & & \texttt{delta\_mag\_fid\_1} & 0.009 & & & \texttt{Gskew\_1} & 0.009\\

\texttt{LinearTrend\_1} & 0.012  & & & \texttt{IAR\_phi\_1} & 0.009  & & & \texttt{Pvar\_2} & 0.007 & & & \texttt{Q31\_1} & 0.009\\

\texttt{SPM\_gamma\_2} & 0.010  & & & \texttt{last\_diffmaglim\_before\_fid\_1}$\dagger$ & 0.009  & & & \texttt{IAR\_phi\_2} & 0.007  & & & \texttt{Autocor\_length\_1} & 0.009\\

\texttt{Pvar\_2} & 0.010  & & & \texttt{PPE} & 0.008  & & & \texttt{GP\_DRW\_tau\_2} & 0.007 & & & \texttt{IAR\_phi\_2} & 0.008\\

\texttt{MHPS\_low\_2} & 0.009  & & & \texttt{Harmonics\_mag\_6\_1} & 0.008  & & & \texttt{GP\_DRW\_tau\_1} & 0.007 & & & \texttt{SF\_ML\_gamma\_1} & 0.008\\

\texttt{SF\_ML\_amplitude\_2} & 0.009  & & & \texttt{MHPS\_low\_2} & 0.008  & & & \texttt{MHPS\_high\_1} & 0.007 & & & \texttt{Amplitude\_2} & 0.007\\

\texttt{($g$-$r$)\_max\_corr} & 0.009  & & &  \texttt{Gskew\_2} & 0.008  & & &  \texttt{MHPS\_low\_1} & 0.006 & & & \texttt{delta\_mag\_fid\_2} & 0.007\\

\hline
 \hline

\end{tabular}
\end{center}
\end{table*}

\subsubsection{Results for the unlabeled ZTF set}\label{results_unlabeled_set}

We now turn to discuss the results obtained when applying the final BRF classifier to all the sources in ZTF with $\geq6$ detections in $g$ or $\geq6$ detections in $r$. Considering the data obtained by ZTF until 2020/06/09, there are 868,371 sources that satisfy this condition, hereafter defined as the ``unlabeled ZTF set''. We define the class of a given object in the unlabeled ZTF set by selecting the class with the maximum probability obtained by the deployed BRF classifier. However, users of this classifier can use the obtained probabilities to make their own probability cuts and select samples for their science. The features, classifications and probabilities obtained for the unlabeled ZTF set with data until 2020/06/09 can be downloaded at Zenodo: \dataset[10.5281/zenodo.4279623]{https://doi.org/10.5281/zenodo.4279623}. 

It is important to note that the classifications obtained by the light curve classifier are updated every day, as new alerts are received. Whenever a new alert is received for a given source, the ALeRCE pipeline recomputes its variability features and provides an updated classification. These updated classifications can be found at the \href{http://alerce.online}{ALeRCE Explorer website}, using the ``light curve classifier'' tab, and specifying the desired class. Considering the results shown in Figure \ref{figure:recall_ndet}, we expect that the quality of the classification for a given source will improve as more detections are added to the light curve. In addition, with new alerts, more objects will satisfy the condition of having $\geq6$ detections in $g$ or $\geq6$ detections in $r$, and thus, the size of the unlabeled ZTF set increases every day. Moreover, with new detections the size of the labeled set will increase, allowing the training of new BRF models. Updates regarding any possible modification to the light curve classifier (e.g., labeled set and models) will be published on the \href{http://alerce.science}{ALeRCE Science website}.

\begin{figure}[htbp]
\begin{center}

 \includegraphics[scale=0.6]{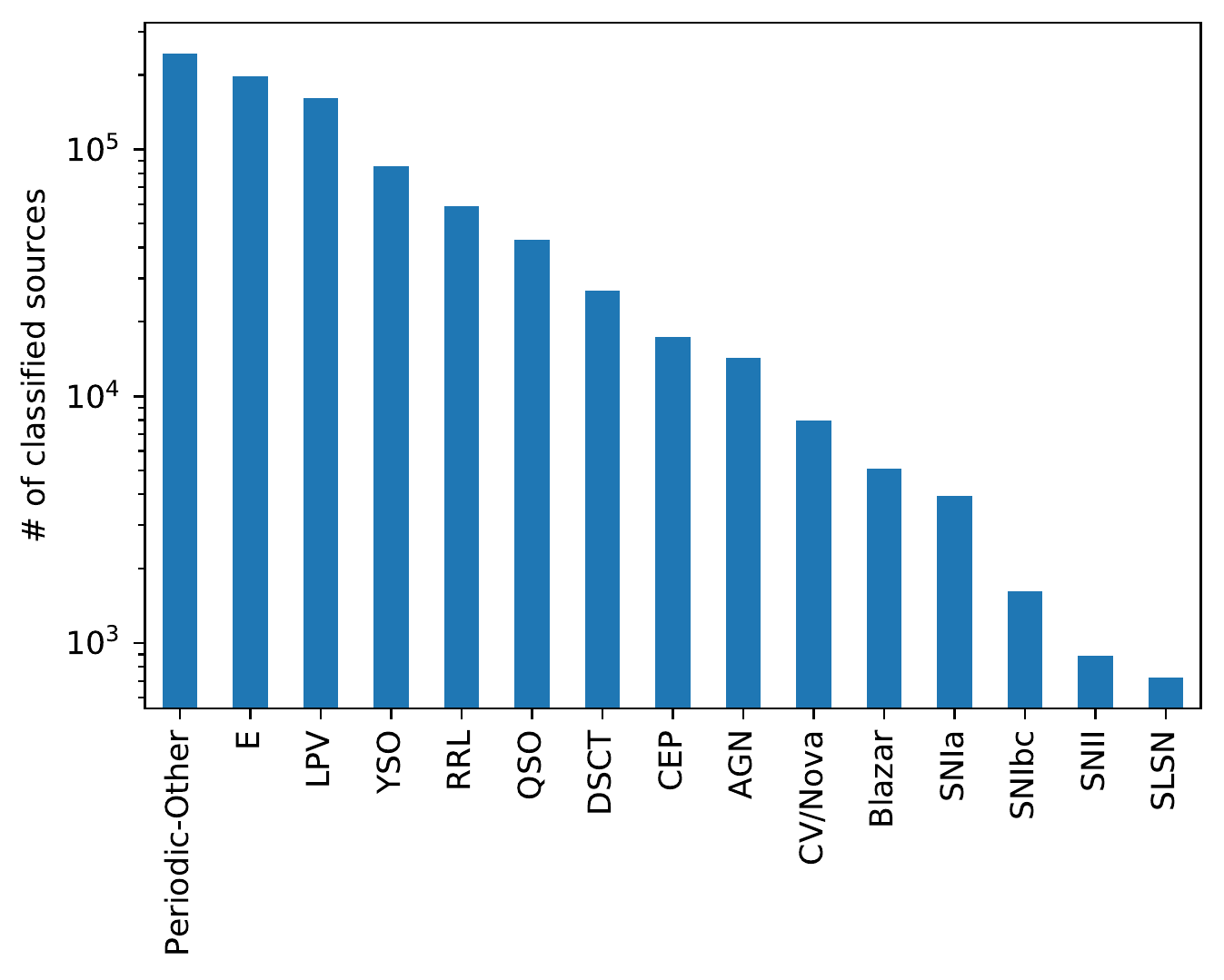}

\caption{Number of candidates per class for all the sources in the unlabeled ZTF set. It can be seen that the number of sources per class for the unlabeled ZTF set does not correlate with the number of sources per class for the labeled set (see Figure \ref{figure:num_classes_ls}).  \label{figure:num_classes}}
\end{center}
\end{figure}

Figure \ref{figure:num_classes} shows the number of candidates per class obtained for the 868,371 sources with enough alerts until 2020/06/09. Compared to Figure \ref{figure:num_classes_ls}, it can be noticed that there is no correlation between the number of sources per class for the unlababel set and the number of sources per class for the labeled set. The Periodic-Other, E, and LPV classes have the highest number of candidates, while the SN classes have the lowest. The distribution of candidates per class is consistent with the astrophysical number densities (i.e., we are likely not misclassifying large numbers of sources). For instance, Blazars are relativistically beamed (and thus seen to have farther distances), but only over very small viewing angles, and hence are expected to be less common than QSOs and AGNs.  In the case of SNe, not factoring in the amount of time a particular SN is above the magnitude limits of the search (the ``control time''), we find ratios of SNe~II/SNe~Ia and SNe~Ibc/SN~Ia of 0.21 and 0.41, respectively. Computing these ratios using the number of such classes reported from ASAS-SN discoveries in \cite{Holoien19} yields 0.36 and 0.09, respectively. The significant differences between the SNe~Ibc/SN~Ia ratios implies that we are strongly overestimating the numbers of SN~Ibc; given the similarities between SN~Ibc and SN~Ia light curves, we are likely classifying a non-negligible fraction of SN~Ia as SN~Ibc. This highlights the importance of including distance estimations to improve the classification of transients. We are currently working to include distance-based features in future versions of the classifier.

\begin{figure*}[htbp]
\begin{center}
\begin{tabular}{cc}
   \includegraphics[scale=0.55]{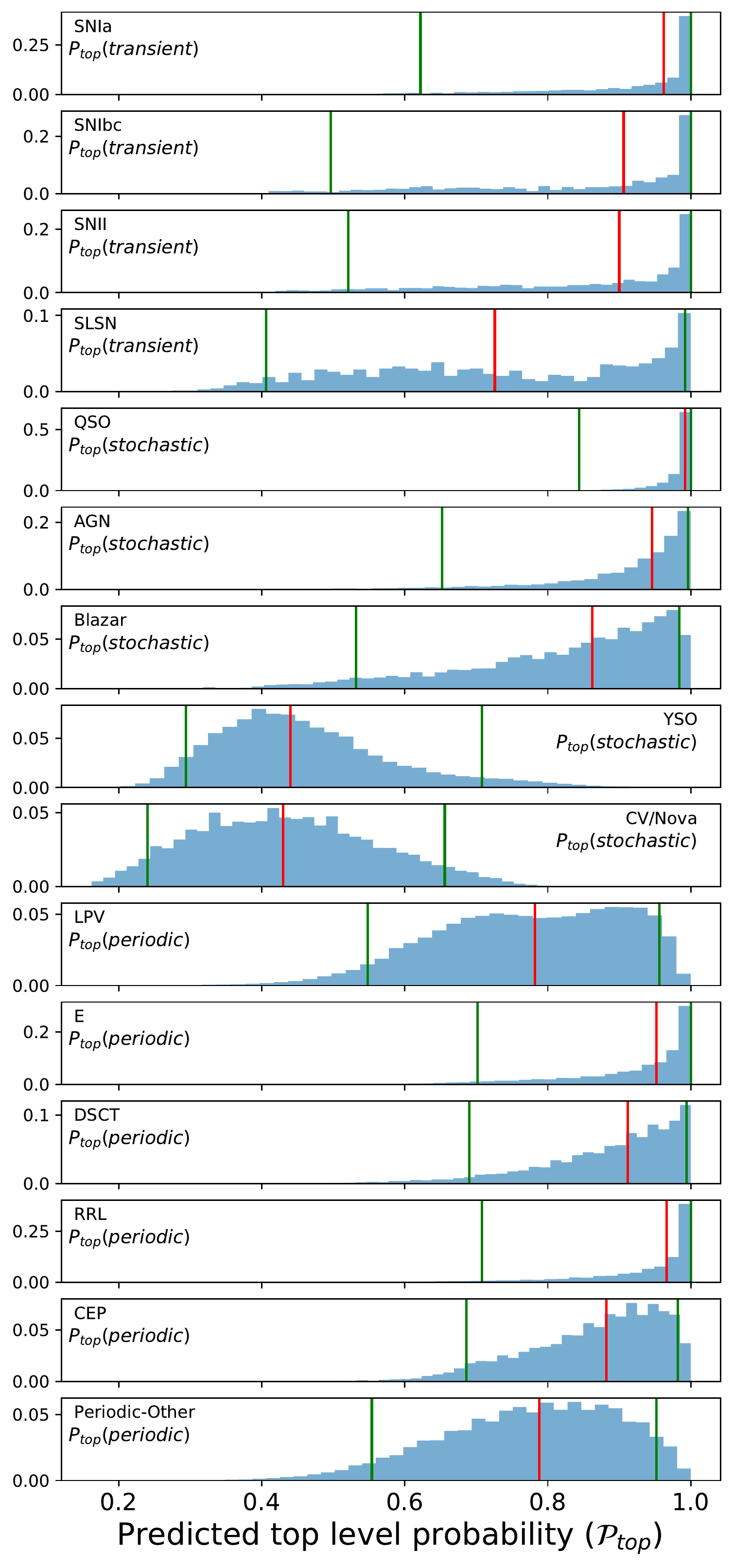} &

  \includegraphics[scale=0.55]{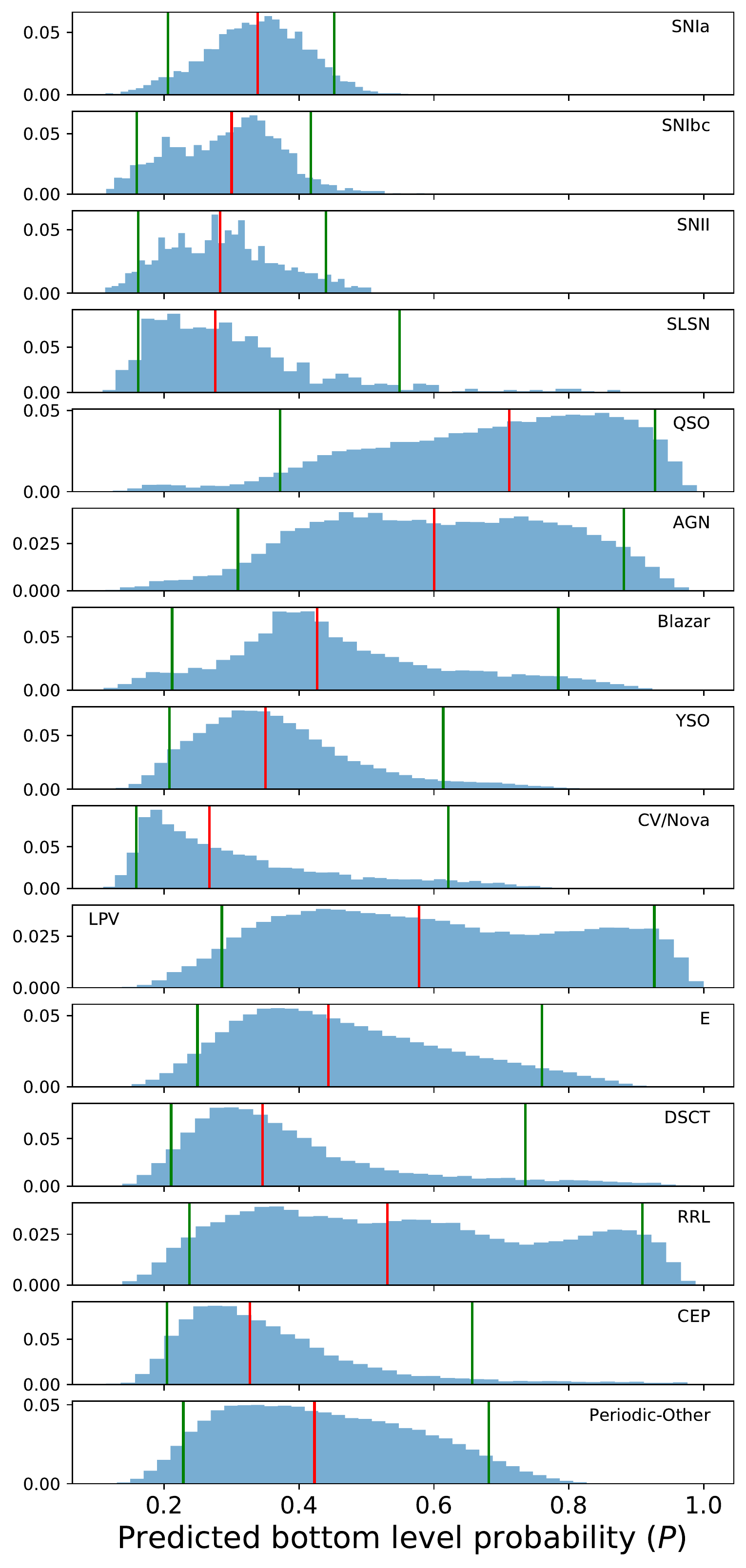} \\

\end{tabular}
\caption{Left: normalized probability distributions of the top level of the deployed BRF classifier, split by subclass, for candidates from the unlabeled ZTF set. The reported values correspond to the probabilities obtained for each class of the top level, as indicated below the class name. Right: normalized probability distributions of the bottom level of the deployed BRF classifier, split by subclass, for candidates from the unlabeled ZTF set. The red lines show the median probability for each class. The green lines show the 5 and 95 percentiles of the probabilities. Some subclasses show broad distributions to low values, implying that they are not so well-represented or characterized by the highest ranked features within the hierarchical class.
\label{figure:prob_dist}}
\end{center}
\end{figure*}

To investigate the quality of the predictions, we plotted the probability distributions of the top and bottom levels on the left and right sides of Figure \ref{figure:prob_dist}. The red lines denote the position of the median probability for each class, and the green lines denote the 5 and 95 percentiles. It is clear from the figure that the distribution of probabilities for the top level are higher compared to the bottom level. For the top level, the classes with the lowest probabilities are CV/Nova, and YSO. For the bottom level, the classes with the highest probabilities are LPV, QSO, and AGN, and the lowest probabilities are for the different classes of SNe, YSO, CV/Nova, and some periodic variables. 

\begin{figure}[htbp]
\begin{center}
\begin{tabular}{c}
   \includegraphics[scale=0.4]{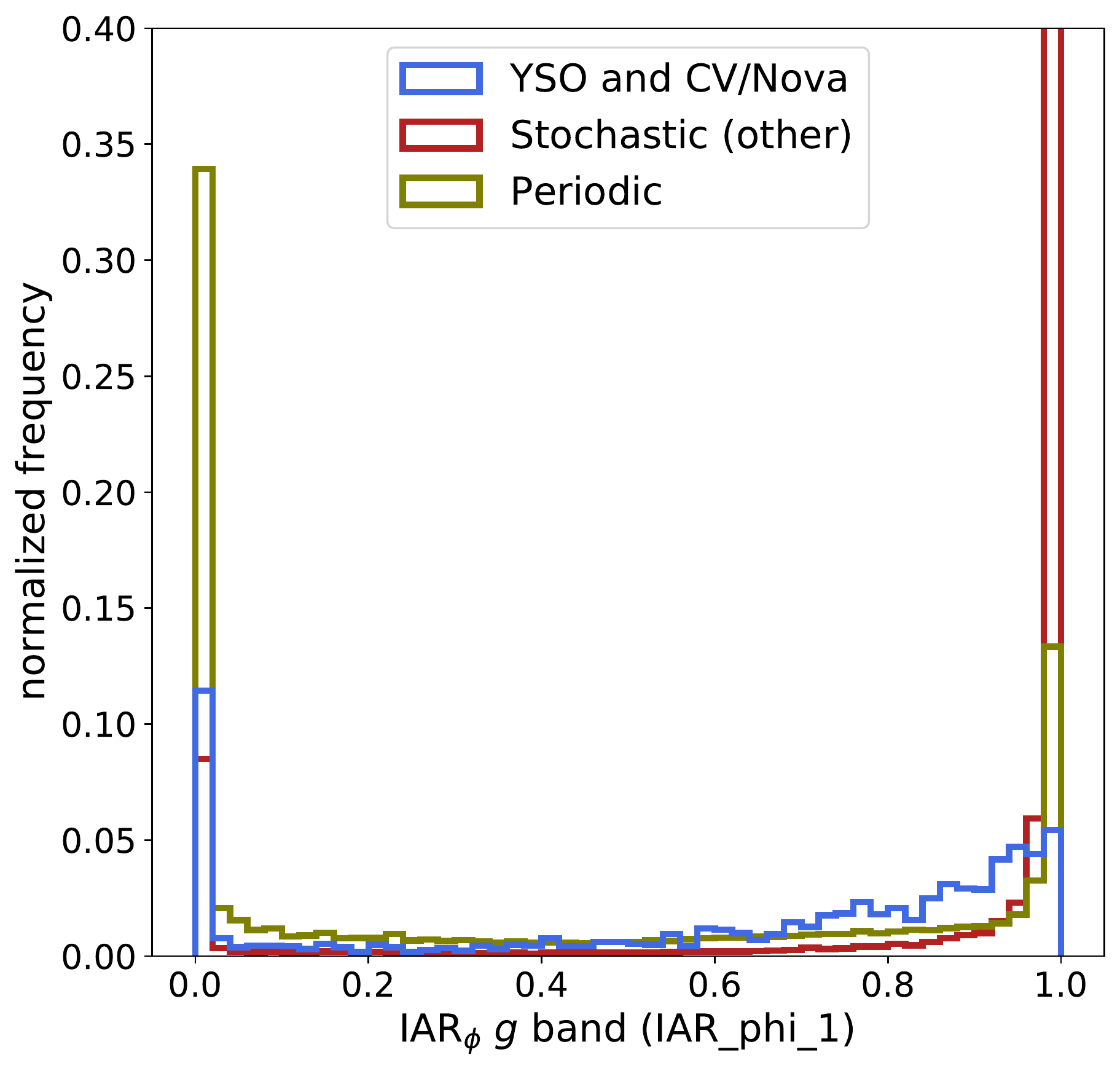} \\

  \includegraphics[scale=0.4]{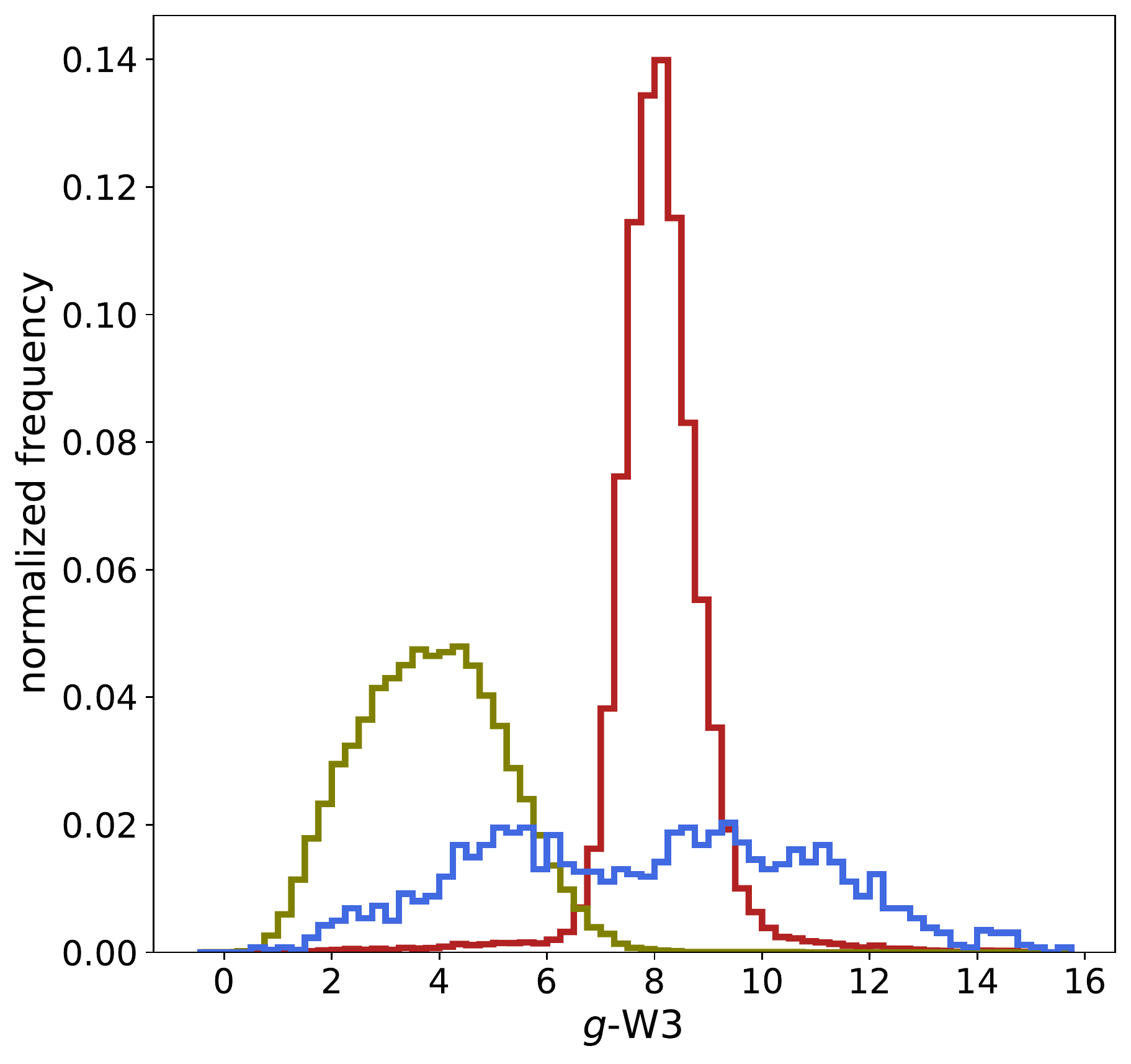} \\

\end{tabular}
\caption{Normalized IAR $\phi$ distribution (\textit{top}) in the $g$ band (\texttt{IAR\_phi\_1}; cut in the y-axes at 0.4), and  normalized \texttt{$g$-W3} distribution (\textit{bottom}), for YSOs and CV/Novae (blue), stochastic sources (red; excluding CV/Novae and YSOs), and periodic sources (yellow). It can be seen that there is an overlap in the \texttt{IAR\_phi\_1} and \texttt{$g$-W3} distributions of CV/Novae, YSOs,  and periodic sources.
\label{figure:cv_yso_dists}}
\end{center}
\end{figure}

The low probabilities obtained for some classes are related with the confusion between classes observed in Figure \ref{figure:conf_mat_BRF}. For instance, in Figure \ref{figure:conf_mat_BRF} we can see that the SN classes present a high confusion among them. On the other hand, the SNIa, SNIbc, SNII and SLSN median probabilities of sources classified as SNII are 0.16, 0.19, 0.28 and 0.19, respectively. The high confusion among the SN classes may be due to the low number of sources in the labeled set, but also to the intrinsic similarities among these classes. For example, the physical mechanism responsible for the main peak of the light curve of SNe Ia and SNe Ib/c is the same, the diffusion of energy deposited by radioactive $^{56}{\rm Ni}$ \citep{2008AIPC.1053..237A}. Indeed, \cite{Villar19} report that 15\% of their Type Ibc SNe are classified as Ia. This might be improved by performing data augmentation using, for example, Gaussian process modeling (e.g., \citealt{Boone19AJ}). On the other hand, the low probabilities observed for CV/Novae and YSOs can be produced by the similarities between their colors and the colors of some periodic sources, and the fact that some CV/Novae and YSOs present very rapid variability compared to the ZTF cadence, that produces light curves with low auto-correlation, and thus low values of the \texttt{IAR\_phi} parameter, which is normally observed for periodic sources (excluding LPVs). These similarities can be seen in Figure \ref{figure:cv_yso_dists}, where we show the distributions of \texttt{IAR\_phi\_1} and \texttt{$g$-W3} for YSOs and CV/Novae (grouped), the rest of the stochastic classes, and periodic sources from the labeled set.

\begin{figure}[htbp]
\begin{center}
  \includegraphics[scale=0.55]{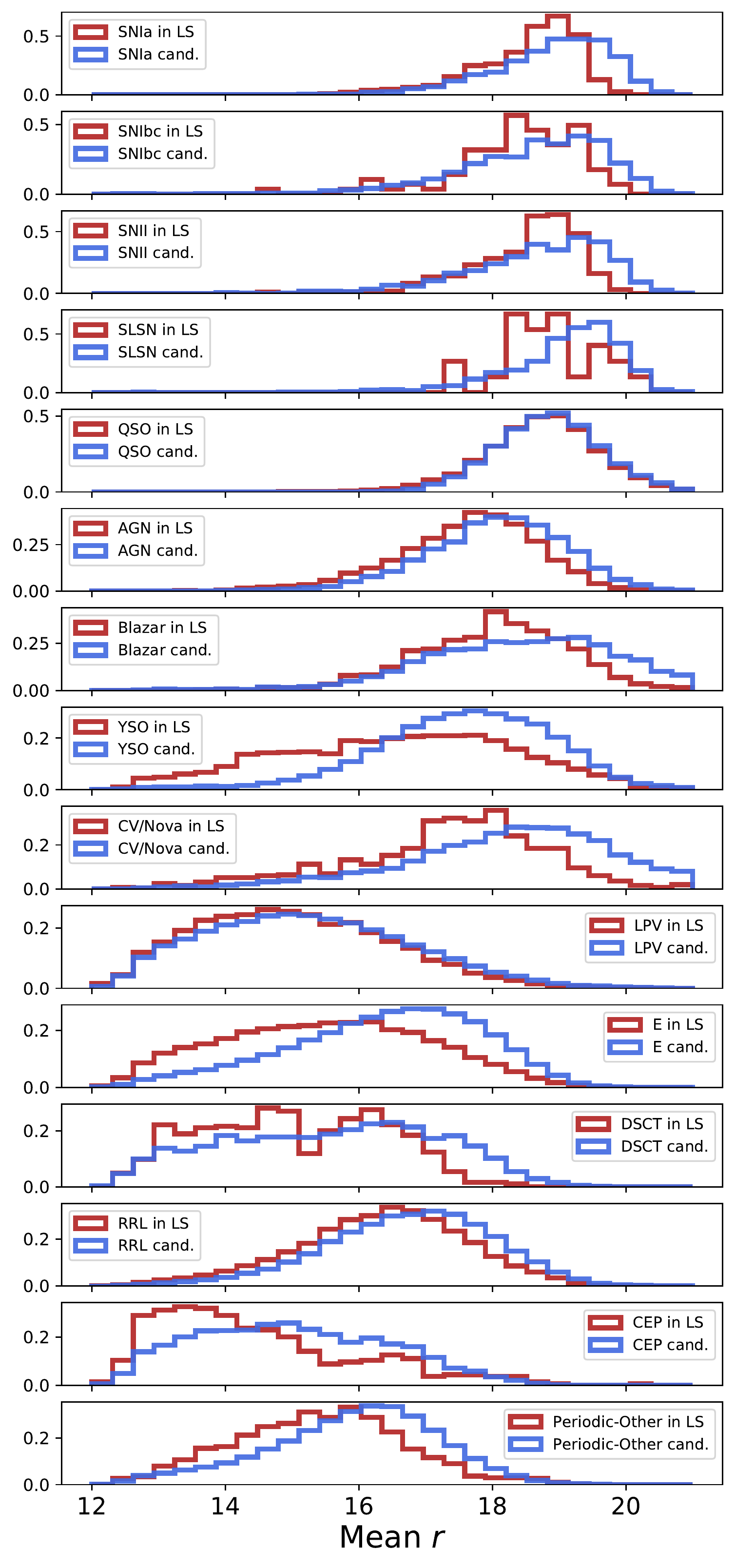} \\ 
 \caption{Normalized magnitude distributions in the $r$ band for sources in the labeled set (LS; red histograms) and candidates from the unlabeled ZTF set (cand.; blue histograms). \label{figure:rmag_dist}}
\end{center}
\end{figure}

Figure \ref{figure:rmag_dist} shows the normalized $r$ band magnitude distribution of the different classes considering sources present in the labeled set and candidates from the unlabeled ZTF set. In general, the distributions of magnitudes of the candidates are similar to or fainter than found among sources from the labeled set. For instance, the SNe classes have candidates that are $\sim0.5$ magnitudes fainter than the labeled set, and the YSO, CV/Nova and E classes have candidates that are $\sim1$ magnitude fainter than the labeled set. These results show that the classifier is able to detect faint and bright candidates, regardless of the luminosity biases present in the labeled set, which can be dominated by the brightest tail of the true magnitude distribution of each class. This will be particularly relevant for surveys like LSST, since, in general, available training set will be $\sim2-3$ magnitudes brighter than the limiting magnitudes of the single images, and thus we would expect that currently existent bright samples will allow us to detect fainter candidates.

\begin{figure*}[htbp]
\begin{center}

 \includegraphics[scale=0.57]{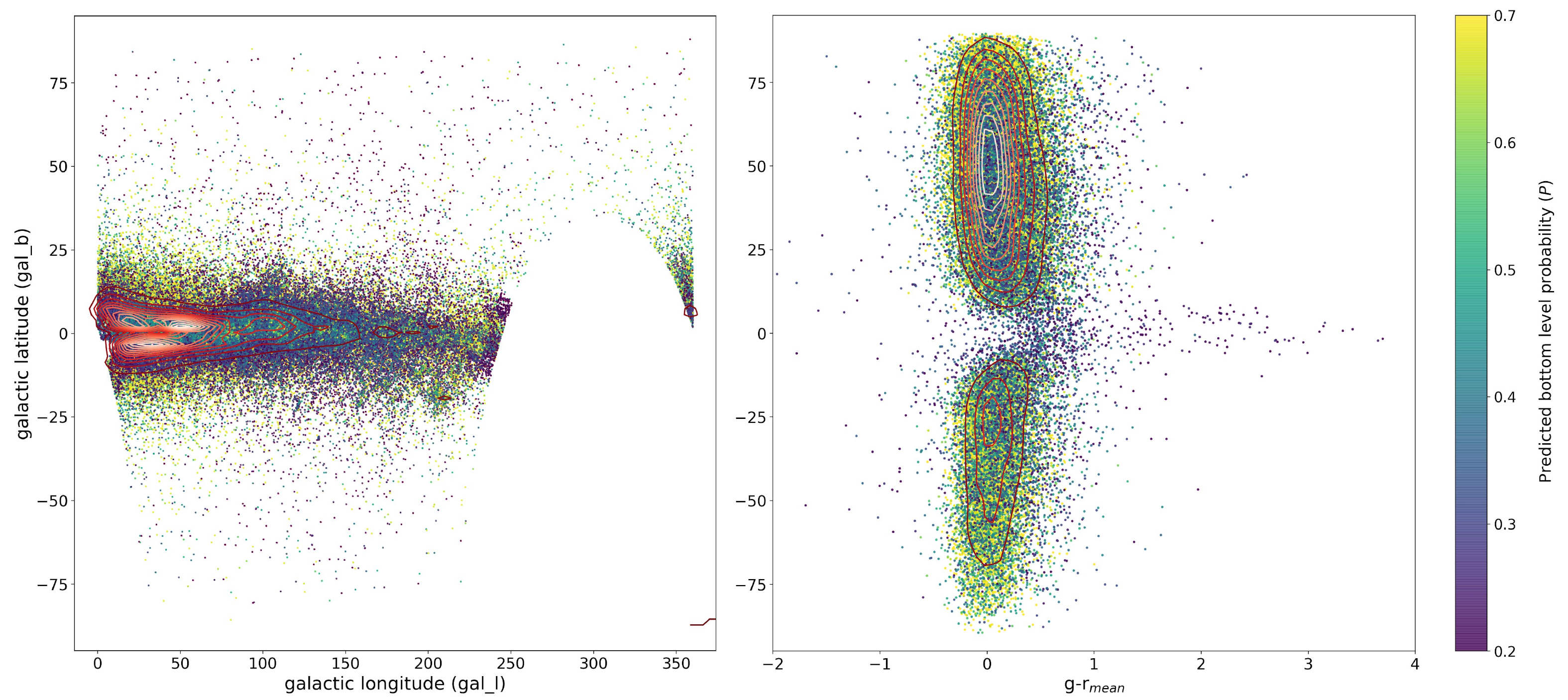}

\caption{Left: Galactic latitude (gal\_b, in degrees) versus Galactic longitude (gal\_l, in degrees) for candidates expected to be mostly observed around the Galactic plane (LPV, CEP, and YSO classes). Right: Galactic latitude (gal\_b, in degrees) versus $g-r_{max}$ for extragalactic candidates (QSO, AGN, Blazar, SNIa, SNIbc, SNII, and SLSN classes). The bottom level probability computed by the deployed BRF classifier are color-coded according to the colorbar to the right. The red contours show the density of points in each plot. It can be seen that most of the extragalactic candidates are located outside the Galactic plane, while most of the YSO, LPV and CEPcandidates are located in the Galactic plane. \label{figure:gal_b_plots}}
\end{center}
\end{figure*}

A simple way to test the performance of the BRF classifier is to verify whether the results obtained when the model is applied to the unlabeled ZTF set are in agreement with what is astrophysically expected from previous works. For instance, younger Galactic targets like YSOs, Classical Cepheids, and LPVs should reside near the Galactic plane (e.g., \citealt{CS15,Mowlavi18}), while  extragalactic sources like AGNs, QSOs, Blazars, and SNe should have roughly isotropic distributions, perhaps with fewer sources near the Galactic plane due to attenuation/reddening by gas and dust (e.g., \citealt{Calzetti00,Padovani17}). On the left side of Figure \ref{figure:gal_b_plots} the sky distribution, in Galactic coordinates, of LPV, CEP, and YSO candidates is shown. It is clear from the figure that most of them are located in the Galactic plane, and that sources located outside the plane have a low BRF probability. This is consistent with the results obtained by previous works (e.g., \citealt{Mowlavi18,Rimoldini19}). The right panel of Figure \ref{figure:gal_b_plots} shows the Galactic latitude versus the $g-r$ color obtained using the mean magnitude of the light curves in each band, for extragalactic candidates (QSO, AGN, Blazar, SNIa, SNIbc, SNII, and SLSN). From the figure we can see that the fraction of extragalactic candidates observed around the Galactic plane is low, and that most of the candidates located in the plane have low probabilities. Moreover, the $g-r$ colors of the extragalactic candidates are consistent with what is expected for these classes, with clear evidence of reddening for the candidates located around the Galactic plane. The sky distribution, in Galactic coordinates, of the extragalactic candidates can be found in Figure \ref{fig:extraga_gal_coords} of the appendix.

\section{Conclusions}\label{discussion}

\subsection{Summary}\label{summary}

In this paper we presented the first version of the ALeRCE light curve classifier. This classifier uses a total of 152 features, including variability features computed from ZTF light curves with $\geq6$ epochs in $g$ or $r$ bands, and colors computed using ZTF and AllWISE photometry (see Section \ref{features}), to classify each source into 15 subclasses, including periodic, transient, and stochastic variable sources (see Section \ref{taxonomy}). The light curve classifier uses a balanced RF classifier (see Section \ref{RF}), constructed with a two-level scheme. The first level (top level) classifies each source as periodic, stochastic or transient. The second level (bottom level) consists of three classifiers that further resolve each hierarchical class into subclasses.

We trained and tested the BRF classifier using a labeled set obtained by cross-matching the ZTF database with different catalogs of transients, stochastic and periodic sources (see Section \ref{trainingset}). For the top level we obtained macro-averaged precision, recall, and F1-score values of 0.96, 0.99, and 0.97, respectively, while for the bottom level we obtained macro-averaged precision, recall, and F1-score values of 0.57, 0.76, and 0.59, respectively.

We used the BRF classifier to classify 868,371 sources from ZTF (unlabeled ZTF set), obtaining results that are in agreement with what we expect astrophysically. For instance, most of the high probability extragalactic candidates are located outside the Galactic plane, and most of the high probability YSO, LPV, and CEP candidates are located in the Galactic plane.

The condition of $\geq6$ detections in $g$ or $r$ normally equates to a timespan between 3 and 30 days since the first detection. Whenever a new detection is received for an object, the ALeRCE pipeline processes it and provides an updated classification in $\sim 1$ second. The light curve classifier provides updated classifications for objects with new ZTF alerts every day. These updated classifications can be found on the \href{https://alerce.online}{ALeRCE Explorer website}, selecting the ``Light curve classifier'' option, and specifying the desired class. Catalogs containing the labeled set, and the features and RF probabilities obtained by the top and bottom levels for the unlabeled ZTF set (up to 2020/06/09) can be downloaded at Zenodo:  \dataset[10.5281/zenodo.4279623]{https://doi.org/10.5281/zenodo.4279623}. In addition, more examples and instructions on how to use the ALeRCE database and classifications can be found on the \href{http://alerce.science}{ALeRCE Science website} and in \cite{Forster20}, and a detailed description of the ALeRCE database can be found in the  \href{https://docs.google.com/spreadsheets/d/1OH3Dz-s8pWy-cY2FuT59miItHMrpIRm1aOVCRTPwLGA/edit?usp=sharing}{database schema}. Finally, the code used to train the deployed BRF classifier can be found on the \href{https://github.com/alercebroker/light_curve_classifier_SanchezSaez_2020}{``light\_curve\_classifier\_SanchezSaez\_2020'' GitHub repository}, and the implementation of this classifier in the ALeRCE pipeline can be found on the \href{https://github.com/alercebroker/lc_classifier}{``lc\_classifier'' GitHub repository}, which version  1.0.1  is archived in Zenodo \citep{lcclassifier}.

\subsection{Final remarks and perspectives}

One of the main challenges found during the development of the ALeRCE light curve classifier was the high imbalance present in the labeled set. For instance,
the transient sources represent 1.4\% of the labeled set, while the periodic sources represent 70.5\%. Each hierarchical class also suffers from high imbalance among its subclasses; for example, in the case of the transient class, SNIa comprise 74.0\% of the sample, while SLSN correspond to only 1.4\%. We addressed this problem by using the balanced RF implementation of the \texttt{imbalanced-learn} Python package, which follows the procedure proposed by \cite{Chen04}. This method uses a downsampling majority class technique to train each tree with a balanced sub-sample. We also tested two other algorithms, GBoost and MLP, but concluded that more work is needed if we want to obtain better results from those. 

Another challenge was to find features useful to separate the different classes. Previous works have normally used features similar to those available in the FATS Python package (e.g., \citealt{Kim14,Martinez-Palomera18}), however our first tests demonstrated that these features were not informative enough to separate the 15 classes considered by the light curve classifier, in particular the stochastic and transient classes. Thus, novel features were designed and implemented for this work, like the \texttt{IAR\_phi} parameter, the MHPS features, and the non-detection features. In addition, during the development of the light curve classifier we realized that some stochastic classes are hard to separate using just variability features. In particular, the separation of YSOs from the other stochastic classes improved significantly once we included AllWISE colors in the set of features.

Furthermore, the computation of reliable periods was quite challenging, particularly considering that the ALeRCE pipeline requires fast computation of features. \cite{Huijse18} demonstrated that very good results can be achieved by  quadratic mutual information (QMI) estimators, however these techniques are computational expensive [$O(n^2)$]. We solve this issue by using the MHAOV periodogram, which provides less reliable periods, but is much faster to compute [$O(n)$]. Periods become increasingly unreliable as the number of datapoints decreases, but the classification of periodic variables can still be accurate, as other features can compensate for the decreasing quality of the periodogram (e.g., features related with the amplitude and the timescale of the variability). ALeRCE is currently working to implement methods for period estimation that are both accurate and fast. Computing the periodogram is expensive for sources with a large number of detections. We are currently exploring so-called ``online'' periodograms, which are updated as new samples arrive, at a fraction of the computational cost of recomputing the period each time from scratch \citep{Zorich20}, as well as other techniques that might work better with eclipsing binary light curves (e.g., \citealt{Kovacs02,Mighell13}).

Moreover, the classification of the different SN classes was particularly challenging. First, the number of SNe in the labeled set was very small compared to other classes, and second,  the light curves of SN classes can present similarities, which makes their separation difficult, as discussed in Section \ref{results_unlabeled_set}. We solved this issue by using the \texttt{BalancedRandomForestClassifier} method from \texttt{imbalanced-learn}, and by including the SPM features, whose definition is a modification of the work of \cite{Villar19}. In the future we plan to test other techniques to improve the separation of SN classes. Previous works have performed Gaussian process regression to model SNe light curves and generate new light curves with different cadences therefrom (e.g., \citealt{Boone19AJ}). Moreover, better results can be obtained if we use information regarding the SN host galaxy (e.g. \citealt{2013ApJ...778..167F,Baldeschi2020}).

In the future we also plan to perform data augmentation to improve the classification of variable objects. For the case of variable stars, light curves can be modeled with Gaussian process or with a combination of harmonics, and then basic transformations can be applied to these models to obtain light curves with different periods and amplitudes (e.g., \citealt{Elorrieta16,Martinez-Palomera18,Castro18,Aguirre19,Hosenie20}). To the best of our knowledge for the case of AGNs, QSOs and Blazars no previous attempts to perform data augmentation have been made. A promising option is to use synthetic light curve generators that consider the physical processes behind the variability (e.g., \citealt{Sartori19}). 

Most of the features used by this classifier can be implemented and used to classify light curves from other data sets. In particular, for the case of LSST, the non-detection features can be adapted to work with the forced photometry that will be provided for each alert (\texttt{DIAForcedSources} in the Data Products Definition Document; \citealt{LSSTDataProducs}). LSST will also benefit from the multiband $ugrizy$ light curves. As we demonstrated in this work, in general the light curve classification improves when both ZTF $g$ and $r$ data are available. For the case of LSST this would be the same, and probably we should even be able to further resolve some of the subclasses presented in this work. For instance, using the $zy$ light curves we should be able to separate local type 1 and type 2 active galactic nuclei, since for low redshift sources, we can detect variability from the dusty torus at these wavelengths (see \citealt{Sanchez17} and references therein), or identify high redshift QSOs, whose emission is expected to be absorbed in the bluer bands. We encourage researches interested in classifying stochastic and transient sources in particular to use the novel (or modified) features presented in this work, like the \texttt{IAR\_phi} parameter, the MHPS features, the SPM features, and the non-detection features. 

It is worth to note that the ALeRCE light curve classifier is being constantly improved, and this work describes its Version 1.0. Future versions of this classifier may include new classes of variable and transient objects, as well as sub-classes of sources already present in the taxonomy (e.g., RRL types ab and c; classical and type II Cepheids; contact, detached, and semi-detached eclipsing binaries; among others). We are also working to find new features and techniques that can improve the performance of the classifier. Future work will report any changes included in the classifier model, like the inclusion of data augmentation, or the use of other classification strategies (e.g., semi supervised training). We recommend the users of this classifier to check the \href{http://alerce.science}{ALeRCE Science website} to get updates related with the different classifiers and the data processing. We are exploring different classification algorithms which are not based on manually designed features, but on automatically derived, recurrent, implicitly extracted features, via deep learning (e.g. \citealt{Naul18,Muthukrishna19,Becker20}). However, up to this point we have found that the former produce better results when applied to real data. Most likely, a combination of simulated and real data will be required to train reliable deep learning classification models in the future, as found by \cite{Carrasco-Davis19}.

\acknowledgments

The authors acknowledge support from: ANID Millennium Science Initiative ICN12\_009; ANID grants PIA AFB-170001 (F.F., I.R., C.V., E.C., D.R., L.S.G., C.S.C., D.R.U.), PIA ACT172033 (P.A.), Basal-CATA AFB-170002 (P.S.S., F.E.B., M.C., D.D.), FONDECYT Regular Nº 1200710 (F.F.), 1190818 (F.E.B.), 1200495 (F.E.B.), 1171678 (P.E.) and 1171273 (M.C.), FONDECYT Initiation Nº 11191130 (G.C.V.), and FONDECYT Postdoctorado Nº 3200250 (P.S.S.), and 3200222 (D.D.), the Max-Planck Society, Germany, through a Partner Group grant (P.A.), Vicerector\'ia de Investigaci\'on de Pontificia Universidad Cat\'olica de Chile (ECI).  

Powered@NLHPC: This research was partially supported by the supercomputing infrastructure of the NLHPC (ECM-02). This work has been possible thanks to the use of AWS-U.Chile-NLHPC credits.

Based on observations obtained with the Samuel Oschin 48-inch Telescope at the Palomar Observatory as part of the Zwicky Transient Facility project. ZTF is supported by the National Science Foundation under Grant No. AST-1440341 and a collaboration including Caltech, IPAC, the Weizmann Institute for Science, the Oskar Klein Center at Stockholm University, the University of Maryland, the University of Washington, Deutsches Elektronen-Synchrotron and Humboldt University, Los Alamos National Laboratories, the TANGO Consortium of Taiwan, the University of Wisconsin at Milwaukee, and Lawrence Berkeley National Laboratories. Operations are conducted by COO, IPAC, and UW. AllWISE makes use of data from WISE, which is a joint project of the University of California, Los Angeles, and the Jet Propulsion Laboratory/California Institute of Technology, and NEOWISE, which is a project of the Jet Propulsion Laboratory/California Institute of Technology. WISE and NEOWISE are funded by the National Aeronautics and Space Administration.

\software{FATS \citep{Nun15}, IAR\footnote{Python and R implementations are available in \url{https://github.com/felipeelorrieta/IAR_Model}} \citep{Eyheramendy18}, Imbalanced-learn \citep{imblearn}, Keras \citep{Keras}, P4J \citep{Huijse18}, Scikit-learn \citep{Pedregosa12}, Tensorflow \citep{Tensorflow}, XGBoost \citep{Chen16}. For graphical representations and data manipulation: Astropy \citep{Astropy13,Astropy18}, Jupyter \citep{Kluyver2016jupyter}, Matplotlib \citep{matplotlib}, Numpy \& Scipy \citep{numpy}, Pandas \citep{pandas}, Python \citep{van1995python}, Seaborn \citep{seaborn}.}

\appendix

\section{Further description of some variability features}\label{features_appendix}

In this section we provide additional description of some of the features listed in Table \ref{table:det_features} (those marked with **). These features correspond to new variants of features included in the FATS package and other works:

\begin{itemize}

\item Damp Random Walk (DRW) parameters: a DRW model is defined by a stochastic differential equation which includes a damping term that pushes the signal back to its mean: $dX(t)=-\frac{1}{\tau_{DRW}}X(t)dt+\sigma_{DRW}\sqrt{dt}\,\epsilon(t)+b\,dt,    \quad      \tau_{DRW},\sigma_{DRW},t>0$. $\tau_{DRW}$ corresponds to the characteristic time for the time series to become roughly uncorrelated, $\sigma_{DRW}$ corresponds to the amplitude of the variability at short timescales ($t \ll \tau_{DRW}$), and $\epsilon(t)$ is a white noise process with zero mean and variance equal to 1. DRW modelling is typically used to describe light curves of active galactic nuclei \citep{Kelly09}. In this case we obtained the $\sigma_{DRW}$ and $\tau_{DRW}$ parameters using Gaussian process regression, with a Ornstein-Uhlenbeck kernel, as in \cite{Graham17}. 
\texttt{GP\_DRW\_sigma} denotes $\sigma_{DRW}$, while \texttt{GP\_DRW\_tau} denotes  $\tau_{DRW}$.

\item Excess Variance $(\sigma_{\text{rms}})$: Measure of the intrinsic variability amplitude in a given band (see \citealt{Sanchez17}, and references therein). $\sigma^2_{\text{rms}}=(\sigma_{LC}^2-\overline{\sigma}_{m}^2)/\overline{m}^2$, where $\sigma_{LC}$ is the standard deviation of the light curve, $\overline{\sigma}_{m}$ is the average photometric error, and $\overline{m}$ is the average magnitude. We denoted $\sigma^2_{\text{rms}}$ as \texttt{ExcessVar}.

\item Multiband Period: The period is estimated using the Multi Harmonic Analysis of Variance (MHAOV) periodogram \citep{mhaov}, which is based on fitting periodic orthogonal polynomials to the data. A single period estimate per light curve is computed by fitting both bands using the MHAOV multiband extension proposed by \cite{Mondrik15}. We denote this period as \texttt{Multiband\_period}. For sources with detections only in $g$ or only in $r$, the \texttt{Multiband\_period} reports the single band period. To avoid overfitting the data when few samples are available we set the number of harmonics to one. This might not capture the best period for non-sinusoidal light curves, e.g., detached and semi-detached eclipsing binaries, returning a harmonic instead. We found that having a harmonic of the true period is in general sufficient to classify non-sinusoidal light curves correctly given that other features such as the \texttt{power rate} are included. We choose MHAOV for this analysis as it provides a good trade-off between performance and computational complexity. This method is now implemented and available in the \texttt{P4J} package \citep{Huijse18}.

\item Harmonics parameters: Harmonic series \citep{Stellingwerf86} are commonly used to model and classify periodic light curves \citep{Debosscher07,sarro2009automated, Richards11, Elorrieta16}. In this work we fit a harmonic series up to the seventh harmonic, according to the expression

\begin{equation}
    \label{harmonic_sin_cos}
    y(t_{j}) = \sum_{k=1}^{7} \left [ A_{k} \cos \left (\frac{2 \pi k t_{j} }{P} \right ) + B_{k} \sin \left (\frac{2 \pi k t_{j} }{P} \right ) \right ] + C,
\end{equation}

where $t_j$ corresponds to the observational time of the $j$-th detection, $P$ is the best candidate period computed from the multiband periodogram as above, and $y(t_{j})$ is the magnitude estimated by the harmonic model. Even though we use the \texttt{Multiband\_period}, the harmonic model is computed using the detections in each band independently. $A_k$, $B_k$ and $C$ for $k = 1, \dots, 7$ are obtained by minimizing the weighted mean square error between the observed magnitudes and the model. 

Note that the model is linear with respect to its parameters, so the latter can be computed using weighted linear regression. The inverse of the square of each observational error is used as a weight, which minimizes contributions from noisier observations. The cost function is given by

\begin{equation}
   \min_{A_k, B_k} \frac{1}{J}\sum_{j=1}^{J}\frac{[\text{mag}(t_{j}) - y(t_{j})]^{2}}{\text{sigma}(t_{j})^{2}},
\end{equation}

where $J$ is the number of observations, $\text{mag}(t_{j})$ is the observed magnitude at time $t_{j}$, and $\text{sigma}(t_{j})$ is the observational error at time $t_{j}$. The solution to the weighted least squares optimization problem is found using the Moore-Penrose pseudoinverse. This solution has the additional property of having the minimum Euclidean norm when the problem is underdetermined \citep{GeneralizedInverses}, which in this case corresponds to having less than 15 observations.

Once the parameters are learnt, equation \ref{harmonic_sin_cos} can be rewritten as 

\begin{equation}\label{eq:harmonics}
    y(t_{j}) = \sum_{k=1}^{7} M_{k} \cos \left (\frac{2 \pi k t_{j} }{P} - \phi_{k} \right ) + C, 
\end{equation}

with $M_k = \sqrt{A_{k}^{2} + B_{k}^{2}}$ and $\phi_{k} = \arctan (B_{k} / A_{k})$. In this way, the harmonics are now described by the amplitude and phase of each component. The model is shifted in time in order to have zero phase in the first harmonic, which is done following the expression $\phi'_{k} = \phi_{k} - k \phi_{1}$, replacing $\phi_k$ by $\phi'_k$ in Eq. \ref{eq:harmonics}.

Finally, the parameters $M_{k}$ for $k = 1, \dots, 7$, $\phi'_{k}$ for $i = 2, \dots, 7$ and the mean square error are used as features, which are denoted \texttt{Harmonics\_mag\_1}, ..., \texttt{Harmonics\_mag\_7}, \texttt{Harmonics\_phase\_2}, ..., \texttt{Harmonics\_phase\_7}, \texttt{Harmonics\_mse}, respectively.

\item $P_{\text{var}}$: Probability that the source is intrinsically variable in a given band (see \citealt{Paolillo04}, and references therein). It considers the $\chi^2$ of the light curve respect to its mean, and calculates the probability $P_{var}=P(\chi^2)$ that a $\chi^2$ lower or equal to the observed value could occur by chance for an intrinsically non-variable source, assuming that for each light curve its $\chi^2$ will follow a probability distribution described by an incomplete gamma function $\Gamma(\nu/2, \chi^2/2)$, where $\nu$ corresponds to the degrees of freedom. We denoted $P_{var}$ as \texttt{Pvar}.

\item Structure Function (SF) parameters: The SF quantifies the amplitude of the variability as a function of the time difference between pairs of detections ($\tau$). In this work we consider the definition provided by \cite{Caplar17}. We model the SF as a power law: $\text{SF}(\tau)=A_{\text{SF}}\left( \frac{\tau}{1\text{yr}}\right)^{\gamma_{\text{SF}}}$, where $\gamma_{\text{SF}}$ corresponds to the logarithmic gradient of the change in magnitude, and $A_{\text{SF}}$ corresponds to the amplitude of the variability at 1~yr. \texttt{SF\_ML\_amplitude} denotes $A_{\text{SF}}$, while \texttt{SF\_ML\_gamma} denotes $\gamma_{\text{SF}}$.

\item Supernova parametric model (SPM): \cite{2019ApJ...884...83V} introduced an analytic model describing SN light curves as a six parameter function, which they used to characterize and classify SN light curves from the Pan-STARRS1 Medium-deep Survey \citep{2016arXiv161205560C}. This model is an extension of previous empirical efforts to analytically describe supernova light curves, including the effects of different explosion times, normalization factors, initial rise timescales, rate of decline after peak, plateau lengths, or tail decay timescales. We introduce two modifications to this model. First we reparametrize the function to always remain positive in a simple validity range by a set of inequalities. After the first modification, the model is the following:

 \begin{align}\label{spm_intermediate_expression}
 F \: & = \: \begin{cases} \begin{array}{cc}\cfrac{A \left(1 - \beta' \frac{t - t_0}{t_1 - t_0}\right)}{1 + \exp{\left(-\frac{t - t_0}{\tau_{\rm rise}}\right)}} & \mbox{if} \; t < t_1  \\ 
\\
  \cfrac{A (1 - \beta') \exp{\left(-\frac{t - t_1}{\tau_{\rm fall}}\right)}}{1 + \exp{\left(-\frac{t - t_0}{\tau_{\rm rise}}\right)}} & \mbox{if} \; t \ge t_1,
  \end{array}
\end{cases}
\end{align}

where we also use $\gamma \equiv t_1 - t_0$ as a parameter instead of $t_1$. This function is positive valued when $A > 0$, $\gamma > 0$, $\tau_{\text{rise}} > 0$, $\tau_{\text{fall}} > 0$ and $0 < \beta' < 1$. 

The second difference with respect to \cite{Villar19} is replacing the piecewise-defined function for a soft transition between the two components. This is done by including a sigmoid function $\sigma(t) = 1 / (1 + \exp(-t))$, which allows a soft transition between zero and one. As the parameter $t_1$ defines the transition between the two pieces of the model in Eq. \ref{spm_intermediate_expression}, it cannot be optimized properly using first-order methods. Our proposed model allows using this technique effectively to learn the parameters of the model, which is given by the following equation:
 
 \begin{align}
   F \: = & \: \cfrac{A \left(1 - \beta' \frac{t - t_0}{t_1 - t_0}\right)}{1 + \exp{\left(-\frac{t - t_0}{\tau_{\rm rise}}\right)}} \cdot  \left[1 - \sigma \, \left( \frac{t-t_1}{3}\right) \right] 
          + \: \cfrac{A (1 - \beta') \exp{\left( - \frac{t - t_1}{ \tau_{\rm fall}} \right)}}{1 + \exp{\left(-\frac{t - t_0}{ \tau_{\rm rise}}\right)}} \cdot \left[\sigma \left(\frac{t - t_1}{3}\right)\right].
\end{align}

In this particular model, for all the sources we use the light curves  based on the difference images (\texttt{lc\_diff}). This is done to avoid the contamination from unrelated host galaxy emission, which can distort the real shape of the SNe light curves. We also subtract from $t$ the MJD value of the first detection observed for a given source.
We computed $A$ (\texttt{SPM\_A}), $\beta'$ (\texttt{SPM\_beta}), $t_0$ (\texttt{SPM\_t0}), $\gamma$ (\texttt{SPM\_gamma}), $\tau_{\text{rise}}$ (\texttt{SPM\_tau\_rise}), and $\tau_{\text{fall}}$ (\texttt{SPM\_tau\_fall}), for each band independently. In addition, we computed the reduced $\chi^2$ of the fit for the light curve, denoted as \texttt{SPM\_chi}. The parameters are found using the function \texttt{curve\_fit} provided by the Scipy library \citep{2020SciPy-NMeth}.

\end{itemize}

\section{One-level Multi-class RF model}\label{on-level-RF}

\begin{figure}
    \centering
    \includegraphics[scale=0.35]{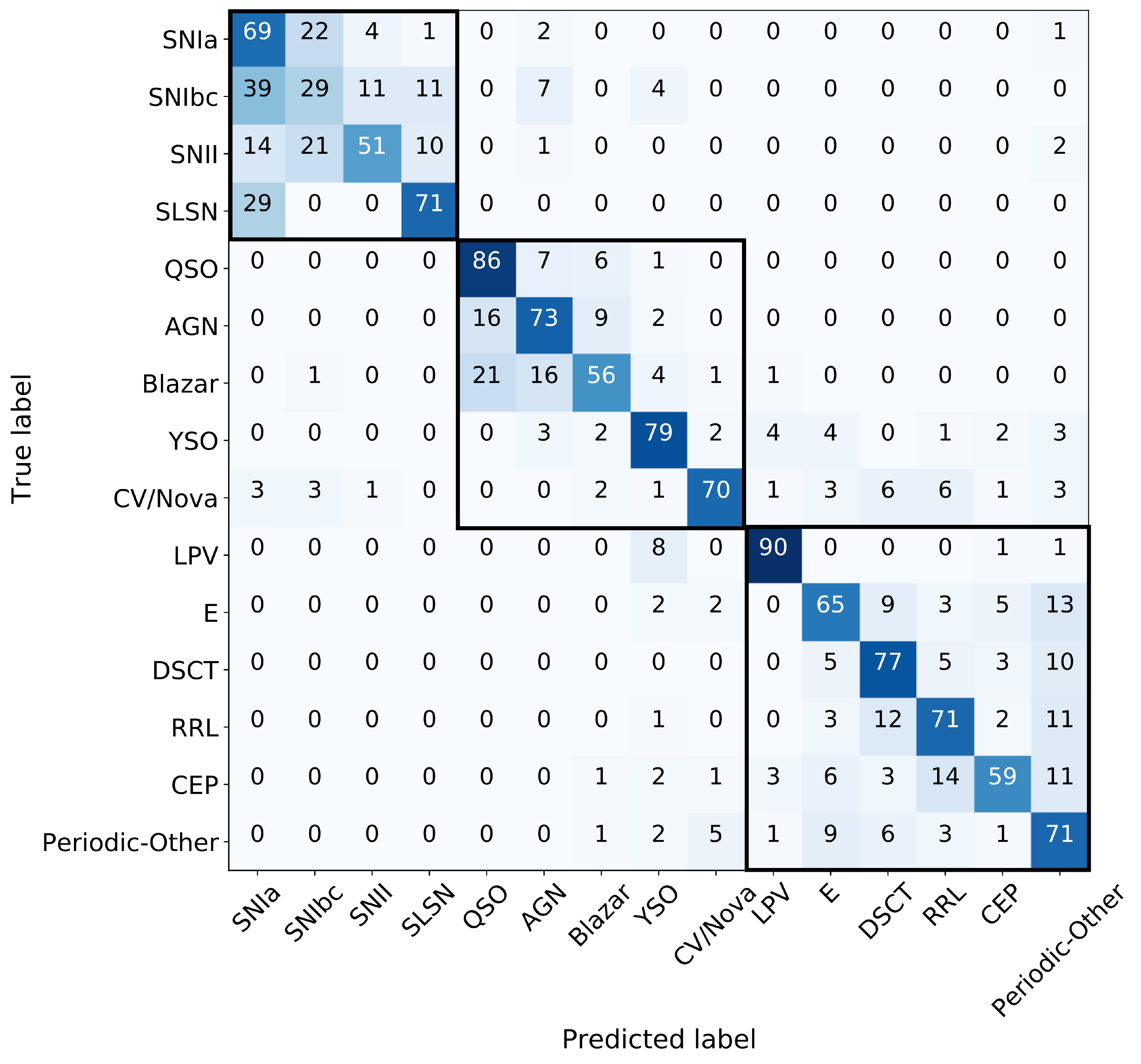}
    \caption{Confusion matrix for a one-level multi-class RF Model. The black squares highlight the three hierarchical classes (from top to bottom, transient, stochastic, and periodic, respectively). To normalize the confusion matrix results as percentages, we divide each row by the total number of objects per class with known labels. We round this percentages to integer values. The performance of this model is poorer compared to the BRF classifier (see Figure \ref{figure:conf_mat_BRF}).}
    \label{fig:conf_mat_nothier}
\end{figure}

The first model tested for the ALeRCE light curve classifier was a simple one-level RF model with 15 classes, implemented using the \texttt{imbalanced-learn} Python package. This model uses 500 trees, maximum depth trees, and maximum number of features equal to the square root of the total number of features. Figure \ref{fig:conf_mat_nothier} shows the confusion matrix obtained by this model. The precision, recall and F1-score obtained are 0.49, 0.68, and 0.50, respectively. Clearly this model has a lower performance compared to the BRF classifier. 

\section{Further analysis of classes with anomalous recall curves}

\subsection{The particular case of AGN, QSO and Blazar}\label{qso_agn}

In this work we present the first attempt to separate different types of active galactic nuclei according to their variability properties. As we mentioned in Section \ref{taxonomy}, we separate active galactic nuclei in the following way:

\begin{itemize}

\item AGN: type 1 Seyfert galaxies (i.e., active galactic nuclei whose emission is dominated by the host galaxy), selected from MILLIQUAS (broad type ``A''), and from \cite{Oh15}. 

\item QSO: type 1 core-dominated active galactic nuclei (i.e., active galactic nuclei whose emission is dominated by their active nuclei), selected from MILLIQUAS (broad type ``Q'').

\item Blazar: BL Lac objects and Flat Spectrum Radio Quasars (FSRQ), selected from ROMABZCAT and MILLIQUAS.

\end{itemize}

\begin{figure*}[htbp]
\begin{center}
\begin{tabular}{cc}
  \includegraphics[scale=0.5]{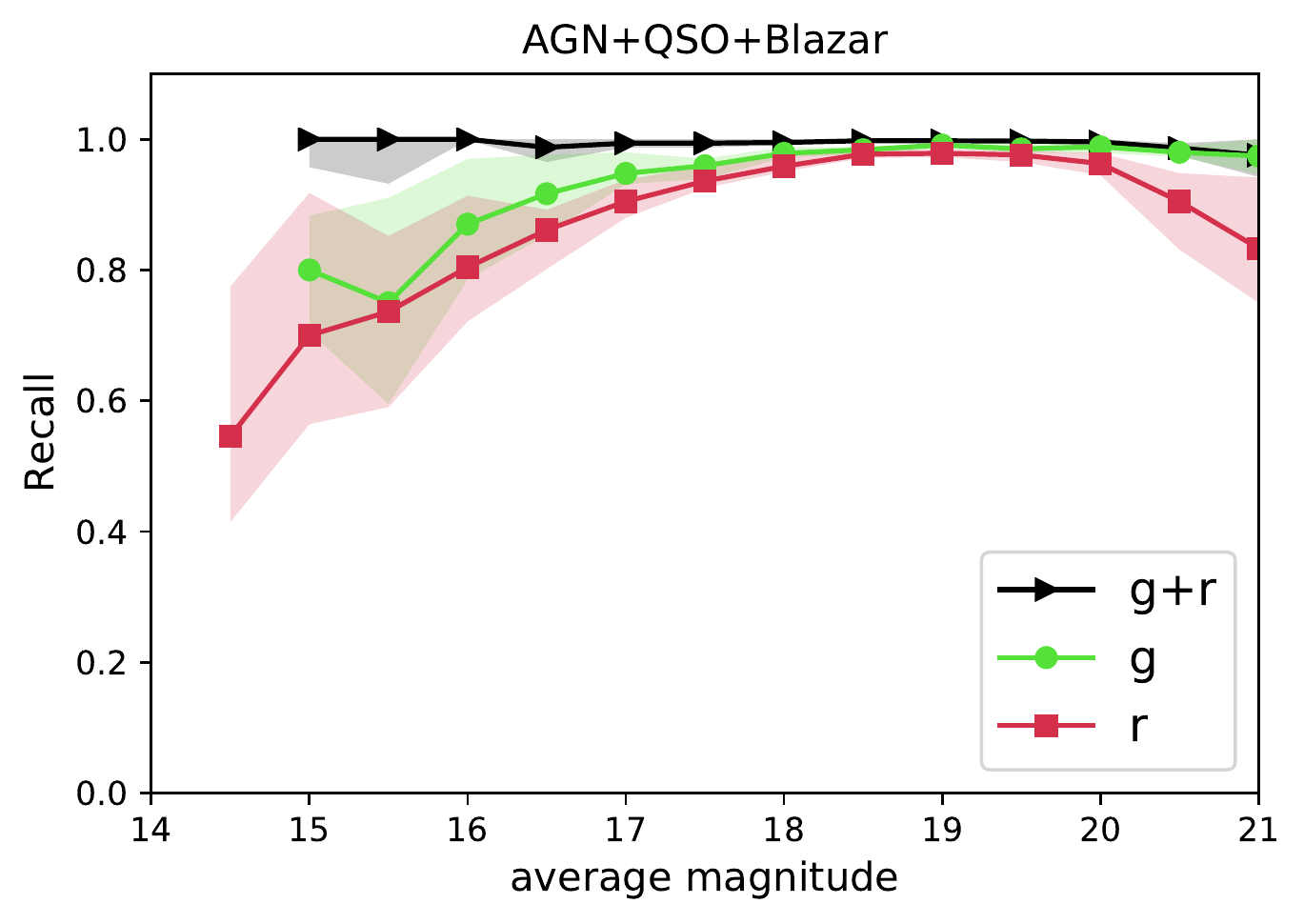} & \includegraphics[scale=0.5]{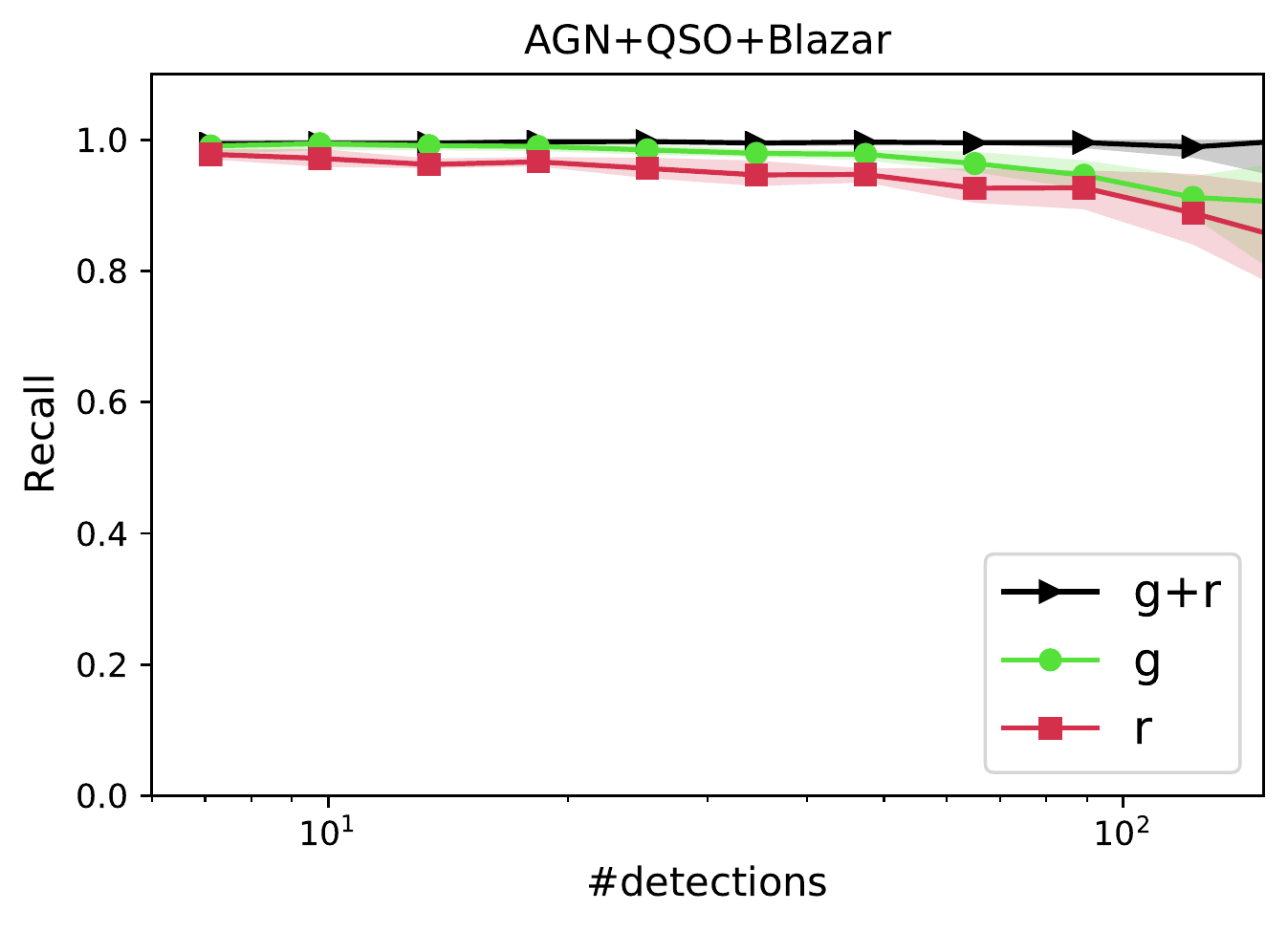} \\ 

\end{tabular}
\caption{Similar to Figures \ref{figure:recall_mag} and \ref{figure:recall_ndet}, but treating  AGNs, QSOs, and Blazars as a single class. When grouped together, the recall is much better behaved than for any individual active galactic nuclei class, highlighting remaining difficulties distinguishing between these classes. \label{figure:recall_for_agn}}
\end{center}
\end{figure*}

In Figure \ref{figure:recall_ndet} we showed the recall curves for each class as a function of the number of detections, and we could observe that these curves decrease for QSOs and AGNs when more detections are available, particularly in the $r$ band. These results are puzzling, considering that we would normally expect to improve the classification of a given class as more detections are included in the light curves. In order to better understand the origin of these results, we show in Figure \ref{figure:recall_for_agn} the recall curves as a function of average magnitude and number of detections for AGNs, QSOs and Blazars grouped as a single class. In this case, the recall curves are around 0.8 and 1.0 for every bin of magnitude (specially for $g>16$ or $r>16$) and number of detections. From this we can infer that the light curve classifier has a very good performance selecting active galactic nuclei as a single class, but  some issues still remain regarding the separation of AGNs, QSOs, and Blazars. There are two main explanations for these results: a) the method cannot properly separate QSOs from AGNs and Blazars, or b) there are sources in the labeled set with incorrect labels.  

\begin{figure*}[htbp]
\begin{center}

 \includegraphics[scale=0.53]{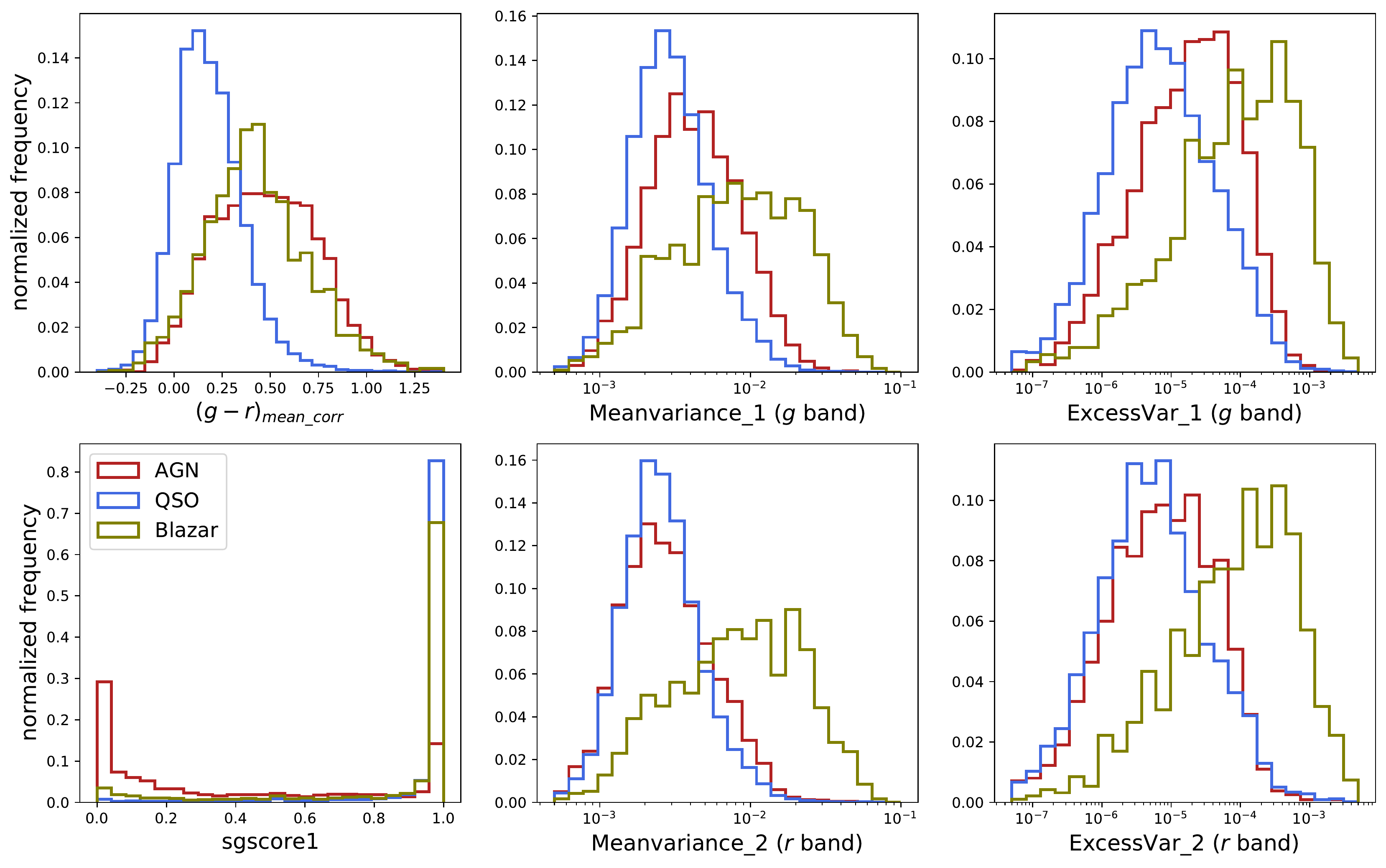}

\caption{Distribution of \texttt{($g$-$r$)\_mean\_corr}, \texttt{Meanvariance\_1}, \texttt{ExcessVar\_1}, \texttt{sgscore1}, \texttt{Meanvariance\_2}, and \texttt{ExcessVar\_2}, for QSOs (blue), AGNs (red), and Blazars (yellow) classes from the labeled set. These features (among others) can be used to separate these classes of active galactic nuclei. Subscripts ``\_1'' and ``\_2'' refers respectively to $g$ and $r$ bands. \label{figure:agn_sep}}
\end{center}
\end{figure*}

A possible way to explore how well the light curve classifier can discriminate among AGNs, QSOs, and Blazars, is to check whether the features available in the light curve classifier can separate these three populations. Figure \ref{figure:agn_sep} shows six different features used by the classifier, \texttt{($g$-$r$)\_mean\_corr}, \texttt{Meanvariance\_1}, \texttt{ExcessVar\_1}, \texttt{sgscore1}, \texttt{Meanvariance\_2}, and \texttt{ExcessVar\_2}, for QSOs, AGNs, and Blazars from the labeled set (most of these features are in the top 30-ranked features shown in Table \ref{table:features_ranking}). From the figure we can see that these three classes have different color distributions, different morphologies, and also different variability properties. AGNs and Blazars tend to be redder than QSOs (see \texttt{($g$-$r$)\_mean\_corr}), Blazars and AGNs tend to have larger amplitudes (see \texttt{Meanvariance\_1} and \texttt{ExcessVar\_1}), and AGNs tend to have more extended morphologies compared to QSOs and Blazars. These are just some examples of features that can be used to separate the three classes above mentioned. After a visual inspection of the feature distribution of AGNs, QSOs, and Blazars, we found that more than 30 features can be used to separate them, including for instance \texttt{PercentileAmplitude}, \texttt{Q31}, \texttt{GP\_DRW\_sigma}, \texttt{GP\_DRW\_tau}, \texttt{MHPS\_low}, \texttt{MHPS\_high}, \texttt{SF\_ML\_amplitude}, among others. From this we can infer that the light curve classifier should be able to separate these three populations. 

In addition, Figure \ref{figure:agn_sep} highlights that the $g$ band features seem to separate better the AGN and QSO classes compared to same features in the $r$ band. This behavior is also seen in other features, like \texttt{PercentileAmplitude}, \texttt{GP\_DRW\_sigma}, \texttt{Std}, among other features related with the amplitude of the variability. These differences might be produced by the combined effect of having a higher contamination from the host in the $r$ band and intrinsically lower amplitude of the variability in the $r$ band, due to the well known anti-correlation between amplitude of the variability and the wavelength of emission (see \citealt{Sanchez17} and references therein). From this, we can understand the differences observed in the $g$ and $r$ band recall curves of AGNs and QSOs shown in Figure \ref{figure:recall_ndet}, which should be produced by these differences in the features distributions.

On the other hand, one key source of confusion between AGN and QSO is the definition of this division criteria in the labeled set.  MILLIQUAS uses a luminosity-based division to separate AGNs and QSOs (see Section 5 of \citealt{Flesch15}), while the variability amplitude features are also affected by the ratio between core and host luminosities. Our variability-based criteria appears to separate host-dominated from core-dominated sources, in rough correspondence with low and high core luminosities, respectively.
In addition, it is hard to make strict cuts to separate the AGN, QSO, and Blazar classes, considering their observational properties (e.g., luminosity, redshift, orientation). In particular, the dividing line between AGN and QSO has never been defined properly (e.g., \citealt{Netzer13,Padovani17}), and thus different catalogs will provide different classifications for the same object. Note that since the redshift of the sources in the unlabeled set is generally unknown, luminosity cannot be used as a feature by the classifier. Our aim is to separate host-dominated from core-dominated sources to favor the identification of different populations of active galactic nuclei by the light curve classifier, and use the AGN and QSO classifications as representative of these classes in the extremes of the luminosity distribution, but expect that there will be strong mixing of these classes close to the dividing luminosity used in the labeled set.

In order to see whether there are sources in the labeled set with incorrect QSO or AGN labels, we crossmatched our labeled set with the SIMBAD database. There are 26168 QSOs in the labeled set (all obtained from MILLIQUAS), and 1590 of them are classified as Seyfert or AGN by SIMBAD (6\%), with 1580 having a reported redshift $<1$. On the other hand, there are 4667 AGNs in the labeled set, and 830 (17\%) of them are classified as QSO in SIMBAD. Therefore, there are 2420 sources in the labeled set with inconsistent classification in MILLIQUAS and SIMBAD. The light curve classifier classifies 920 of these sources as QSO and 1319 as AGN. 

To understand better this discrepancy in the classification of some sources, we crossmatched our labeled set with the catalog of spectral properties of 
Quasars from SDSS DR14 provided by \cite{Rakshit20}, as well as with the catalog of \cite{Oh15}, in order to obtain bolometric luminosities (L$_{\text{bol}}$), BH masses (BH$_{\text{mass}}$), Eddington ratios (L/L$_{\text{Edd}}$) and redshifts, for our sample of QSOs and AGNs. We excluded Blazars from this analysis, since they are not properly identified and characterized in these catalogs. 

\begin{figure*}[htbp]
\begin{center}

 \includegraphics[scale=0.53]{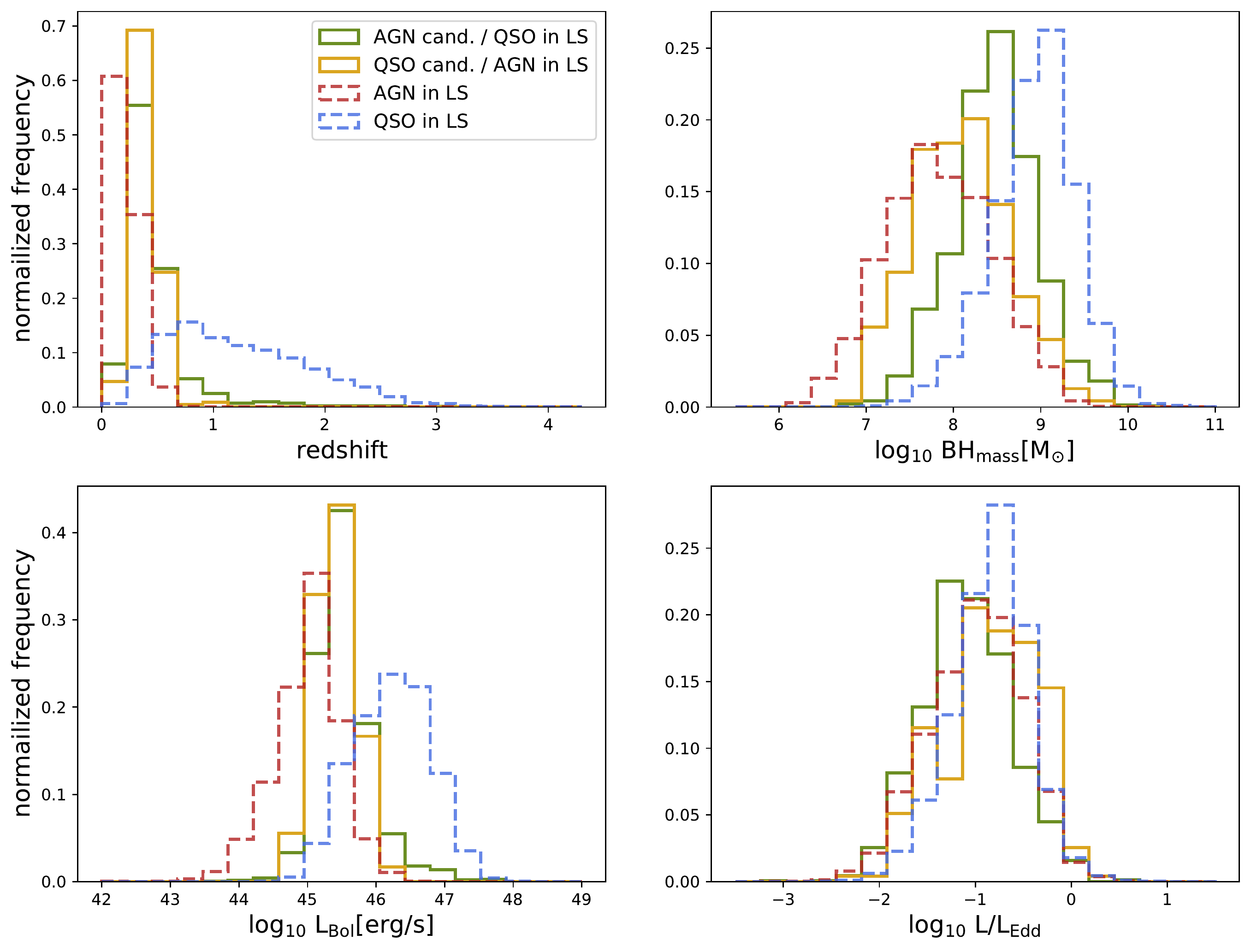}

\caption{Distribution of redshift, BH$_{\text{mass}}$, L$_{\text{bol}}$, and L/L$_{\text{Edd}}$ of AGNs (red dashed line) and QSOs (blue dashed line) from the labeled set, and missclassified AGN (green solid line) and QSO (yellow solid line) candidates. The spectral properties where obtained from catalogs that make use of SDSS spectra \citep{Oh15,Rakshit20} \label{figure:sdss_dist}}
\end{center}
\end{figure*}

We show in Figure \ref{figure:sdss_dist} the distribution of these spectral properties for AGNs and QSOs from the labeled set and for misclassified sources (i.e., AGNs from the labeled set classified as QSO, and vice versa). It is clear from the figure that AGNs and QSOs from the labeled set have spectral property distributions that are not strictly separated. From  this we can conclude that the luminosity-based division performed by MILLIQUAS is not physically so meaningful. One issue, for instance, is to what extent the host contributes to the total magnitude used to classify the source as AGN or QSO. In addition, Figure \ref{figure:sdss_dist} shows that the misclasified sources have similar distributions of redshift and luminosity. Moreover, these distributions lie exactly in the range of values where the properties of AGNs and QSOs from the labeled set overlap (redshift$\sim0.5$ and L$_{\text{bol}}\sim 10
^{45.5}$ [erg/s]).

From these results, we can infer that the decrease in the recall as a function of the number of detections observed for QSOs and AGNs is produced by the discrepancy in the QSO/AGN classification obtained from different catalogs, which is produced by the difficulty of making strict cuts to separate AGNs and QSOs. When more epochs are available, it is easier for the light curve classifier to perform a correct variability-based classification, and therefore, identify an original inconsistent classification provided by a given catalog. We propose that by using the variability properties of active galactic nuclei we can more easily separate them as core-dominated or host-dominated, or in other words, as bright QSOs or low luminosity AGNs. 

\subsection{The particular case of RRL}\label{RRL_case}

\begin{figure*}[htbp]
\begin{center}
\begin{tabular}{cc}
\includegraphics[scale=0.38]{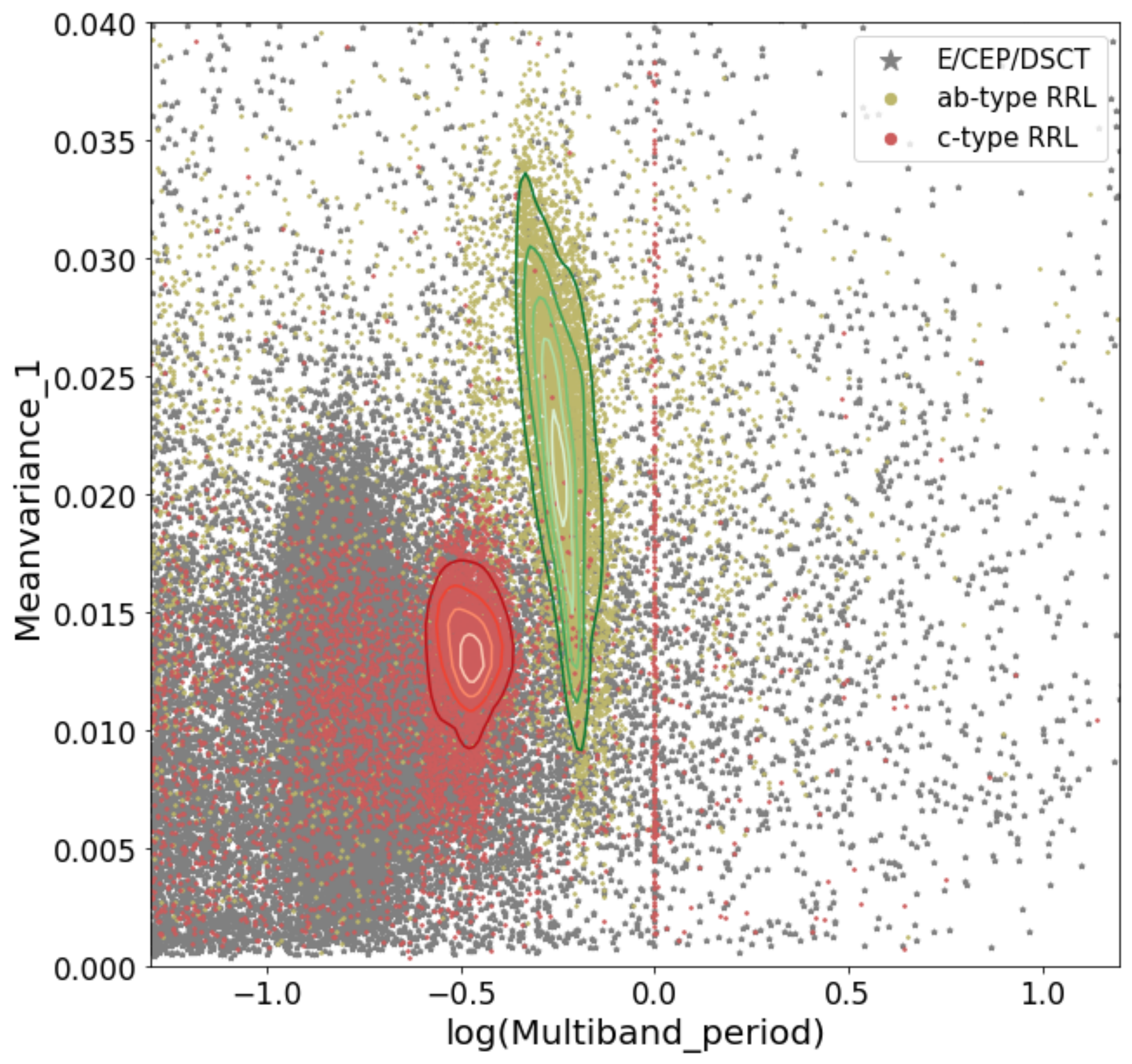} &
  \includegraphics[scale=0.38]{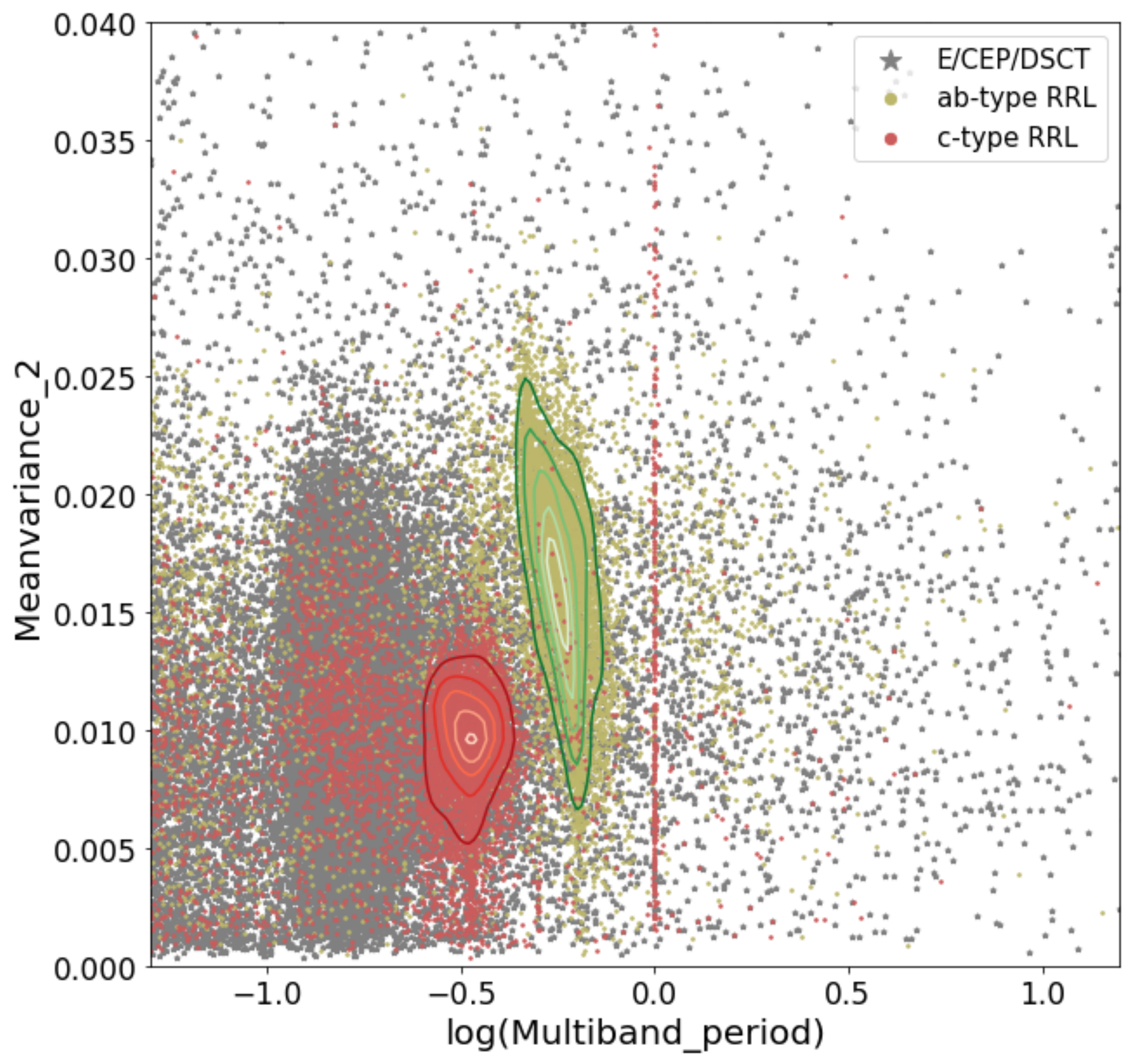} \\
\end{tabular}
\caption{Logarithm of the Multiband period versus the \texttt{Meanvariance\_1} ($g$ band, left) and the  \texttt{Meanvariance\_2} ($r$ band, right), for bright RRLs (mean $g<16$) split according to their sub-classes `ab' (green) and `c' (red), and for E, CEP, and DSCT (grey). We show a zoom in the area where most of the RRLs lie. The contours show the density of points of each RRL class. \label{figure:period_vs_meanvariance}}
\end{center}
\end{figure*}

In Section \ref{performance} we claimed that the low recall values obtained when only the $g$ band photometry is used for bright RRL can be explained by the differences in the variability features of the RRL sub-types. Figure \ref{figure:period_vs_meanvariance} shows the \texttt{Multiband\_period} versus the \texttt{Meanvariance} measured in the $g$ and $r$ bands, for bright RRLs (mean $g<16$) split into their sub-classes `ab' and `c', and for E, CEP, and DSCT (grouped as a single class). From the figure we can notice that ab-type RRL tend to have larger \texttt{Meanvariance} in the $g$ band, which helps to distinguish them from E/CEP/DSCT, while c-type RRL have values of \texttt{Meanvariance} in both bands similar to those of E/CEP/DSCT, which makes it difficult to tell them apart. The RRL class in our labeled set is dominated by ab-type RRL ($\sim$80\%), and thus the light curve classifier identifies those more easily. However, for sources with $g\leq15$, the fraction of ab-type RRL decreases to 64\%, which explains the low recall values obtained for this regime of brightness. 

In spite of all this, the fraction of RRL with only $g$ band photometry is low (see Figure \ref{figure:band_dist}), and thus, the low performance obtained when we only consider the $g$ band features for bright RRL does not substantially affect the presented results. As a comparison, when both bands are available for bright RRL, we classify \%87.5 of them as RRL, but when we hide the $r$ band features, leaving only the $g$ band features, we recover only \%17.2 of the bright RRL.

\subsection{The particular case of CV/Nova}\label{CV_case}

In Section \ref{performance} we show that the recall curves of the CV/Nova class are close to zero when only the $g$ band is available. We noticed that in this case most of the CV/Novae are misclassified as periodic by the top level of the classifier (i.e., the class with the highest probability is periodic), and that most of them are misclassified as CEP or RRL by the bottom level. For these sources, the second or third classes with the highest probabilities in the bottom level is CV/Nova, and the probability of the CV/Nova class returned by the Stochastic classifier [$P_S(CV/Nova)$] is larger than the probability of being CEP or RRL returned by the Periodic classifier [$P_P(CEP)$ or $P_P(RRL)$]. This confusion of the CV/Nova class with the periodic classes can also be seen in Figures \ref{figure:conf_mat_BRF} and \ref{figure:conf_max_deployed}. Therefore, the problem is produced in the first level of the BRF model. 

\begin{figure*}[htbp]
\begin{center}
\begin{tabular}{cc}
\includegraphics[scale=0.35]{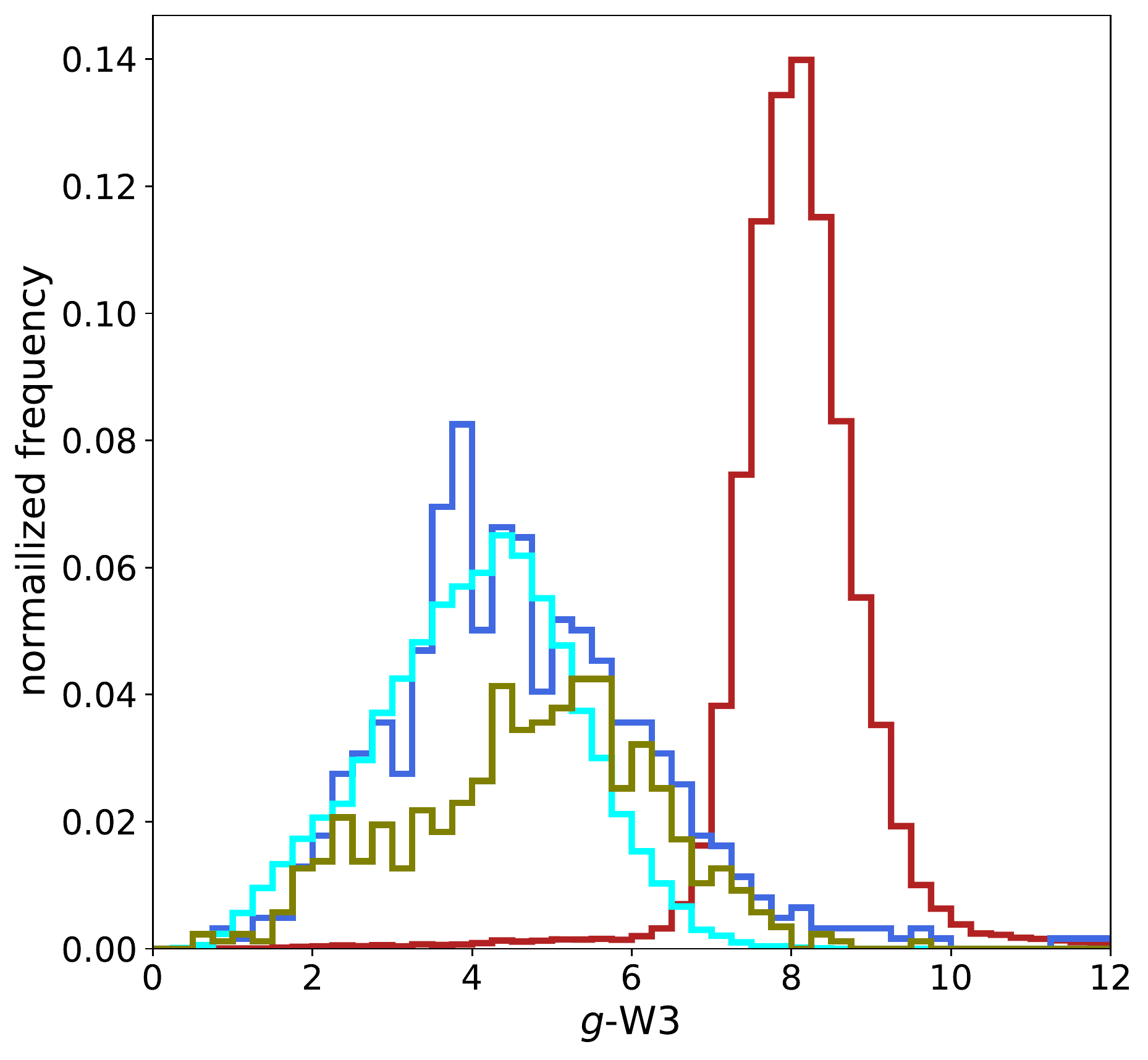} &
  \includegraphics[scale=0.35]{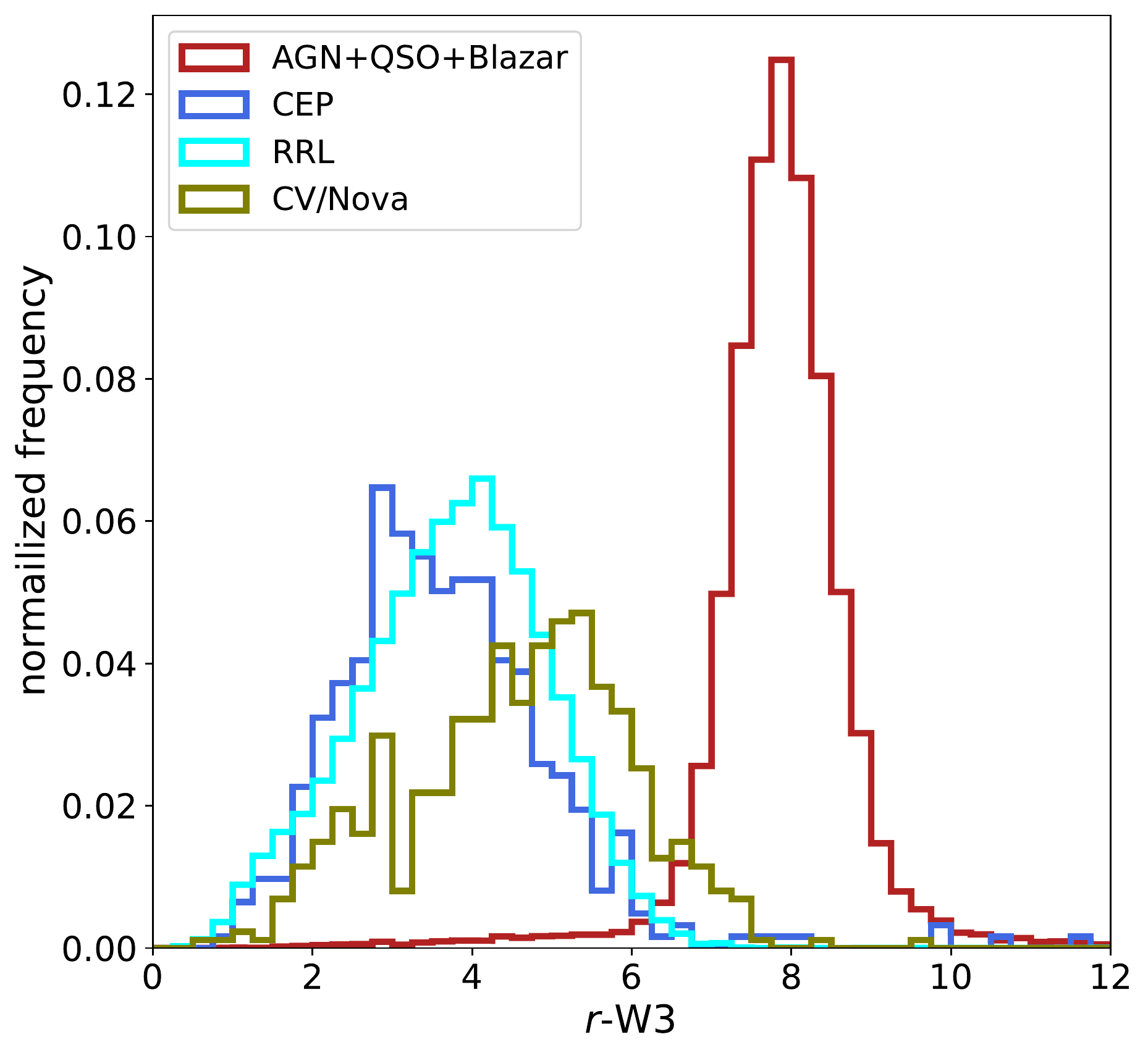} \\
\end{tabular}
\caption{Normalized $g-$W3 (left) and $r-$W3 (right) distributions, for active galaxies (red), CEP (blue), RRL (cyan), and CV/Nova (yellow). \label{figure:cv_nova}}
\end{center}
\end{figure*}

We inspected the feature distributions in both bands for the CV/Nova, CEP, and RRL classes, in order to understand why the results obtained when only the $g$ band is available are so different from the results obtained when only the $r$ band is available. We did not find large differences among the $g$ and $r$ feature distributions, except for the $g-$W3 and $r-$W3 colors, where the confusion of CV/Nova with CEP and RRL is larger for the case of $g-$W3, as can be seen in Figure \ref{figure:cv_nova}. However, these differences are not large enough to totally explain the low recall curves obtained for the CV/Nova class. Another possible explanation is that there is a large fraction of Cepheids with photometry only in the $g$ band (see Figure \ref{figure:band_dist}). This may be playing a role in the obtained results, since for the BRF model a lack of features in the $r$ band and AllWISE+ZTF colors similar to those of the periodic classes could imply that the source is periodic.

Despite all this, more than 96\% of the CV/Novae have photometry available in both bands, thus the low recall obtained for the $g$ band is not a relevant issue. We plan to explore in future work better ways to classify CV/Novae. In particular, we will focus our efforts on classifying sub-classes of CVs and Novae. 

\section{Sky distribution of the extragalactic candidates}\label{gal_coords_extragalactics}

\begin{figure}
    \centering
    \includegraphics[scale=0.35]{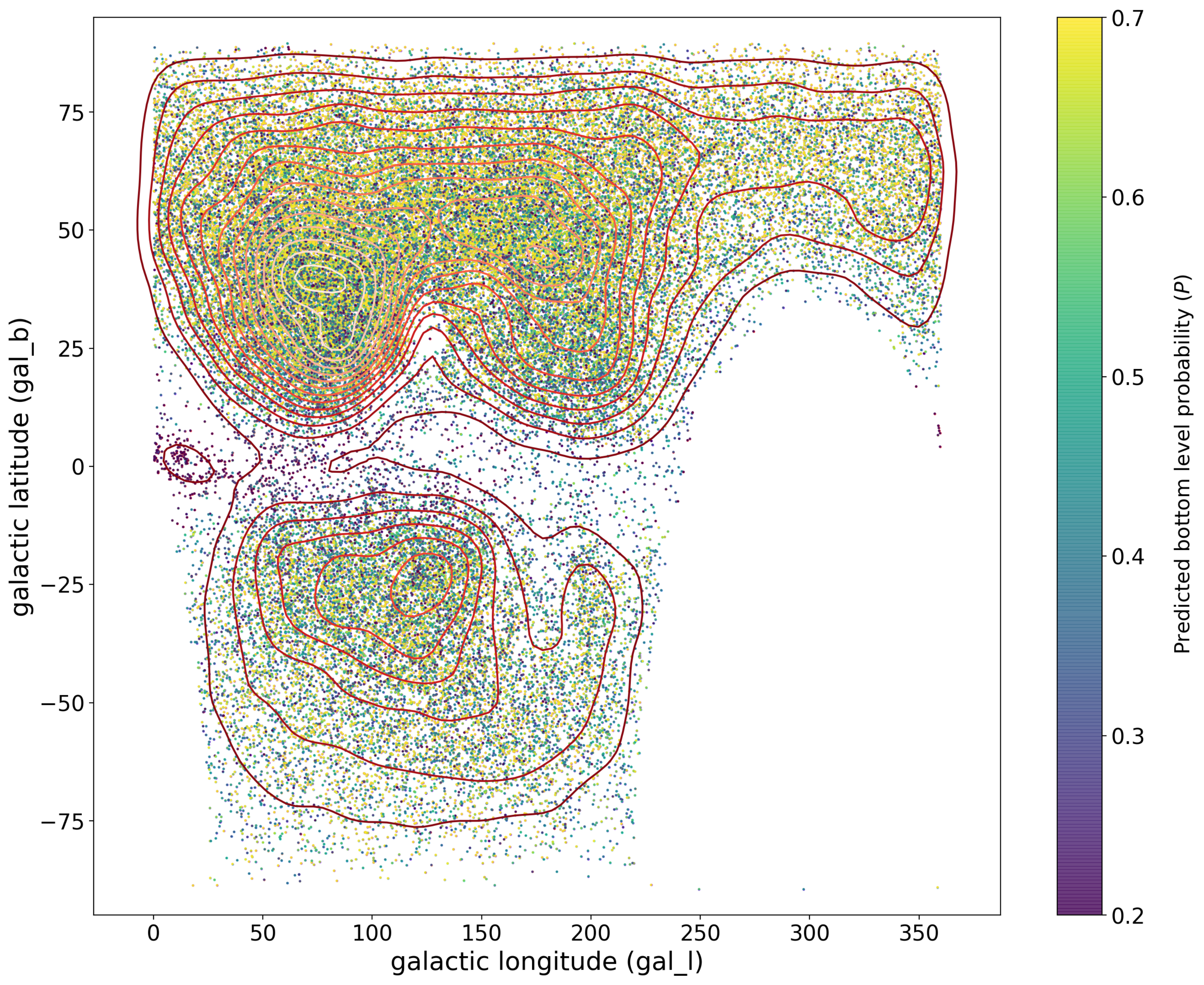}
    \caption{Galactic latitude (gal\_b, in degrees) versus Galactic longitude (gal\_l, in degrees) for extragalactic candidates (QSO, AGN, Blazar, SNIa, SNIbc, SNII, and SLSN classes). The contours show the density of points in the plot. The bottom level probability computed by the deployed BRF classifier are color-coded according to the color bar to the right.}
    \label{fig:extraga_gal_coords}
\end{figure}

Figure \ref{fig:extraga_gal_coords} shows the sky  distribution (in Galactic  coordinates) of extragalactic candidates (QSO, AGN, Blazar, SNIa, SNIbc, SNII, and SLSN). It is expected that only a few extragalactic candidates are observed in the Galactic plane, which is confirmed in Figure \ref{fig:extraga_gal_coords}. 

\bibliography{bibliography.bib}
\bibliographystyle{aasjournal}



\end{document}